\documentclass[fleqn,usenatbib]{mnras}

\usepackage{fix-cm}
\usepackage{newtxtext,newtxmath}
\usepackage{soul}
\usepackage{subcaption}

\usepackage[T1]{fontenc}

\DeclareRobustCommand{\VAN}[3]{#2}
\let\VANthebibliography\thebibliography
\def\thebibliography{\DeclareRobustCommand{\VAN}[3]{##3}\VANthebibliography}

\usepackage{graphicx}	
\usepackage{amsmath}	
\usepackage{orcidlink}

\title[Plasma Instabilities in Transients]{Plasma Instabilities Dominate Radioactive Transients Magnetic Fields: \\
The self-confinement of leptons in Type Ia and Core-Collapse Supernovae, and Kilonovae}

\author[D. D. Desai et al.]{
Dhvanil~D.~Desai\orcidlink{0000-0002-2164-859X}$^{1}$\thanks{E-mail: dddesai@hawaii.edu},
Colby~C.~Haggerty\orcidlink{0000-0002-2160-7288}$^{1}$,
Benjamin~J.~Shappee\orcidlink{0000-0003-4631-1149}$^{1}$,
Michael~A.~Tucker\orcidlink{0000-0002-2471-8442}$^{2,3}$,
\newauthor
Zachary~Davis\orcidlink{0000-0002-0959-9991}$^{1}$,
Chris~Ashall\orcidlink{0000-0002-5221-7557}$^{1}$,
Laura~Chomiuk\orcidlink{0000-0002-8400-3705}$^{4}$,
Keyan~Gootkin\orcidlink{0000-0003-0922-138X}$^{1}$,
Damiano~Caprioli\orcidlink{0000-0003-0939-8775}$^{5,6}$,
\newauthor
Antoine~Bret$^{7,8}$,
Hayk~Hakobyan\orcidlink{0000-0001-8939-6862}$^{9,10}$ \\
$^{1}$ Institute for Astronomy, University of Hawaii, 2680 Woodlawn Drive, Honolulu, HI 96822 \\
$^{2}$ Center for Cosmology and AstroParticale Physics, 191 W Woodruff Ave, Columbus, OH 43210 \\
$^{3}$ Department of Astronomy, The Ohio State University, 140 W 18th Ave, Columbus, OH 43210 \\
$^{4}$ Center for Data Intensive and Time Domain Astronomy, Department of Physics and Astronomy, Michigan State University, East Lansing, MI 48824 \\
$^{5}$ Department of Astronomy and Astrophysics, University of Chicago, 5640 S Ellis Ave, Chicago, IL 60637 \\ 
$^{6}$ Enrico Fermi Institute, University of Chicago, 5640 S Ellis Ave, Chicago, IL 60637 \\
$^{7}$ ETSI Industriales, Universidad de Castilla-La Mancha, 13071, Ciudad Real, Spain \\
$^{8}$ Instituto de Investigaciones Energ\'eticas y Aplicaciones Industriales, Campus Universitario de Ciudad Real, 13071, Ciudad Real, Spain \\
$^{9}$ Computational Sciences Department, Princeton Plasma Physics Laboratory (PPPL), Princeton, NJ 08540 \\
$^{10}$ Physics Department \& Columbia Astrophysics Laboratory, Columbia University, 538 West 120th Street, New York, NY 10027
}

\date{Accepted XXX. Received YYY; in original form ZZZ}

\pubyear{\the\year{}}

\begin{document}
\label{firstpage}
\pagerange{\pageref{firstpage}--\pageref{lastpage}}
\maketitle

\begin{abstract}
The light curves of radioactive transients, such as supernovae and kilonovae, are powered by the decay of radioisotopes, which release high-energy leptons through $\beta^+$ and $\beta^-$ decays. These leptons deposit energy into the expanding ejecta. As the ejecta density decreases during expansion, the plasma becomes collisionless, with particle motion governed by electromagnetic forces. In such environments, strong or turbulent magnetic fields are thought to confine particles, though the origin of these fields and the confinement mechanism have remained unclear. Using fully kinetic particle-in-cell (PIC) simulations, we demonstrate that plasma instabilities can naturally confine high-energy leptons. These leptons generate magnetic fields through plasma streaming instabilities, even in the absence of pre-existing fields. The self-generated magnetic fields slow lepton diffusion, enabling confinement and transferring energy to thermal electrons and ions. Our results naturally explain the positron trapping inferred from late-time observations of thermonuclear and core-collapse supernovae. Furthermore, they suggest potential implications for electron dynamics in the ejecta of kilonovae. We also estimate synchrotron radio luminosities from positrons for Type Ia supernovae and find that such emission could only be detectable with next-generation radio observatories from a Galactic or local-group supernova in an environment without any circumstellar material.
\end{abstract}




\section{Introduction} \label{sec:intro}
Radioactive decay powers some of the most luminous transients in the universe, including Type Ia supernovae (SNe~Ia), core-collapse supernovae (CC~SNe), and kilonovae (KNe) \citep[e.g.,][]{Arnett82,WoosleyWeaver86,Arnett96,Metzger10,Roberts11,Seitenzahl13,Barnes16}. In SNe, the decay chain $^{56}$Ni~$\rightarrow$~$^{56}$Co~$\rightarrow$~$^{56}$Fe generates the bulk of the radioactive luminosity and gives the light curve its shape \citep[e.g.,][]{Arnett82}. Similarly, in KNe, the decay of $r$-process elements powers their distinctive light curves \citep[e.g.,][]{Metzger10,BarnesKasen13,TanakaHotokezaka13,Grossman14}. The radioactive decay heats the ejecta by emission of $\gamma$-rays and high-energy $\beta$-decays producing electrons (in KNe) and positrons (in SNe).

A challenge in understanding these radioactive transients lies in determining the mechanism of lepton confinement and energy deposition within the expanding ejecta. As the ejecta of these transients expand and cool, collisions become too infrequent to modify the trajectories of leptons. In the presence of a weak or no magnetic field, most high-energy particles would escape once the ejecta becomes optically thin to collisions. In a collisionless plasma, magnetic fields dictate the diffusion length of charged particles, potentially trapping them based on the structure and strength of the fields \citep[][]{Colgate80,ChanLingenfelter93,Milne99}. 

Some degree of positron confinement is required to explain the late-time evolution of SNe. For SNe~Ia, the late-time light curves show no evidence for radioactive-decay energy escaping the ejecta \citep[e.g.,][]{Stritzinger07,Leloudas09,Graur16,Shappee17b,Graur19,Chen23} even out to $\sim2400$ days \citep{Tucker22} and their nebular spectra requires energy deposited at low velocities, suggesting local positron confinement \citep{Ashall24}. Similar arguments are made for CC~SNe \citep[e.g.,][]{Jerkstrand12,Jerkstrand15,Dessart20,Dessart21}, but interpretation is complicated by the diverse landscape of plausible progenitors and the increased likelihood of emission from SN ejecta interacting with circumstellar material (CSM). For KNe, current light curve models require lepton confinement \citep[e.g.,][]{KasenBarnes19} but we note that the theoretical landscape for KNe remains broad and relatively untested.

SNe~Ia, in particular, are ideal environments for understanding lepton propagation in radioactively-powered ejecta. Despite the observational necessity, the mechanism behind the continued positron confinement in SNe~Ia is unclear. Previous studies have suggested that extremely strong pre-existing magnetic fields ($> 10^6\,\mathrm{G}$) can explain local positron trapping at late times \citep[$>300$ days;][]{Milne99,Penney14,Hristov21}, but implies initial white dwarf (WD) surface fields of $>10^9\,\mathrm{G}$, surpassing all observed field strengths \citep[$10^3 - 10^9\,\mathrm{G}$;][]{Schmidt03,Kulebi09,Ferrario15}. Another option is the delayed deposition of positron energy, which is predicted to flatten the light curve \citep{KushnirWaxman20}, but has yet to be observationally confirmed.

Plasma streaming instabilities offer a promising solution for positron confinement without the need for strong pre-existing magnetic fields. Plasma streaming instabilities draw energy from the bulk motion of one species and channel it into amplifying electromagnetic perturbations over time, disrupting the plasma's equilibrium and leading to strong and often turbulent fields. These instabilities play a crucial role in the dynamics of high-energy astrophysical phenomena, but are often overlooked in transient simulations. They are instrumental in processes ranging from electron scattering in the solar wind \citep[][]{RobergClark18,Cattell11} and intracluster medium of galaxy clusters \citep[][]{RobergClark18} to forward shock of $\gamma$-ray bursts \citep[][]{Sironi11,Golant25}. Among the most pertinent plasma streaming instabilities arising from a net electron or positron current are filamentation instabilities \citep[e.g.,][]{Weibel59,Bell78,Bell04,bret09,Reville12,Caprioli13,Gupta21}. Such instabilities can grow exponentially from the density perturbations in the anisotropic flow of particles even without an initial magnetic field and saturate when the electron or positron gyro-radius (Larmor radius) becomes comparable to their inertial length. They can efficiently trap energetic particles by generating local magnetic fields that slow down or halt particle diffusion \citep[][]{Riquelme11,Schroer21,Gupta21}.

In this work, we study plasma streaming instabilities in radioactive transients using simulations, taking SNe~Ia as fiducial examples. Using fully kinetic particle-in-cell (PIC) simulations, we demonstrate how these instabilities arise and evolve in the presence of high-energy leptons, leading to self-generated magnetic fields that confine these particles even in the absence of an initial magnetic field. 
In Section~\ref{sec:theory}, we establish the theoretical groundwork of plasma instabilities generated by high-energy leptons. In Section~\ref{sec:PIC}, we describe our PIC simulations and the results. In Section~\ref{sec:implications}, we apply these findings to radioactive transients such as SNe~Ia, CC~SNe, and KNe. Finally, in Section~\ref{sec:conclusion}, we summarize our conclusions.

\section{Linear Theory} \label{sec:theory}
To estimate the impact of plasma instabilities on energetic leptonic confinement in transients, we first need to identify the timescale of the fastest-growing unstable modes. To do this, we use a linear theory model that approximates how energetic leptons might escape from a system. Given the considerable uncertainties of the system, including the properties of the progenitors and explosion mechanisms, we will linearize a reduced set of fluid equations for a toy system with several simplifying assumptions.

We will consider a 3-species plasma made up of the heavy ions from the ejecta, the thermal background electrons, and the energetic positrons generated from radioactive decay. We consider only high-energy positrons for simplicity in this section, but the generalization to all high-energy leptons is straightforward. We anticipate that electron/positron scale fluctuations will likely grow fastest based on the driving species; accordingly, we take the heavy ions to be cold, immobile, and unresponsive to the instability (an assumption that is supported in Section~\ref{subsec:fiducial}). The positrons drift in the $+x$ direction as a cold beam with Lorentz factor $\gamma$ and corresponding bulk velocity $u_+ = c\sqrt{1 - 1/\gamma^2}$, with a number density of $n_p$. The cold background electron population is also drifting in the $+x$ to enforce quasi-neutrality and make sure that there is no initial net current ($\mathbf{J} = q_i n_i u_i - e n_e u_e + e n_p u_p = 0$). We assume that the excess charge from the positrons is balanced by excess background electrons ($n_i = n_e - n_p$), which sets the electron drift speed to $u_e = (n_p/n_e)u_p$. For the radioactive transients we consider, $n_p/n_e$ can vary considerably and so for the simulations we consider a range of $n_p/n_e$ from $0.01$ to $0.2$. We consider a system with no initial magnetic fields, which is assumed to be the configuration in which positron confinement is least likely to occur.

\begin{figure}
    \centering
    \includegraphics[width=\linewidth]{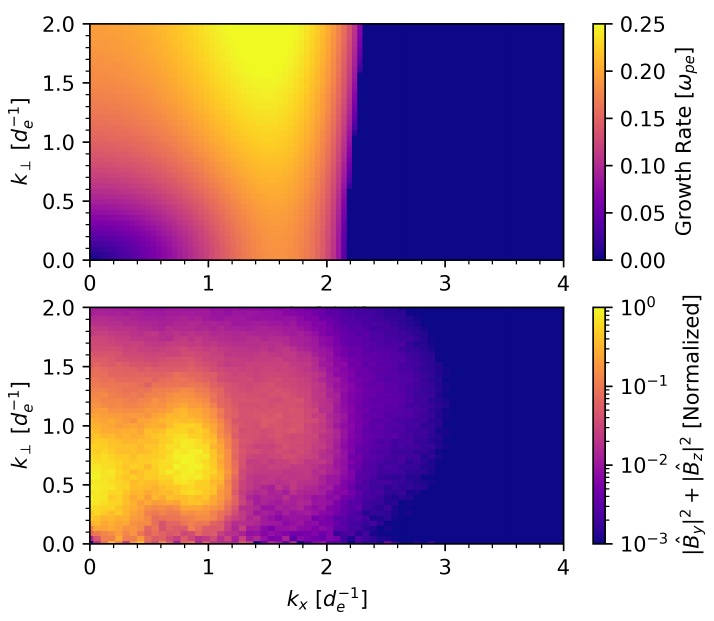}
    \caption{\textit{Top}: The linear theory growth rate of the perpendicular magnetic field ($y$ and $z$) as a function of wave number, $k_x$ (parallel) and $k_{\perp} = \sqrt{k_y^2 + k_z^2}$ for the fiducial simulation in units of inverse of the background electron plasma frequency and inverse inertial length, respectively. \textit{Bottom}: The power spectrum for the magnetic field in the fiducial simulation over the same $k$ range. The power spectrum was calculated at the end of the linear growth phase of the instability ($\sim 35\,\omega_{\mathrm{pe}}^{-1}$).}
\label{fig:fft_modes}
\end{figure}

We linearize a simplified set of equations for the electron and positron equations of motion which are derived from the $0^{\text{th}}$ and $1^{\text{st}}$ moments of the Vlasov equation \citep[][]{Vlasov61}
\begin{equation}
\begin{aligned}
&\frac{\partial n_\alpha}{\partial t} = \mathbf{\nabla} \cdot (n_\alpha \mathbf{u_\alpha}),\\
&\frac{\partial \gamma_\alpha \mathbf{u_\alpha}}{\partial t} + (\mathbf{u_\alpha} \cdot \mathbf{\nabla})\, \gamma_\alpha \mathbf{u_\alpha} 
= \frac{q_\alpha}{m_\alpha}\left ( \mathbf{E} + \frac{u_\alpha}{c}\times \mathbf{B} \right),
\end{aligned}
\end{equation}
where the $\alpha$ subscript corresponds to either the electron or positron species with corresponding charge ($q_\alpha$) and mass ($m_\alpha$). Note that we have neglected the pressure term, which closes the equations and allows us to neglect the energy equation in the calculation. We then linearize these equations and Maxwell's equations by expanding terms to $0^{\text{th}}$ and $1^{\text{st}}$ order, taking $\mathbf{E_0} = \mathbf{B_0} = 0$ and with $0^{\text{th}}$ order flows for the positions and electrons in the $\hat{x}$ direction. We solve for first order perturbations of the form $\propto \exp{i(k_x x + k_\perp z - \omega t) }$, replacing $\mathbf{\nabla} \rightarrow i\mathbf{k}$ and $\partial/\partial t \rightarrow -i\omega$. The equations are then rearranged to only depend on $\mathbf{E_1}$, yielding the dispersion matrix, and the corresponding dispersion equation. The equations are normalized\footnote{This normalization is the same as the code used for simulations in Section~\ref{sec:PIC}, thus allowing for easier comparison.} with velocities in units of the speed of light ($c$), times in units of the inverse electron plasma frequency for the number density of the background ions $\omega_{pe} \equiv \sqrt{4\pi (n_e - n_p) e^2/m_e}$, and lengths in units of the corresponding electron inertial length $d_e \equiv c/\omega_{pe}$. The plasma frequency $\omega_{pe}$ represents the natural oscillation frequency of electrons in a plasma due to electrostatic interactions, and $d_e$ is the characteristic length scale over which electrons can move without being influenced by the electric field. The dimensionless equations are then computed using the procedure described in \citet{BretPoP2010} and the code described in \citet{BretCPC}, and the imaginary/unstable modes are determined as a function of $k_x$ and $k_\perp$.

An example solution to the dispersion relation with $\gamma_p = 2$ and $n_p = 0.2 n_e$, our fiducial model for SNe~Ia (see Section~\ref{subsubsec:setup}), is shown in the top panel of Figure~\ref{fig:fft_modes}. The figure shows the unstable mode growth rate as a function of $k_x$ (the direction of the electron/positron flow) and $k_\perp$. It shows that there are many unstable modes, with the perpendicular and oblique modes growing the fastest, with $k_{x} d_e \sim 1.5$ and $k_{\perp} d_e \sim 3 $. The instability is suppressed on a smaller scales, parallel to the flow direction, with the instability turning off above $k_x d_e > 2$. The fastest growing modes should grow on e-folding times comparable to $\sim 4/\omega_{pe}$. While this time scale will vary dramatically with homologous expansion of the ejecta (as $\omega_{pe} \propto \sqrt{n_e}$), it will be on the order of $\sim 10^{-9}\,\mathrm{s}$, near the $^{56}$Ni peak in SNe~Ia with $n_i \sim 10^{9}\,\mathrm{cm^{-3}}$. This timescale is short relative to both the expansion time ($\sim$ days) and the positron collision time, which is typically orders of magnitude greater than the inverse electron plasma frequency within a few hours of explosion. Thus, we expect instabilities to develop on short timescales in these systems.

In the linear model, modes with larger $k_\perp$ grow faster for the relevant parameter space. This is due to the assumptions that went into the linearized equations, particularly that all of the distributions were cold and that the pressure term could be ignored, which omits the damping that can occur at higher wave numbers. While this result is idealized, we will show in the next section that the modes' approximate obliquity and growth rates agree well with simulations. While the analytical model predicts that instabilities grow, it cannot determine the amplitude of the saturated magnetic field, which is ultimately responsible for energetic lepton confinement. Investigating the non-linear process requires fully kinetic plasma simulations.

\section{Particle-in-Cell Simulations} \label{sec:PIC}
To investigate the generation and saturation of instabilities arising from the flow of relativistic leptons, we use \texttt{tristan-mp v2}\footnote{Code on GitHub: \\\url{https://github.com/PrincetonUniversity/tristan-mp-v2}; \\ Documentation: \\\url{https://princetonuniversity.github.io/tristan-v2/}} \citep[][]{Spitkovsky05,tristanv2}, a publicly available, fully parallelized PIC code specifically designed for astrophysical plasma simulations. The PIC method solves the coupled Vlasov-Maxwell system of equations for a collisionless plasma. Maxwell's equations 
\begin{equation} \label{eq:maxwell}
\begin{aligned}
&\nabla \cdot \mathbf{E}=4 \pi \rho, \;& \nabla \times \mathbf{E}&=-\frac{1}{c} \frac{\partial \mathbf{B}}{\partial t}, \\
&\nabla \cdot \mathbf{B}=0, \;& \nabla \times \mathbf{B}&=\frac{4 \pi \mathbf{J}}{c}+\frac{1}{c} \frac{\partial \mathbf{E}}{\partial t} ,
\end{aligned}
\end{equation}
relate the evolution of electric ($\mathbf{E}$) and magnetic ($\mathbf{B}$) fields to the charge density ($\rho$) and current density ($\mathbf{J}$). The divergence equations are used as constraints on the system.
For a given species $j$ (such as positrons or electrons) in a collisionless plasma, the Vlasov equation 
\begin{equation} \label{eq:vlasov}
\frac{\partial f_j}{\partial t} + \mathbf{v}_j \cdot \nabla f_j - \frac{q_j}{m_j}\left(\mathbf{E}+\frac{\mathbf{v}_j}{c} \times \mathbf{B}\right) \cdot \nabla_{\mathbf{v}} f_j = 0 
\end{equation}
describes the time evolution of the distribution function $f_j$ in phase space. Here, $q_j$ and $m_j$ are the charge and mass of the particles, and $\mathbf{v}_j$ is their velocity vector. The term $\mathbf{v}_j \cdot \nabla f_j$ describes advection in real space, indicating changes in the distribution function as the particles move. The term $- \frac{q_j}{m_j}\left(\mathbf{E}+\frac{\mathbf{v}_j}{c} \times \mathbf{B}\right) \cdot \nabla_{\mathbf{v}} f_j$ describes advection in velocity space due to particle acceleration by Lorentz forces.

PIC simulations iteratively advance the system in time. Rather than simulating the full distribution function, a PIC code samples macro-particles from the phase-space distribution. These macro-particles are placed on a Cartesian grid, and their positions and velocities are updated using the Vlasov equation (Eq.~\ref{eq:vlasov}). The updated positions and velocities are then interpolated onto a grid to determine average charge and current densities. These densities are used to update the electric and magnetic fields using Maxwell's equations (Eq.~\ref{eq:maxwell}). The updated fields, in turn, determine the new positions and velocities of the macro-particles.

PIC simulations offer significant advantages over traditional magnetohydrodynamic (MHD) simulations, particularly in scenarios involving kinetic processes. While MHD simulations treat the plasma as a fluid, assuming it can be described by macroscopic quantities like density, pressure, and bulk velocity, PIC simulations keep track of the full velocity-space distribution functions, providing a detailed representation of the plasma's microscopic behavior. This particle-level accuracy makes PIC simulations particularly useful for studying non-thermal phenomena, such as shock acceleration, magnetic reconnection, and the generation of instabilities in relativistic plasmas. In astrophysical contexts where high-energy particles and complex electromagnetic interactions are prevalent, the ability to capture fine-scale kinetic effects is crucial. PIC simulations, therefore, enable a precise investigation of the underlying processes driving plasma dynamics, which is not possible for the fluid approach of MHD codes. However, this comes at the cost of computation time and the complexity of the simulations.

\subsection{The Fiducial Simulation} \label{subsec:fiducial}
\subsubsection{Simulation Setup} \label{subsubsec:setup}
Our simulations are performed on a 3D grid with dimensions of $l_x = 512\,\Delta x$, $l_y = 512\,\Delta x$, and $l_z = 512\,\Delta x$, where $\Delta x = d_e/5$ is the size of each cell and $d_e = c/\omega_{\mathrm{pe}}$ is the electron inertial length. The resulting box size of $102.4\,d_e$ ensures that the box spans at least several wavelengths of the fastest-growing modes of the instabilities discussed in Section~\ref{sec:theory}. The time step $\Delta t$ is determined by the speed of light and grid spacing with $\Delta t = 0.09\,\omega_{\mathrm{pe}}^{-1}$, where $\omega_{\mathrm{pe}}$ is the electron plasma frequency. Since $\omega_{\mathrm{pe}}$ depends on the number density $n_e$, and $d_e = c/\omega_{\mathrm{pe}}$, choosing a value for the number density sets the physical units of the simulation.

To relate these simulation parameters to real scenarios, consider a typical ion density of $n_i \approx 10^{7}\,\mathrm{cm^{-3}}$ at $\sim100$ days after explosion for SNe~Ia (see Section~\ref{subsubsec:Ia_overview}). Assuming the electron density roughly equal to the ion density, the electron plasma frequency is $\omega_{\mathrm{pe}} \approx 1.8 \times 10^{8}\,\mathrm{s^{-1}}$ and the inertial length is $d_e \approx 168\,\mathrm{cm}$. The physical size of the simulation box is $l_x = l_y = l_z \approx 172\,\mathrm{m}$, and the time step is $\Delta t \approx 0.5\,\mathrm{ns}$. We run the simulation for a total of 6000 steps, corresponding to a total physical time of $\approx 3\,\mathrm{\mu s}$. These parameters are chosen to accurately capture the micro-physics of the plasma while ensuring computational efficiency.

The fiducial setup assumes $n_i (100\,\mathrm{d}) \approx 10^{7}\,\mathrm{cm^{-3}}$ and consists of three particle species: electrons ($e^-$), positrons ($e^+$), and ions. In the case of radioactive transients, high-energy electrons or positrons emerge from the decay of the synthesized radioactive material, which we will call the `decay leptons' or `high-energy leptons' to distinguish them from the free electrons (see Section~\ref{subsec:equations}). We assume that the ions are all singly ionized \citep{Li12}, although this assumption can be extended to different charge states. To reduce computation time, we use an artificial mass ratio of $m_i/m_e = 25$. This mass ratio was chosen to ensure that ion length scales fit within the simulation box, which is $20.5 \,d_i$ in size, allowing the ions to couple effectively with the plasma. We tested other mass ratios and confirmed that this assumption does not affect the results. For a reasonable trade-off between sufficient particle density in each cell and computational cost, we use 10 particles per cell in our simulations.

In our simulations, the high-energy leptons are relativistic positrons originating from the decay of $^{56}$Ni. Due to radioactive decay, the positrons have energies around $\sim$1 MeV (see Table~\ref{tab:decay_data_SNe} and Section~\ref{subsec:equations}), corresponding to a relativistic Lorentz factor of $\gamma \approx 2$. The positrons are assumed to be beam-like with a bulk velocity of $\beta_p \equiv v_p/c \approx 0.866$\footnote{A note on the assumption of beam-like positrons is provided in Appendix~\ref{app:beam_ass}.}. The positron density is $n_p = f_p n_i$ where $f_p$ is the fraction of positron-producing decays (i.e. the $\beta^+$ branching ratio) from $n_i$ initial ions. We assume a fiducial positron fraction of $f_p = 0.2$, motivated by the $^{56}$Ni decay channel. To maintain quasi-neutrality, the electron density must be $n_e = (1+f_p)n_i$. Additionally, we assume the initial current density $\boldsymbol{J}_0 = 0$. To achieve this, the electrons must have a bulk velocity of $\beta_e = \frac{f_p}{1+f_p} \beta_p$ in the same direction as the positrons, while the heavy ions are initially stationary. The electrons, positrons, and ions are initialized with relativistic Maxwell-J\"uttner distributions and a low temperature of $k_B T_0 = 10^{-4}mc^2$.

Our fiducial setup employs simple periodic boundary conditions: when a particle exits the simulation box on one side, it re-enters the box from the opposite side. This approach conserves the total number of particles in the box at all times.

\subsubsection{Simulation Results} \label{subsubsec:results}
We begin by examining a fiducial simulation to illustrate the evolution of the magnetic field and particle dynamics. The parameters for this simulation are designed to represent a subsection of a pure radioactive sphere of $^{56}$Ni with a high-energy positron fraction of $f_p = 0.2$ and an initial positron kinetic energy of $E_p = 0.5\,\mathrm{MeV}$. A broader suite of models is presented in Section~\ref{subsec:suite}.

\begin{figure}
    \centering
    \includegraphics[width=\columnwidth]{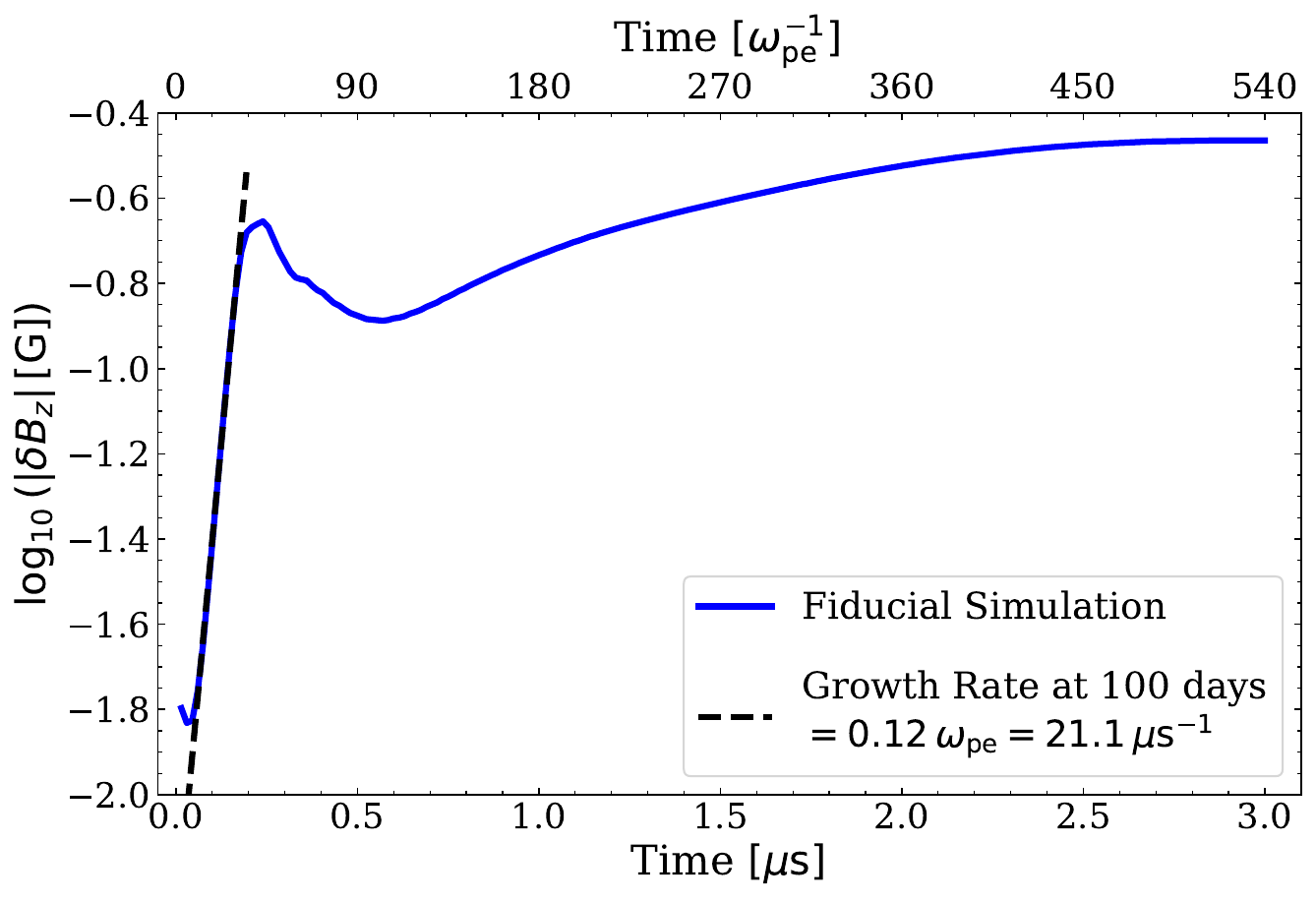}
    \caption{Perpendicular magnetic field averaged over the simulation box for the fiducial SN~Ia simulation at 100 days post-explosion with a density of $n_i \approx 10^{7}\,\mathrm{cm^{-3}}$ (solid blue line). The dashed black line shows the growth rate of the instability by fitting the exponential part of the growth, and is given in the legend. Due to the symmetry of the problem, $B_y = B_z$.}
    \label{fig:Bz_vs_t}
\end{figure}

To quantify the fluctuations in the magnetic field, we calculate the standard deviation of each magnetic field component. This standard deviation, $\delta B_i$, where $i$ indicates the component direction, provides an estimate of the amplitude of these fluctuations. Figure~\ref{fig:Bz_vs_t} shows the evolution of the $\boldsymbol{\hat{z}}$ component of the magnetic field ($B_z$). The bulk flow direction, which aligns with the axis of symmetry, is along $\boldsymbol{\hat{x}}$, as the current flows in this direction.

Initially, all the magnetic fields ($B_x$, $B_y$, and $B_z$) are zero at $t = 0$. The parallel magnetic field ($B_x$) remains consistent with noise throughout the simulation. The perpendicular fields ($B_y$ and $B_z$) rapidly develop fluctuations due to particle noise, growing exponentially until $t \approx 40\,\omega_{\mathrm{pe}}^{-1} = 0.22\,\mu \mathrm{s}$. By fitting an exponential model to the growth phase, we estimate a growth rate of $0.12\,\mathrm{\omega_{pe}}$.

We further compare with theory by examining the power spectrum of the perpendicular magnetic fields in the bottom panel of Figure~\ref{fig:fft_modes}. This spectrum, represented by the Fourier transform (denoted by a `$\, \hat{\ }\, $') of the perpendicular field components, $|\hat{B}y|^2 + |\hat{B}z|^2$, is shown as a function of wave number $k = 2\pi/\lambda$. Consistent with theoretical predictions, the magnetic fluctuations are largely in oblique modes, with wave numbers comparable to the electron inertial length. The modes with the highest power have $k_\perp \sim k_x \sim 1/d_e$, corresponding to a theoretical growth rate of $\sim 0.125\,\omega_{\mathrm{pe}}$, which agrees well with the simulation results. We note some disagreement between the simulation and theory in the power spectrum; this likely stems from the oversimplified linear theory model, which neglects the effects of ions, as well as both kinetic and thermal effects. Despite these limitations, the linear theory captures the mode growth in reasonable agreement with the simulation.

\begin{figure*}
    \centering
    \includegraphics[width=\textwidth]{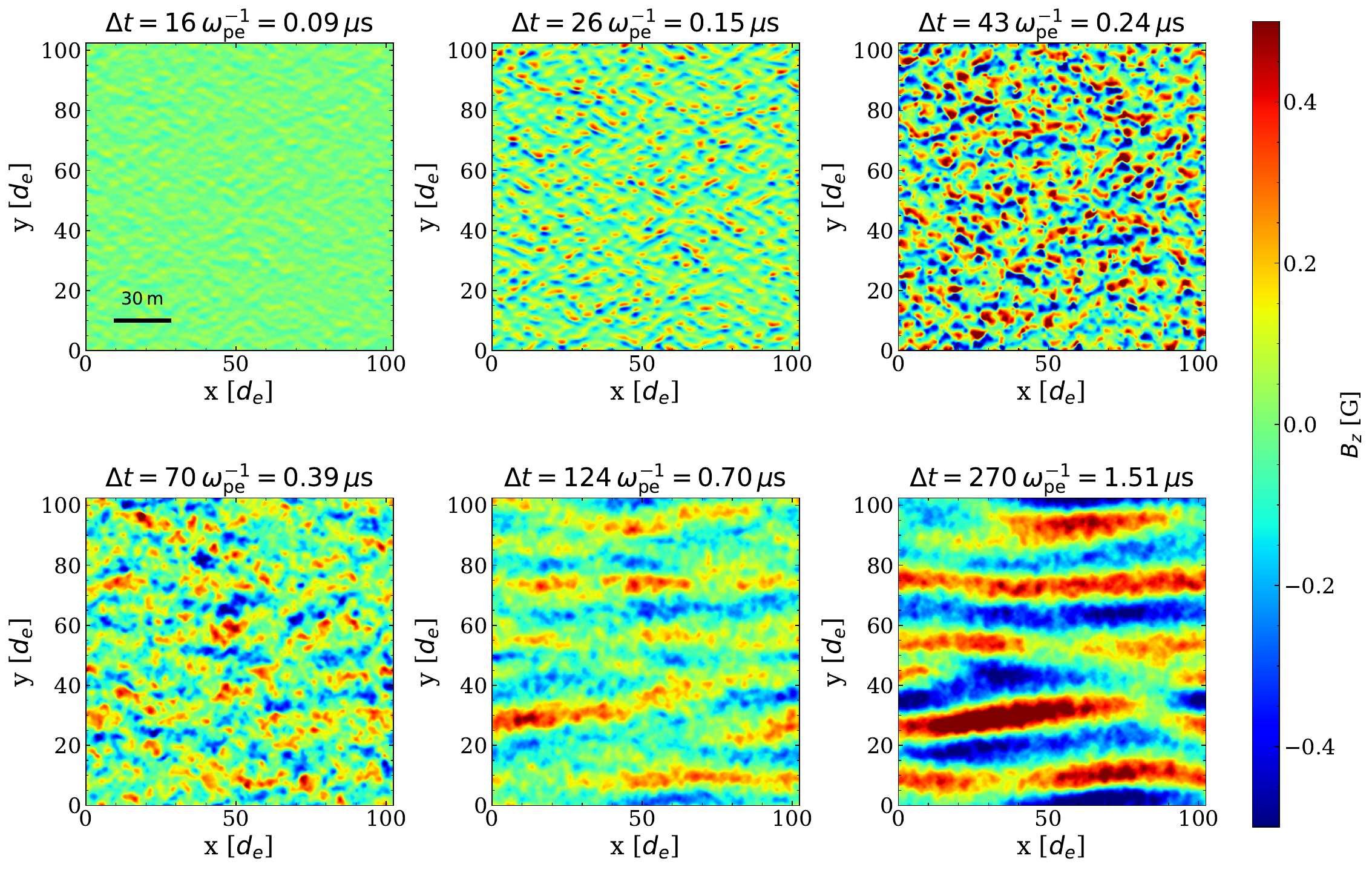}
    \caption{Magnetic field $B_z$ in a slice through the $xy$-plane for a set of times for the fiducial SN~Ia simulation at 100 days post-explosion with a density of $n_i \approx 10^{7}\,\mathrm{cm^{-3}}$. Physical length scale is given in the first panel and is identical for all panels. The panel for $t=0$ is not shown because the magnetic field is zero across the grid by definition. At time $t=26\,\mathrm{\omega_{pe}^{-1}} = 0.15\,\mu \mathrm{s}$, the magnetic fields are growing exponentially as a result of the instability. After time $t \approx 43\,\mathrm{\omega_{pe}^{-1}} = 0.24\,\mu \mathrm{s}$, the exponential growth of instability has stopped, and the fields evolve into a roughly constant pattern. Their characteristic wavelength increases until the field saturates ($t \approx 450\,\mathrm{\omega_{pe}^{-1}} = 2.5\,\mu \mathrm{s}$). Similar evolution is seen in all of our simulations discussed in Section~\ref{subsec:suite}.}
    \label{fig:Bz2d_vs_t} 
\end{figure*}

Following the saturation of the initial streaming instability, the magnetic fields enter a second growth phase up to $\approx 0.34\,\mathrm{G}$ after $t \approx 450\,\omega_{\mathrm{pe}}^{-1} = 2.5\,\mu\mathrm{s}$. This growth, occurring in the non-linear regime, is driven by the remaining free-streaming energy after the initial saturation. The dynamics of magnetic field growth and saturation in this regime are likely important to the physics of energy transfer between leptons and hadrons, warranting further exploration in future studies. 

From this simple simulation with an initial magnetic field of zero, the system generates a saturated magnetic field in microseconds perpendicular perpendicular to the bulk flow. Since positrons are constrained to move along magnetic field lines, the perpendicular fields suggest that the positrons will remain fixed at the same mass coordinates in a radially symmetric explosion. To move outward, they must diffuse, indicating that they would deposit their energy locally instead.

Figure~\ref{fig:Bz2d_vs_t} shows the $\boldsymbol{\hat{z}}$ (perpendicular to the positron flow) component of the magnetic field, $B_z$, at each grid point in a slice of the simulation box in the $xy$-plane at several time steps. A panel for $t=0$ is not shown because the magnetic field is initially zero across the grid by definition. This magnetic field is for the fiducial simulation, which assumes a density at $t=100\,\mathrm{days}$ for SNe~Ia, so we call the field $B_z (100\,\mathrm{d})$. We see $B_z (100\,\mathrm{d})$ starting at zero and develop fluctuations which grow in time and eventually saturate. The fluctuations at $t = 40\,\mathrm{\omega_{pe}^{-1}} = 0.22\,\mu \mathrm{s}$ have a characteristic wavelength of $\sim 5\,d_e = 0.84\,\mathrm{m}$ when the exponential growth of instability stops. The wavelength of the instability at this time is small, but as the field saturates, the wavelength grows and becomes more coherent. By time $t = 450\,\mathrm{\omega_{pe}^{-1}} = 2.5\,\mu \mathrm{s}$, the fields have saturated at a value of $B_z (100\,\mathrm{d}) \approx 0.34\,\mathrm{G}$ and a merged pattern is formed with $B_z (100\,\mathrm{d})$ varying in $\boldsymbol{\hat{y}}$. The bulk flow of particles is in the $\boldsymbol{\hat{x}}$ direction; due to the symmetry of the problem, we see a similar effect for the magnetic field in the $\boldsymbol{\hat{y}}$ direction. After saturation, the characteristic wavelength of the fluctuations is $\sim 20\,d_e = 33.6\,\mathrm{m}$, which is of the same order as the Larmor radius of positrons ($r_{L,p} \sim 50\,\mathrm{m}$), and therefore more effective at scattering them. 

Supporting the persistence of plasma-generated magnetic fields, \citet{GarasevDerishev16} show that a continuously injected, anisotropic particle distribution can produce long-lived, large-scale magnetic fields in relativistic collisionless shocks. Other studies of particle acceleration in relativistic shocks have also shown the effects of self-generated magnetic fields from filamentation instabilities \citep[e.g.,][]{Sironi11,Sironi13}. This aligns closely with our findings and supports the conclusion that magnetic fields generated by plasma instabilities can persist long-term. Once the instability saturates, the magnetic field will remain in the ejecta, influencing particle confinement and energy transport throughout its evolution.

\begin{figure}
    \centering
    \includegraphics[width=\columnwidth]{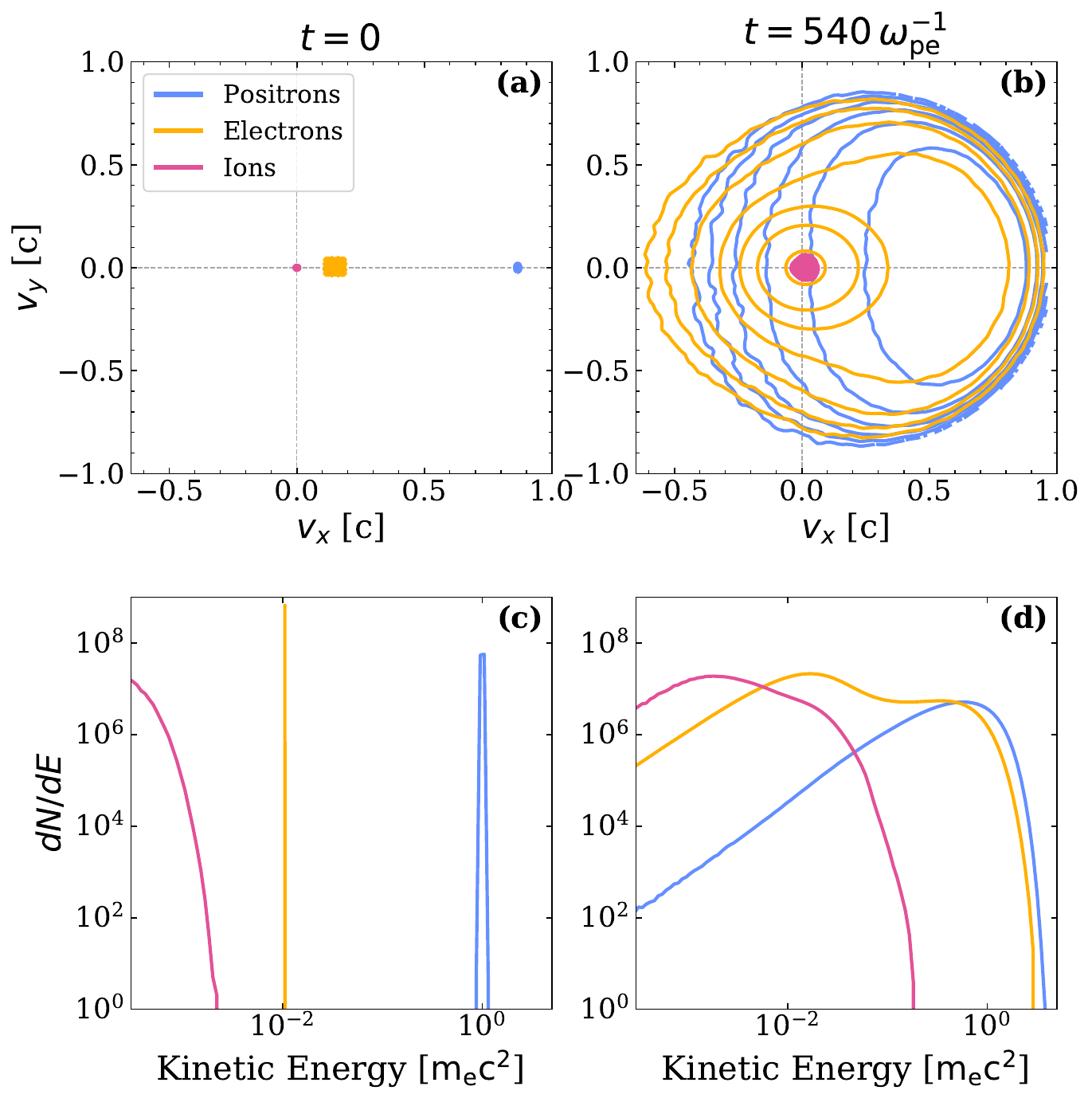}
    \caption{(a) The initial and (b) final ($t = 540\,\mathrm{\omega_{pe}^{-1}}$) velocity distributions for the positrons, electrons, and ions for the fiducial SN~Ia simulation. The velocity of $v_{x,y} = 0$ is marked by the dashed gray lines, indicating the point at which the particles would be at rest. (c) The initial and (d) final ($t = 540\,\mathrm{\omega_{pe}^{-1}}$) kinetic energy distributions for the positrons, electrons, and ions. By the end of the simulation ($3\,\mathrm{\mu s}$), the positrons have transferred significant energy into the electrons and ions. }
    \label{fig:init_final_vel_energy}
\end{figure}

To examine how the particles are affected by the growing magnetic field, we show the initial and final distributions of particles in the 2D velocity space of $v_x$ and $v_y$, in Figure~\ref{fig:init_final_vel_energy}a and b. Due to the symmetry of the problem, $v_y$ and $v_z$ have the same statistical properties. The contours show the distributions of positrons, electrons, and ions. The inner levels of the contour represent a higher density of particles in velocity-space. 

Figure~\ref{fig:init_final_vel_energy}a shows the initial velocity distributions of the three particle species. The ions are cold and hence their distribution is centered at $v_{x,i} = 0$ and $v_{y,i} = 0$ with a small spread. The electrons have a much broader velocity distribution because they are comparably cold but less massive.
Their distribution is centered at a velocity of $v_{x,e} = \frac{1}{6} v_{x,p}$ and $v_{y,e} = 0$. The positrons are the most energetic of the three populations since they are a hot beam drifting at $v_{x,p} = 0.866\,c$, corresponding to a Lorentz factor of $\gamma=2$. Figure~\ref{fig:init_final_vel_energy}b shows the final velocity distributions. The final distribution of ions is broader than the initial one, indicating some energy transfer to the ions. The positrons have scattered significantly in the perpendicular directions and their average drift velocity has decreased to $v_{x,p} \approx 0.67\,c$. In a longer simulation run up to $t = 2500\,\mathrm{\omega_{pe}^{-1}} = 13.9\,\mathrm{\mu s}$, the average drift velocity of the positrons continues to decrease, reaching values $v_{x,p} \ll c$. There is a second sub-population of electrons that has velocities in the $\boldsymbol{\hat{x}}$ direction that are much higher than initial. This illustrates the collisionless transfer of energy from the positrons to the electrons, due to the instability. 

The energy transfer between positrons and electrons is also illustrated with the kinetic energy distributions shown in Figure~\ref{fig:init_final_vel_energy}c and d in units of $mc^2$. Initially, the ions start at rest with little spread while the electrons are initialized with a small drift velocity, leading to a total kinetic energy higher than that of the ions. By the time the magnetic field saturates ($t > 450\,\mathrm{\omega_{pe}^{-1}}$), the positron distribution has broadened toward lower energies, whereas the electrons have gained energy with a secondary peak at higher energies also seen in Figure~\ref{fig:init_final_vel_energy}b. This emphasizes that the positrons are losing energy to the electrons and slowing down as a result of this instability. 

Figure~\ref{fig:init_final_vel_energy}d also shows a significant population of higher-energy ions. Although our simulations used an artificially low ion-to-electron mass ratio ($m_i/m_e = 25$; Section~\ref{subsubsec:setup}), tests with $m_i/m_e = 100$ yielded similar instability growth rates and ion kinetic energies. While the energy transfer to the ions persists with larger mass ratio, the simulations are still far from the realistic value of $m_i/m_e \approx 10^5$ for $^{56}$Fe.

\begin{figure}
    \centering
    \includegraphics[width=\columnwidth]{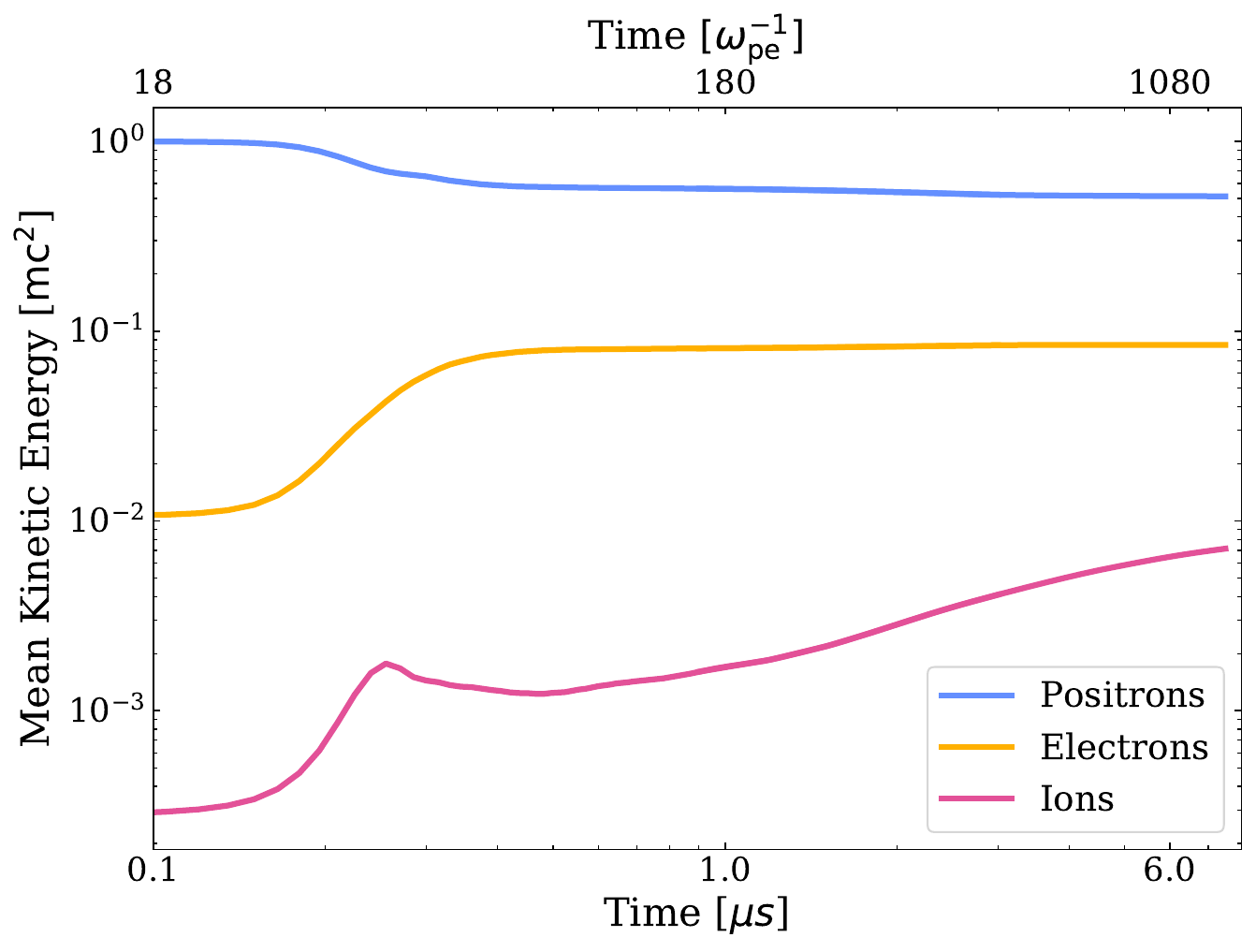}
    \caption{Mean kinetic energy of positrons, electrons, and ions as a function of time for the fiducial parameters. This simulation goes past the fiducial $3\,\mathrm{\mu s}$ up to $7.5\,\mathrm{\mu s}$ to show the increasing ion kinetic energies.}
    \label{fig:mean_KE}
\end{figure}

Figure~\ref{fig:mean_KE} confirms that the energy lost by the positrons is transferred to the electrons and ions. Almost $50\%$ of positron kinetic energy is deposited in other species only in $7.5\,\mathrm{\mu s}$. While the instability energizes ions, we do not attempt to quantify the specific mechanisms governing this energy transfer in this study. In a physical system, the energy in ions would eventually radiate away as photons, assuming collisional excitation of the ions. However, modeling this radiative process is beyond the scope of our current work.

\subsection{Suite of Simulations} \label{subsec:suite}
To probe the range of non-linear effects in the simulations due to varying high-energy lepton fractions (from radioactive decay), we conduct a suite of simulations with different high-energy positron fractions (high-energy electrons are discussed for KNe in Section~\ref{subsubsec:KN_overview}). This allows us to infer the magnetic field saturation levels corresponding to each positron fraction. For simplicity, the previously defined positron fraction $f_p$ can be generalized to the total decay lepton fraction.

At every point in the evolution of a transient, the saturated magnetic field resulting from plasma instability depends on both the instantaneous ion number density $n_i$ and the decay lepton fraction $f_p$ in the ejecta. This is because the instability saturates almost instantaneously (within $<3\,\mu\mathrm{s}$) compared to the much longer timescale for the evolution of a transient. Since $f_p$ evolves with time as parent nuclei decay, we vary this parameter in our simulations and measure the resulting saturated magnetic field. 

\begin{figure}
    \centering
    \includegraphics[width=\columnwidth]{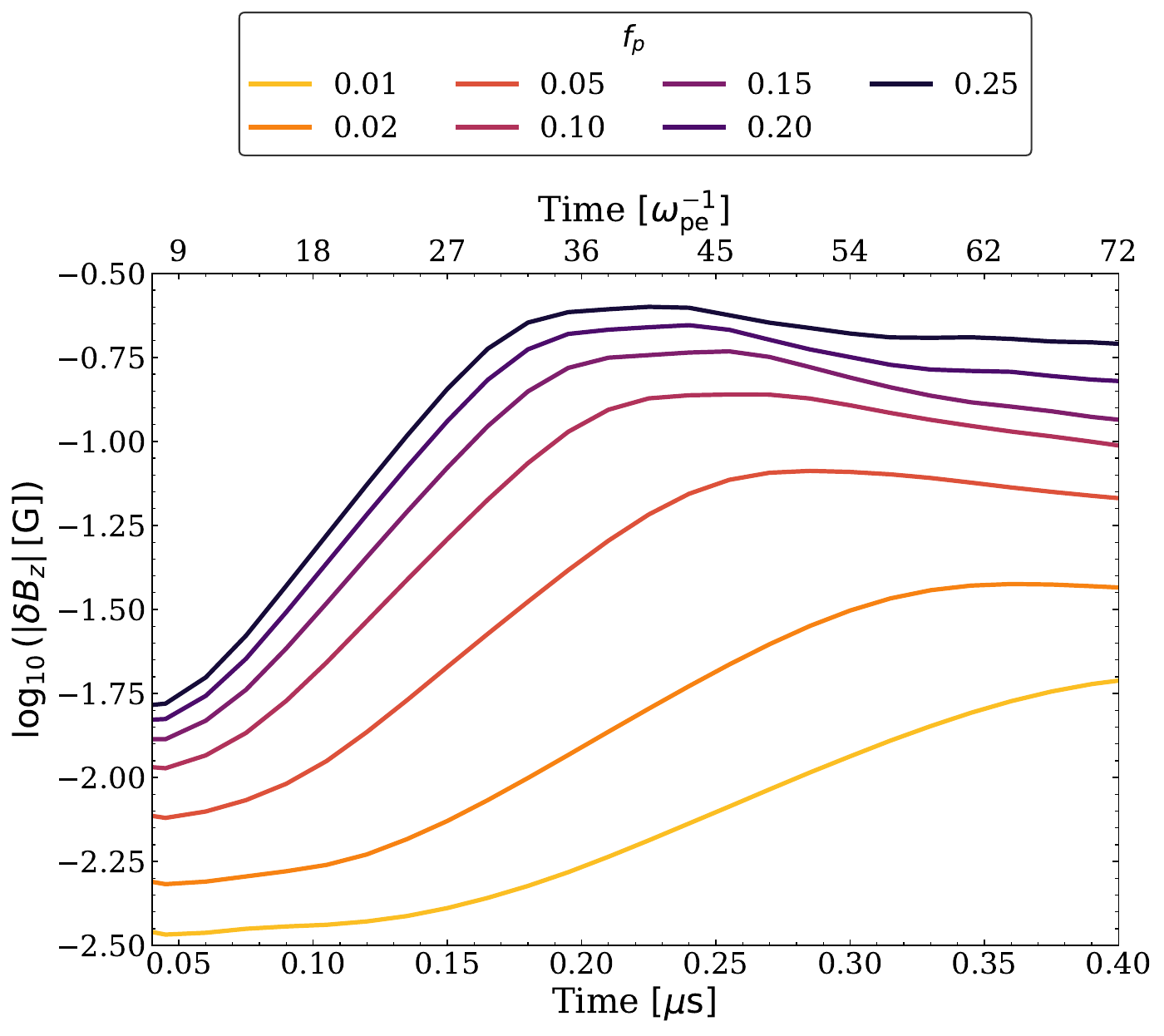}
    \caption{Magnetic field as a function of time for various high-energy lepton fractions $f_p$ at 100 days post-explosion for a typical SN Ia with a density of $n_i \approx 10^{7}\,\mathrm{cm^{-3}}$. The decay lepton energy is kept constant at $E_p = 0.5\,\mathrm{MeV}$. The relation between the magnetic field saturation and lepton fraction is given in Eq.~\ref{eq:Bsat_vs_f_p}.}
    \label{fig:Bsat_vs_f_p}
\end{figure}

Figure~\ref{fig:Bsat_vs_f_p} illustrates how the saturation of the magnetic field varies with $f_p$. By performing a simple power-law fit to the saturated magnetic field at its peak after exponential growth, we find that the scaling is $B_{\mathrm{sat}} \propto f_p^{0.79 \pm 0.02}$. The variation in the saturated magnetic field due to changing the energies of the decay leptons ($E_p$) is very similar to Figure~\ref{fig:Bsat_vs_f_p} for energies of $\{ 0.001, 0.01, 0.1, 0.5 \}\,\mathrm{MeV}$. As the decay lepton energy decreases, so does the peak of the magnetic field, with $B_{\mathrm{sat}} \propto E_p^{0.77 \pm 0.03}$.
Considering the $\sqrt{n_e}$ dependence of the magnetic field strength, and assuming that the electron and ion densities are roughly equal, we find that the saturated magnetic fields from our simulations can be well described by
\begin{eqnarray} \label{eq:Bsat_vs_f_p}
B_{\mathrm{sat}} (t) \approx 0.22\,\mathrm{G}\, \sqrt{\frac{n_i (t)}{10^{7}\,\mathrm{cm^{-3}}}}\, \left( \frac{f_p (t)}{0.2} \right)^{0.79} \left( \frac{E_p}{0.5\,\mathrm{MeV}} \right)^{0.77}.
\end{eqnarray}
We ran a simulation varying $f_p$ and $E_p$ simultaneously and found consistent results with varying them separately, indicating that they can be assumed as separable in Eq.~\ref{eq:Bsat_vs_f_p}.
These relationships highlight the significant influence of decay lepton properties on the saturated magnetic field. We directly connect these findings to astrophysical transients in the following section.

\section{Implications for Radioactive Transients} \label{sec:implications}
Here we consider how these plasma instabilities affect the late-time evolution of radioactively powered transients. First we outline the important decay chains and relevant derivations in Section~\ref{subsec:equations}. Thermonuclear/SNe Ia are considered in Section~\ref{subsec:SNeIa}, Type II SNe in Section~\ref{subsec:SNeII} and Stripped-Envelope supernovae (SE~SNe) in Section~\ref{subsec:SESNe}, and we finish with KNe in Section~\ref{subsec:KNe}.

\subsection{Radioactive Decay and Positrons} \label{subsec:equations}
\subsubsection{Radioactive Decay in Transients}
The luminosity of most SNe~Ia and KNe past $\sim 1$ day primarily results from the radioactive decay of isotopes produced during their explosion \citep[e.g.,][]{Arnett82,LiPaczynski98}. In SNe, most of the synthesized mass comprises nuclear species with mass numbers ranging from $A \approx 12 - 70$, which are generally stable or proton-rich nuclides \citep[e.g.,][]{Arnett96,Seitenzahl13}. For these nuclei, fission and alpha decay are not significant. Radioactive decay in SNe progresses along an isobar towards more neutron-rich nuclides through two primary mechanisms: electron capture and positron emission. In electron capture, an inner atomic electron is captured by a nuclear proton, resulting in the emission of an electron neutrino. Positron emission, on the other hand, involves the decay of a nuclear proton into a neutron, accompanied by the emission of a positron and an electron neutrino. Both processes involve transitions to various nuclear levels in the daughter nucleus, with probabilities determined by statistical branching ratios \citep[e.g.,][]{Arnett96,Seitenzahl13}. 

The decay channels are more complex in KNe, which are powered by radioactive decay of the heavy $r$-process elements \citep[$A \approx 100 - 250$, e.g.,][]{Metzger10,Roberts11,Barnes16} synthesized during neutron star mergers. Some of the heavier $r$-process elements can undergo spontaneous fission and alpha decay. Most $r$-process elements undergo a series of $\beta^-$-decays, where a neutron in the nucleus decays into a proton, electron, and antineutrino, moving the nucleus towards a more stable configuration. Both $\beta^+$-decay in SNe and $\beta^-$-decay in KNe emit high-energy leptons with energies of order $\sim$1 MeV \citep[e.g.,][]{Metzger10,Kasen17}.

\begin{table}
    \centering
    \begin{tabular}{c|c|c|c}
        \hline
        \hline
        Nucleus & Half Life & $\beta^+$ Branching Ratio & Mean $\beta^+$ Energy \\
                & [d]       & [\%]                      & [keV] \\
        \hline
        \hline
        $^{56}$Ni & $6.075$ & $7 \times 10^{-4}$ & $100$ \\
        $\downarrow$ &        &                  & \\
        $^{56}$Co & $77.236$ & $19.7$ & $610$ \\
        $\downarrow$ &        &                  & \\
        $^{56}$Fe & stable & --- & --- \\
        \hline
        $^{57}$Ni & $1.483$ & $43.6$ & $354$ \\
        $\downarrow$ &        &                  & \\
        $^{57}$Co & $271.74$ & $0.0$ & --- \\
        $\downarrow$ &        &                  & \\
        $^{57}$Fe & stable & --- & --- \\
        \hline
        $^{55}$Co & $0.730$ & $76.0$ & $570$ \\
        $\downarrow$ &        &                  & \\
        $^{55}$Fe & $1002$ & $0.0$ & --- \\
        $\downarrow$ &        &                  & \\
        $^{55}$Mn & stable & --- & --- \\
        \hline
        $^{44}$Ti & $21{,}600$ & $0.0$ & --- \\
        $\downarrow$ &        &                  & \\
        $^{44}$Sc & $0.1684$ & $94.3$ & $630$ \\
        $\downarrow$ &        &                  & \\
        $^{44}$Ca & stable & --- & --- \\
        \hline
    \end{tabular}
    \addtocounter{footnote}{-1}
    \caption{$\beta^+$-decay properties for radionuclides relevant for SNe. The nuclear decay data presented here are obtained from the Chart of Nuclides database at the National Nuclear Data Center\protect\footnotemark.}
    \label{tab:decay_data_SNe}
\end{table}

\footnotetext{\url{http://www.nndc.bnl.gov/chart/}}

The decay chains contributing most of the radioactive energy in SNe are listed in Table~~\ref{tab:decay_data_SNe}.
Not every step in the decay chain produces positrons through the $\beta^+$ decay channel. The decays of $^{57}$Co, $^{55}$Fe, and $^{44}$Ti have no positron channel and all decay proceeds via electron capture. The branching ratio of the $\beta^+$-decay channel for $^{56}$Ni to $^{56}$Co is significantly smaller than that for $^{56}$Co to $^{56}$Fe.

In addition to positrons from $\beta^+$ decay, another source of high-energy leptons are Internal Conversion (IC) and Auger electrons \citep[e.g.,][]{Seitenzahl09}. IC electrons are emitted when an excited nucleus transfers its energy directly to an orbital electron, ejecting it from the atom, while Auger electrons result from the filling of inner electron vacancies by outer electrons, with the excess energy released as another electron rather than as X-ray photons. These electrons typically have energies $<10$ keV, much lower than the $\sim$MeV positrons from $\beta^+$ decay. As a result, the magnetic fields generated by these lower-energy electrons through plasma instabilities are more than an order of magnitude weaker than those produced by $\sim$MeV positrons (see Eq.~\ref{eq:Bsat_vs_f_p}), and therefore we neglect them in the following sections.

\subsubsection{Rate of Positrons in SNe} \label{subsubsec:positrons}
Here, we simplify the equations by assuming only the $^{56}$Ni decay chain. This is warranted for SNe~Ia, where the $^{56}$Ni region has only trace amounts of other elements \citep[e.g.,][]{Mazzali15, Scalzo19}. The situation is more complicated for CC~SNe which experience significant mixing during the explosion \citep[e.g.,][]{Muller20}. $^{56}$Ni and its daughter nucleus $^{56}$Co remain the primary radionuclides until $10^4$ days after explosion, where $^{44}$Ti becomes important \citep[e.g.,][]{Jerkstrand11,Seitenzahl14}. 

The number of $^{56}$Ni ions as a function of time is
\begin{eqnarray}
N_{\mathrm{Ni}}(t) = N_0 e^{-t/\tau_{\mathrm{Ni}}} ,
\end{eqnarray}
where $N_0$ is the initial number of ions synthesized in the explosion and $\tau_{\mathrm{Ni}} = 8.8\,\mathrm{days}$ is the e-folding time for $^{56}$Ni decay. Assuming no $^{56}$Co at time $t=0$ and $\tau_{\mathrm{Co}} = 111.4\,\mathrm{days}$, the number of $^{56}$Co ions is
\begin{eqnarray}
N_{\mathrm{Co}}(t) =  \frac{\tau_{\mathrm{Co}} N_0}{\tau_{\mathrm{Co}} - \tau_{\mathrm{Ni}}} \left[ e^{-t/\tau_{\mathrm{Co}}} - e^{-t/\tau_{\mathrm{Ni}}} \right] .
\end{eqnarray}
The first step of this chain produces negligible positrons via $\beta^+$ decay because the branching ratio is only $\eta_{\mathrm{Ni}} = 7 \times 10^{-6}$, but the second step produces positrons 19.7\% of the time ($\eta_{\mathrm{Co}} = 0.197$). The rate of positron production from the first decay is 
\begin{eqnarray}
\frac{dN_{p,\mathrm{Ni}}}{dt} = \frac{\eta_{\mathrm{Ni}}N_0}{\tau_{\mathrm{Ni}}}e^{-t/\tau_{\mathrm{Ni}}}
\label{eq:Ni_positron_diffeq}
\end{eqnarray}
and the rate from the second decay is
\begin{eqnarray}
\frac{dN_{p,\mathrm{Co}}}{dt} = \frac{\eta_{\mathrm{Co}}N_0}{\tau_{\mathrm{Co}} - \tau_{\mathrm{Ni}}} \left(e^{-t/\tau_{\mathrm{Co}}} - e^{-t/\tau_{\mathrm{Ni}}}\right) .
\label{eq:Co_positron_diffeq}
\end{eqnarray}

\begin{figure}
    \centering
    \includegraphics[width=\columnwidth]{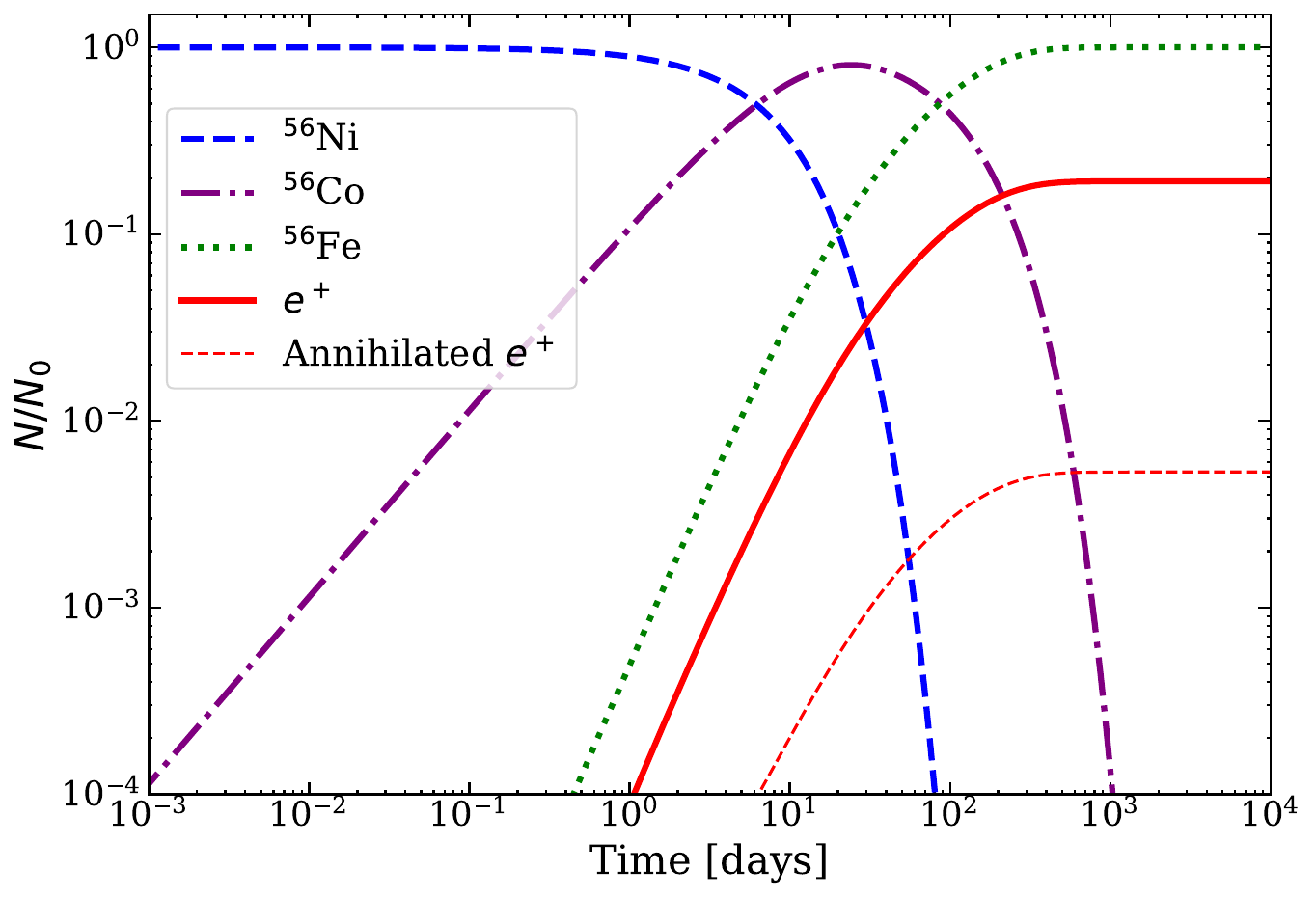}
    \caption{Relative fractions of different isotopes of the $^{56}$Ni decay chain and the positrons produced in the decay steps as a function of time. The solid red curve for the relative number of positrons also includes their destruction through electron-positron annihilation, which is significant at $t \lesssim 1$~day. The dashed red line shows the relative fraction of positrons that were annihilated, assuming our fiducial parameters (Section~\ref{subsec:fiducial}) for a typical SN~Ia (Section~\ref{subsubsec:Ia_overview}).}
    \label{fig:positron_number_from_Ni56decay_annihilation}
\end{figure}

Next we consider electron-positron annihilation cross section as a function of the center-of-momentum gamma factor between electron and positron. Annihilation can significantly reduce the number of positrons at early times when densities are high. The annihilation rate depends on the electron number density $n_e(t)$, center-of-momentum velocity $v_{\mathrm{com}}$, and annihilation cross section $\sigma_{pe}$ \citep{Svensson82}
\begin{eqnarray}
\nu_{\mathrm{ann}} = n_e(t) v_{\mathrm{com}} \sigma_{pe} .
\end{eqnarray}
So the change in the number of positrons due to annihilation is
\begin{eqnarray}
\frac{dN_{p,\mathrm{ann}}}{dt} = - N_p(t) n_e(t) v_{\mathrm{com}} \sigma_{pe} .
\label{eq:Ann_positron_diffeq}
\end{eqnarray}
Assuming the electron number density is the same as the ion number density ($n_e \approx n_i$), the final rate of positron production is
\begin{align}
\frac{dN_p}{dt} &= N_0 \left[ \frac{\eta_{\mathrm{Ni}}}{\tau_{\mathrm{Ni}}}e^{-t/\tau_{\mathrm{Ni}}} + \frac{\eta_{\mathrm{Co}}}{\tau_{\mathrm{Co}} - \tau_{\mathrm{Ni}}} (e^{-t/\tau_{\mathrm{Co}}} - e^{-t/\tau_{\mathrm{Ni}}}) \right] \notag \\
&- v_{\mathrm{com}} \sigma_{pe} n_{i}(t) N_p(t) .
\label{eq:posi_diffeq_final}
\end{align}
Here, $N_0$ and $n_i$ are the parameters that will vary for the different types of transients discussed in the following sections. Figure~\ref{fig:positron_number_from_Ni56decay_annihilation} shows the relative fractions of each isotope in the $^{56}$Ni decay chain along with the relative fraction of positrons including the effects of annihilation. After $\sim1$~day, the densities are low enough that electron-positron annihilation becomes less significant. Assuming homologous expansion, the positron number density is obtained by dividing $N_p$ by the ejecta volume at time $t$
\begin{equation}
    n_p (t) = N_p (t) / V_{\mathrm{ej}}(t) .
    \label{eq:n_p}
\end{equation}

\subsection{Type Ia SNe} \label{subsec:SNeIa} 
\subsubsection{Overview and Physical Properties} \label{subsubsec:Ia_overview}
SNe~Ia are thermonuclear explosions of carbon/oxygen (C/O) WDs \citep[e.g.,][]{hoyle60}. Observational studies have measured the $^{56}$Ni mass ($M_{\mathrm{Ni}}$) in SNe~Ia to range between $0.2$ and $1.6\,M_\odot$, with a median around $0.6\,M_\odot$ \citep[e.g.,][]{Iwamoto99,Seitenzahl09,Scalzo14,Childress15,Scalzo19} and the total ejecta mass ($M_{\mathrm{ej}}$) is between $0.8$ and $2.2\,M_\odot$, with a median around $1.4\,M_\odot$ \citep[e.g.,][]{Scalzo14,Scalzo19}. 

Although the progenitor systems of SNe Ia are uncertain, all plausible scenarios involve a WD and a companion star \citep[for reviews see][]{Maoz14,Ruiter20}. Assuming a near-Chandrasekhar mass for the WD prior to explosion, the range of radii for SNe Ia progenitors can be estimated. The mass-radius relation for WDs is relatively well-defined, as WDs evolve at roughly constant radii \citep[e.g.,][]{Fontaine01,Tremblay17,Saumon22}. We adopt a typical progenitor radius of $R_0 = 7000\,\mathrm{km}$, with a range between $3000$ and $10{,}000\,\mathrm{km}$ to span the radii of magnetic WDs \citep[e.g.,][]{Karinkuzhi24}.

The number density of ions predominantly determines the evolution of plasma-generated magnetic fields (see Eq.~\ref{eq:Bsat_vs_f_p}). We model the temporal evolution of the mean number density as 
\begin{equation}
    n_i(t) = \frac{3M}{4\pi m_i ( R_0 + v_{\mathrm{ej}} t)^3} ,
    \label{eq:n_i}
\end{equation}
where $M$ is the total mass with an average ion nuclei mass $m_i$, a constant ejecta velocity $v_{\mathrm{ej}}$, and an initial radius of $R_0$. This assumes homologous expansion, where the ejecta radius evolves linearly as a function of time ($R_{\mathrm{ej}} = R_0 + v_{\mathrm{ej}} t$) starting with an initial pre-explosion radius $R_0$ at $t=0$.

The masses of $^{57}$Ni and $^{55}$Co are estimated to be $20-100$ times lower than those of $^{56}$Ni \citep{Shappee17b,Graur18,Li19,Tucker22,Tiwari22}. Given their lower mass and the fast decay chains of $^{57}$Ni and $^{55}$Co, the positrons produced are mostly annihilated due to the high ejecta densities, rendering their impact on the positron rate negligible. $^{57}$Co and $^{55}$Fe decay on much longer timescales and dominate the late-time light curves of SNe~Ia \citep[e.g.,][]{Tucker22}. However, they are not a source of high-energy positrons as they decay exclusively by electron capture. At very late times ($> 10^4\,\mathrm{days}$), the very small $^{44}$Ti mass of $M_{\mathrm{Ti}} < 10^{-5}\,M_\odot$ expected from SNe~Ia \citep[e.g.,][]{Iwamoto99,Seitenzahl09} becomes the primary remaining decay chain, and at that time, the ion density is $n_i \sim 4\,\mathrm{cm^{-3}}$ (see Eq.~\ref{eq:n_i}). 

The isotopic composition of the ejecta is often stratified, with $^{56}$Ni predominantly concentrated near the center, while lighter elements are distributed toward the outer layers \citep[e.g.,][]{Ashall14,Ashall16,Sasdelli14,Mazzali15}. However, some models suggest that macroscopic mixing can result in a more extensive distribution of $^{56}$Ni throughout the ejecta \citep[e.g.,][]{Seitenzahl13,Piro16}. For our analysis, we consider two limiting scenarios: one where $^{56}$Ni is centrally concentrated, with lighter elements confined to the outer layers, and another where $^{56}$Ni is homogeneously mixed throughout the ejecta. The first scenario aligns with expectations for normal SNe~Ia, which are the most common \citep[e.g.,][]{Desai24}, while the second is relevant to the less-common, low-luminosity Iax subclass of SNe~Ia, thought to originate from incomplete deflagrations \citep[e.g.,][]{Jha17,Camacho-Neves23}. Although future work is needed to self-consistently incorporate plasma streaming instabilities into realistic explosion models, these simplified scenarios provide robust boundary conditions for exploring the range of possible outcomes.

\subsubsection{Magnetic Fields} \label{subsubsec:Ia_Bfields}
\begin{figure*}
    \centering
    \includegraphics[width=\textwidth]{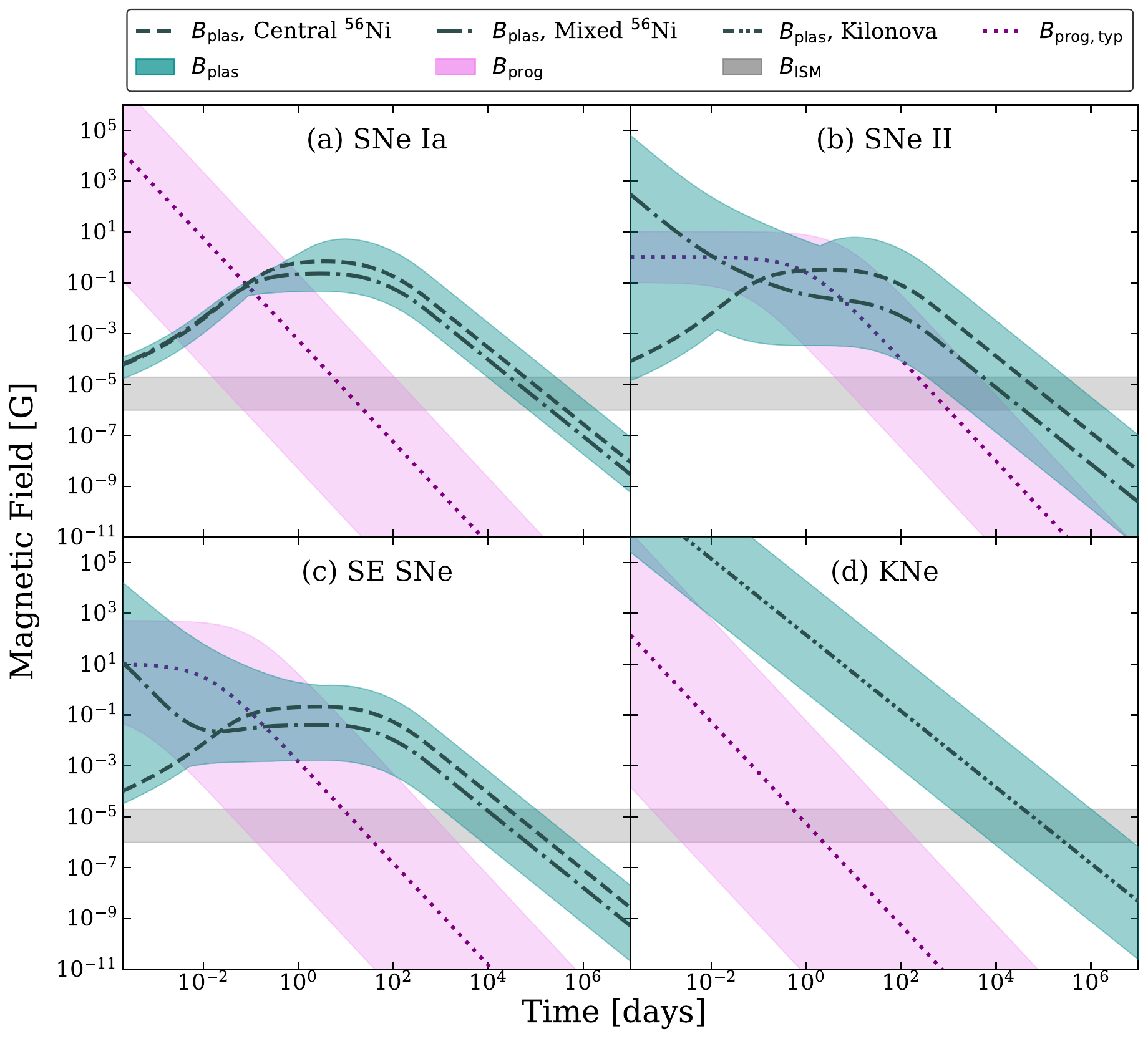}
    \caption{Magnetic field evolution from plasma instabilities compared to progenitor and ISM fields for SNe Ia, SNe II, SE SNe, and KNe. The instability saturates almost instantaneously ($<3\,\mu\mathrm{s}$), making the magnetic field at any time dependent on ejecta density and decay lepton fraction (Eq.~\ref{eq:Bsat_vs_f_p}). Dark cyan bands show saturated fields from the plasma instability, which generally exceed progenitor fields for typical transient parameters for most of the evolution. Panels (a), (b), and (c) compare two field evolution assumptions: dashed lines for centrally concentrated $^{56}$Ni and dash-dotted lines for mixed $^{56}$Ni. In panel (d), the dash-dot-dotted line shows the evolution for typical KN parameters. Violet bands show the range of progenitor fields decaying as $B_{\mathrm{prog}} \propto 1/R_{\mathrm{ej}}^{2} \propto 1/t^2$, starting from pre-explosion surface values, with dotted lines indicating evolution for typical progenitor parameters. Gray bands show the range of ISM magnetic fields \citep{Han17}.}
    \label{fig:B_sat_prog}
\end{figure*}

The evolution of the plasma-generated magnetic field in SNe~Ia is tied to the ion number density, which decreases as the ejecta expand. The ion number density is described by Eq.~\ref{eq:n_i}, which predicts the saturated magnetic field as a function of time, according to Eq.~\ref{eq:Bsat_vs_f_p}.

In the scenario where $^{56}$Ni is centrally concentrated, we estimate the ion number density relevant for the plasma instability, considering only the $^{56}$Ni region without any mixing. Here, we assume a typical ejecta velocity for Fe-group elements of $v_{\mathrm{ej}} = 6500\,\mathrm{km\,s^{-1}}$ \citep[e.g.,][]{Mazzali15,Flors20,Wilk20,Liu23}, with a range between $2000$ and $10{,}000\,\mathrm{km\,s^{-1}}$ \citep{Seitenzahl13,Blondin18}, which is lower than that of the outer layers. Although $^{56}$Ni can be found with up to $20{,}000\,\mathrm{km\,s^{-1}}$, the bulk of $^{56}$Ni has velocity of $<10{,}000\,\mathrm{km\,s^{-1}}$ \citep[e.g.,][]{Mazzali15,Ashall19}. We also use a typical $^{56}$Ni mass of $M_{\mathrm{Ni}} = 0.6\,\mathrm{M_{\odot}}$, an ion mass of $m_i = 56 m_p = 9.3 \times 10^{-23}\,\mathrm{g}$ (corresponding to $A \sim 56$), and a progenitor radius of $R_0 = 7000\,\mathrm{km}$ in Eq.~\ref{eq:n_i} to obtain $n_i (t)$.

In the case of complete mixing, where $^{56}$Ni is evenly distributed throughout the ejecta, the isotope is diluted within the ejecta material. In our fiducial assumption, $^{56}$Ni is the sole isotope generating high-energy positrons, so with mixing, the positron fraction $f_p$ is lower. For this scenario, we assume an ejecta velocity of $v_{\mathrm{ej}} = 11{,}000\,\mathrm{km\,s^{-1}}$, with a range between $8000$ and $16{,}000\,\mathrm{km\,s^{-1}}$ \citep{Zhang20,Pan24}, which incorporates most of the partially-burned region rich in intermediate-mass elements \citep[Ca, Si, S; ][]{Mazzali07,Pan24}. We use a total ejecta mass of $M_{\mathrm{ej}} = 1.4\,\mathrm{M_{\odot}}$, an ion mass of $m_i = 44 m_p = 7.3 \times 10^{-23}\,\mathrm{g}$ (corresponding to $A \sim 44$, weighted mean of all ejecta ions), and a progenitor radius of $R_0 = 7000\,\mathrm{km}$ to calculate $n_i (t)$.

Figure~\ref{fig:B_sat_prog}a illustrates this evolution for SNe~Ia, where the dark cyan band represents the range of possible magnetic field strengths for our adopted range of parameters (i.e., $M_{\mathrm{ej}}$, $M_{\mathrm{Ni}}$, $v_{\mathrm{ej}}$, and $R_0$). The dashed line shows the magnetic field evolution for a set of typical SNe~Ia parameters for the case of centrally concentrated $^{56}$Ni, and the dash-dotted line shows the evolution for $^{56}$Ni mixed within the ejecta. Both evolve similarly, except, centrally concentrated models have higher ion densities and thus a slightly higher magnetic field.

Initially, at $t \lesssim 1$\,day, the ion density is high, leading to rapid positron-electron annihilation, which suppresses the growth of the plasma instability. As the ejecta expand and the density decreases, the plasma instability begins to dominate, leading to an increase in the saturated magnetic field, which peaks between 1 and 10 days post-explosion. After this peak, the instantaneous saturated magnetic field follows a logarithmic slope of approximately $-1.5$, reflecting the ongoing expansion and the decreasing ion density. The plasma instability at each moment in time is assumed to be independent of its past behavior and depends solely on the particle densities at that time (based on Eq.~\ref{eq:Bsat_vs_f_p}), not on the previous magnetic fields. A fully self consistent treatment is beyond the scope of this work.

The surface magnetic fields of the progenitor WDs are expected to range from $10^3$ to  $10^9\,\mathrm{G}$ \citep{Schmidt03,Kulebi09,Ferrario15}. Assuming a dipole field and the the expansion conserving magnetic flux of the surface fields, the field strength decreases with radius as $B_{\mathrm{prog}} \propto R_{\mathrm{ej}}^{-2}$. Given the additional assumption of homologous expansion ($R_{\mathrm{ej}} \propto v_{\mathrm{ej}}t$), the progenitor magnetic field would decline over time as $B_{\mathrm{prog}} \propto t^{-2}$ at a specific ejecta velocity. However, it is important to note that these estimates are based on surface magnetic fields, and the interior magnetic field strength and structure of the progenitor remains uncertain \citep[e.g.,][]{Drewes22}. This uncertainty could influence the magnetic field strength and configuration within the expanding ejecta, but from our simulations, we expect plasma streaming instabilities to dominate the magnetic fields in the ejecta of SNe~Ia. Even at the peak of the light curve ($\sim20$ days post-explosion), plasma instabilities would generate magnetic fields roughly $10^2 - 10^3$ times stronger than those expected from the strongest observed surface fields \citep[][]{Ferrario15}.

This evolution is shown in Figure~\ref{fig:B_sat_prog}a as the lighter violet band, which encompasses the range of possible progenitor magnetic field evolution, with the typical parameters shown as the dotted line. The saturated magnetic field generated by the plasma instability exceeds the progenitor magnetic field within hours of explosion for our typical cases, and by $t \approx 1$\,day even for the most extreme scenarios. Furthermore, this field remains stronger than the typical interstellar medium (ISM) magnetic field until around $10^5$\,days.

With this magnetic field evolution, prompt synchrotron emission from the high-energy positrons is expected to peak in the $100\,\mathrm{MHz}$ to $1\,\mathrm{GHz}$ range, $\sim20$ days after the explosion. However, for radio observatories such as LOw Frequency ARray (LOFAR), Square Kilometre Array Observatory (SKAO) and future radio facilities to be able to detect radio signatures from high-energy positrons, SNe~Ia must be (1) Galactic or local-group SNe, and (2) in extremely clean environments with no CSM. More details on synchrotron emission from SNe~Ia positrons and its detectability are given in Appendix~\ref{app:synch}.

\subsection{Type II SNe} \label{subsec:SNeII}
\subsubsection{Overview and Physical Properties} \label{subsubsec:II_overview}
SNe~II are the most common subclass of CC~SNe that originate from the explosive death of massive stars \citep[$M_{\mathrm{ZAMS}} > 8\,\mathrm{M_{\odot}}$, e.g.,][]{hoyle60,Woosley02,Smartt09}. They are typically categorized by the presence of hydrogen in their spectra, distinguishing them from hydrogen-deficient SE~SNe and thermonuclear SNe~Ia. The light curves of SNe~II are notably diverse, but often exhibit a plateau phase (Type IIP) that persists for several weeks, during which the luminosity remains nearly constant \citep[e.g.,][]{Popov93}. This plateau results from the recombination of ionized hydrogen in the extended outer layers of the ejecta, effectively creating a diffusion barrier that slows the escape of photons. The plateau phase is followed by a decline as the supernova transitions to the radioactive decay phase, primarily powered by the decay of $^{56}$Ni to $^{56}$Co and subsequently to stable $^{56}$Fe \citep[e.g.,][]{Arnett82,WoosleyWeaver86}.

In addition to $^{56}$Ni, other isotopes synthesized during explosion significantly contribute to the overall composition of the SNe~II ejecta. The ejecta are stratified, with the heavier elements such as nickel and titanium located in the inner regions, while intermediate-mass elements like oxygen and silicon are found in the middle layers, and hydrogen and helium dominate the outermost layers \citep[e.g.,][]{WoosleyWeaver95,Thielemann96,Rauscher02}. The central regions, where $^{56}$Ni is most abundant, are particularly relevant for our analysis. However, as the ejecta become increasingly mixed with other isotopes \citep[e.g.,][]{Utrobin04}, the positron fraction within the ejecta varies accordingly, influencing the overall plasma dynamics of the SN. Therefore, similar to the discussion of SNe~Ia in Section~\ref{subsubsec:Ia_overview}, we consider two extreme cases for SNe~II as well. One where $^{56}$Ni is centrally concentrated and another where it is mixed with the ejecta. SNe~II in general have a greater amount of mixing than SNe~Ia, and are more aspherical \citep[e.g.,][]{Vasylyev23,Nagao24}. 

Key observational properties, such as $M_{\mathrm{Ni}}$ and $M_{\mathrm{ej}}$, are typically inferred from the light curves of SNe~II. The nickel mass can vary widely, with reported values ranging from $10^{-3}$ to $0.3\,\mathrm{M_{\odot}}$, and a median value of approximately $0.03\,\mathrm{M_{\odot}}$ \citep{Hamuy03,Muller17,Anderson19,Rodriguez21}. The ejecta mass is typically $10\,\mathrm{M_{\odot}}$ with a range of $8$ to $30\,\mathrm{M_{\odot}}$ \citep[e.g.,][]{Martinez22}. Ejecta velocities ($v_{\mathrm{ej}}$) usually determined from spectral features like the Fe{\sc ii} $\lambda5169$ absorption line span a range from $1000$ to $9000\,\mathrm{km\,s^{-1}}$, with a median velocity of $4000\,\mathrm{km\,s^{-1}}$ \citep{Hamuy03,Gutierrez17}. Velocities of outer layers of ejecta are much faster, estimated with H and He lines at a median of $7000\,\mathrm{km\,s^{-1}}$, with a range of $2000$ to $14{,}000\,\mathrm{km\,s^{-1}}$ \citep{Gutierrez17}.

The progenitors of SNe~II are red supergiant (RSG) stars, which have been identified in pre-explosion imaging \citep[e.g.,][]{VanDyk12a,VanDyk12b,Smartt15,Maund17,Healy24}. The radii of these RSGs typically range from $100$ to $1500\,\mathrm{R_{\odot}}$ \citep{Levesque10,Levesque17}, with the radii of their iron core before collapse ranging from $1000$ to $2000\,\mathrm{km}$ \citep{Woosley02,Foglizzo15}. For example, the radius of Betelgeuse, a well-studied RSG, is measured to be $\sim 862\,\mathrm{R_{\odot}}$ \citep[e.g.,][]{Levesque05,Smith09}. We adopt $862\,\mathrm{R_{\odot}}$ as a representative radius for typical RSG progenitors and $\sim1500\,\mathrm{km}$ for the iron core prior to explosion in the case of centrally concentrated nickel.

\subsubsection{Magnetic Fields} \label{subsubsec:II_Bfields}
Following the same reasoning outlined in Section~\ref{subsubsec:Ia_Bfields}, the evolution of the plasma-generated magnetic field in Type II supernovae (SNe II) can be derived as a function of time using Eq.\ref{eq:Bsat_vs_f_p} with the ion number density defined by Eq.\ref{eq:n_i} using SNe~II parameters from Section~\ref{subsubsec:II_overview}. 

For the scenario where $^{56}$Ni is centrally concentrated, we assume a typical ejecta velocity for Fe-group elements of $v_{\mathrm{ej}} = 4000\,\mathrm{km\,s^{-1}}$, which is lower than that of the outer layers. We also use a typical $^{56}$Ni mass of $M_{\mathrm{Ni}} = 0.03\,\mathrm{M_{\odot}}$, an ion mass of $m_i = 56 m_e = 9.3 \times 10^{-23}\,\mathrm{g}$ (corresponding to $A \sim 56$), and radius of the initial iron core of $R_0 \sim 1500\,\mathrm{km}$ in Eq.~\ref{eq:n_i} to obtain $n_i (t)$.

For the case of complete $^{56}$Ni mixing, we assume an ejecta velocity of $v_{\mathrm{ej}} = 7000\,\mathrm{km\,s^{-1}}$, a total ejecta mass of $M_{\mathrm{ej}} = 10\,\mathrm{M_{\odot}}$, an ion mass of $m_i = 3.2 \times 10^{-23}\,\mathrm{g}$ (corresponding to $A \sim 19$, weighted mean of all ejecta ions), and a progenitor radius of $R_0 = 862\,\mathrm{R_{\odot}}$ to calculate $n_i (t)$ using Eq.~\ref{eq:n_i}.

The temporal evolution of magnetic field strength for SNe~II under these scenarios is illustrated in Figure~\ref{fig:B_sat_prog}b. The dark cyan band represents the range of possible magnetic field strengths generated by plasma instabilities under different SNe~II physical parameter combinations (i.e., $M_{\mathrm{ej}}$, $M_{\mathrm{Ni}}$, $v_{\mathrm{ej}}$, $R_0$). The dashed line shows the magnetic field evolution for a set of typical SNe~II parameters for the case of centrally concentrated $^{56}$Ni, while the dash-dotted line shows the evolution for $^{56}$Ni mixed within the ejecta. Although both scenarios show a similar trend, centrally concentrated $^{56}$Ni results in slightly higher magnetic field strengths due to the increased ion density near the center of the ejecta.

\citet{Tessore17} have measured the surface magnetic field of several RSG stars, finding field strengths on the order of a Gauss with variation of up to $\sim 10\,\mathrm{G}$. Betelgeuse, one of the most extensively studied RSG stars, has a surface magnetic field ranging from $0.1$ to $2\,\mathrm{G}$ \citep{Mathias18}. Based on these observations, we adopt a range of $0.1$ to $10\,\mathrm{G}$ for the progenitor magnetic fields, with a typical value of $1\,\mathrm{G}$. Similar to the discussion in Section~\ref{subsubsec:Ia_Bfields}, the progenitor surface magnetic field is expected to decline over time following a $B_{\mathrm{prog}} \propto t^{-2}$ relationship, as the ejecta expand. The violet band in Figure~\ref{fig:B_sat_prog}b shows this decay for a range of initial progenitor field strengths and progenitor radii. The dotted line represents the field evolution for a typical RSG progenitor with an initial surface field of $1\,\mathrm{G}$. Initially, the progenitor field evolves slowly until the ejecta expansion ($v_{\mathrm{ej}} t \gg R_0$) dominates, after which the field strength decreases more rapidly.

The magnetic field generated by plasma instabilities in the ejecta begins with a large spread before $\sim1$ day due to the wide range of possible explosion parameters. While the ranges of plasma-generated and progenitor magnetic fields overlap initially, the plasma-generated field surpasses the progenitor surface field within $\sim 1$ day in our typical cases (dashed and dash-dotted lines in Figure~\ref{fig:B_sat_prog}b). This rapid increase in difference aligns with the timescale on which ion densities decrease enough to reduce positron-electron annihilation, thereby allowing the instability to sustain itself and amplify the magnetic field. As the ejecta continue to expand, the ion density decreases, and consequently, the instantaneous saturated magnetic field weakens. This decrease follows a power-law decline, as indicated by the slope of approximately $-1.5$ in the logarithmic plot in Figure~\ref{fig:B_sat_prog}b. The plasma-generated magnetic field remains dominant over the ISM magnetic field until roughly $10^5$ days post-explosion, after which the ISM field strength becomes comparable.

It is important to note that these estimates are based on surface magnetic fields of the progenitor star. The internal magnetic field configuration of an RSG progenitor could differ substantially, potentially leading to variations in the magnetic field strength and geometry within the expanding supernova ejecta. Observations of internal fields in massive stars are limited, with the most massive stars showing internal fields of tens of kG for stars with $M = 1-2\,\mathrm{M_{\odot}}$ \citep[][]{Hatt24}, though such measurements are unavailable for late-stage massive stars ($M > 8 \,\mathrm{M_{\odot}}$).

In addition to $^{56}$Ni, other decay chains can contribute to the overall positron population. For example, $^{44}$Ti decay becomes significant at later times ($> 10^4$ days) and could modestly increase the positron fraction \citep[e.g.,][]{Magkotsios10}. However, with $M_{\mathrm{Ti}}/M_{\mathrm{Ni}} < 10^{-3}$ and the $\beta^+$ branching ratio of $^{44}$Sc being only about five times larger than that of $^{56}$Co (see Table~\ref{tab:decay_data_SNe}), the positron contribution from $^{44}$Ti is negligible under our current assumptions. Consequently, its impact on the saturated magnetic field is minimal and is not included in our analysis.

Excess luminosity from interaction between the ejecta and CSM hampers conclusions about positron trapping at late times. The emission of SNe~II is expected to be CSM-dominated at some point \citep[][]{Dessart23,RizzoSmith23}, which complicates empirical estimates of the positron trapping and contributes to a flattened light curve at late times \citep[][]{Baer-Way24}. Furthermore, nebular-phase modeling efforts yield mixed results: some reproduce the data well by assuming local positron confinement \citep[e.g.,][]{Dessart23b}, while others require significant positron escape \citep[e.g.,][]{Silverman17}, and still others remain inconclusive \citep[e.g.,][]{Jerkstrand15}.

\subsection{Stripped-Envelope SNe} \label{subsec:SESNe}
\subsubsection{Overview and Physical Properties} \label{subsubsec:SE_overview}
SE~SNe are a subclass of CC~SNe originating from progenitors that have lost their outer hydrogen and, in some cases, helium envelopes before explosion \citep{clocchiatti96,matheson01}. This loss of material can occur due to strong stellar winds or binary interactions \citep[e.g.,][]{podsiadlowski93,woosley95,wellstein99,wellstein01,heger03,podsiadlowski04,pauldrach12,Benvenuto13}. SE~SNe are further categorized based on the presence or absence of hydrogen and helium in their spectra: Type IIb SNe show transient hydrogen features that diminish over time, Type Ib SNe lack hydrogen but show strong helium lines, and Type Ic SNe are devoid of both hydrogen and helium lines, indicating more extensive stripping \citep[][]{filippenko97}.

The progenitors of SE~SNe are believed to be massive stars that have undergone significant mass loss. For SNe~IIb, the progenitors are often identified as blue supergiants (BSGs), yellow supergiants (YSGs), or RSGs, which retain a thin hydrogen envelope at the time of explosion \citep{Aldering94,Crockett08,Maund04,Maund11,VanDyk11,VanDyk14,Smartt15}. In contrast, SNe Ib and Ic are believed to originate from Wolf-Rayet (WR) stars, which have strong stellar winds capable of removing both hydrogen and helium layers, or interacting binaries \citep[e.g.,][]{Dessart11,Georgy12,Smartt15,Yoon17}. However, there is some debate on the viability of WR stars as progenitors of SE~SNe due to limited direct observational evidence \citep[see][]{Yoon12}. The pre-explosion radii of SE~SNe progenitors vary significantly, ranging from about $1$ to $15\,\mathrm{R_{\odot}}$ for WR stars \citep[e.g.,][]{Modjaz09,Grafener12} and extending up to several hundred solar radii for more extended supergiants. In our analysis, we adopt a typical radius of $15\,\mathrm{R_{\odot}}$, representing a larger WR star, and a range of $1$ to $500\,\mathrm{R_{\odot}}$. Similar to SNe~II, we assume iron-core radii before collapse between $1000$ to $2000\,\mathrm{km}$ with a typical value of $\sim1500\,\mathrm{km}$ for the case of centrally concentrated nickel \citep[e.g.,][]{Woosley02}.

The light curves of SE~SNe often exhibit a double-peaked structure. The first peak is commonly attributed to shock cooling, where interaction with the CSM or extended stellar envelope heats the ejecta, leading to additional emission \citep[e.g.,][]{Desai23,Ertini23}. This initial peak precedes the main radioactive-powered peak and is often the result of the shock's energy diffusing through the outer layers of the progenitor star or CSM. The primary peak, occurring later, is mainly powered by the radioactive decay of $^{56}$Ni to $^{56}$Co and subsequently to $^{56}$Fe, similar to SNe~Ia and II \citep[e.g.,][]{Arnett82,Prentice16,Desai23,Ertini23,Rodriguez24}. The amount of $^{56}$Ni synthesized in SE~SNe varies significantly, from $0.01$ to $0.7\,M_{\odot}$, with a typical value of $0.1\,M_{\odot}$, which is generally higher than in SNe~II \citep{Prentice16,Taddia18,Anderson19,Rodriguez21,Rodriguez23,Desai23}. The total ejecta mass $M_{\mathrm{ej}}$ is lower than SNe~II due to the varying amounts of stripping of outer layers, ranging from $0.2$ to $10\,M_{\odot}$, with a typical value of $3\,M_{\odot}$ \citep{Prentice16,Taddia18,Anderson19,Rodriguez23,Desai23}. Ejecta velocities from Fe{\sc ii} $\lambda5169$ range from $4000$ to $12{,}000\,\mathrm{km\,s^{-1}}$, with a typical value of $8000\,\mathrm{km\,s^{-1}}$ and velocities of lighter elements from H$\alpha$ and He{\sc i} $\lambda5876$ range from $5000$ to $20{,}000\,\mathrm{km\,s^{-1}}$, with a typical value of $10{,}000\,\mathrm{km\,s^{-1}}$ \citep{Liu16,Holmbo23}. 

As with SNe~Ia in Section~\ref{subsubsec:Ia_overview} and SNe~II in Section~\ref{subsubsec:II_overview}, we consider the same two extreme cases for SE~SNe: one where $^{56}$Ni is centrally concentrated and another where it is mixed with the ejecta. The former assumes ejecta velocities of heavier elements from Fe{\sc ii} $\lambda5169$ line velocities and the latter assumes ejecta velocities from lighter elements from H$\alpha$ and He{\sc i} $\lambda5876$ line velocities.

\subsubsection{Magnetic Fields} \label{subsubsec:SE_Bfields}
From SNe~Ia to SNe~II to SE~SNe, the complexity of understanding plasma-generated magnetic fields increases. Unlike SNe~II, the progenitors of SE~SNe are not well understood. Additionally, SE~SNe often interact with the CSM, further complicating studies by introducing environmental contamination.

Assuming a simple $^{56}$Ni model with two extreme scenarios, we can gain insights into plasma-generated magnetic fields in the nickel region. For the first case of central $^{56}$Ni, we adopt typical values of $M_{\mathrm{Ni}} = 0.1\,M_{\odot}$, $m_i = 56 m_e = 9.3 \times 10^{-23}\,\mathrm{g}$ (corresponding to $A \sim 56$), $v_{\mathrm{ej}} = 8000\,\mathrm{km\,s^{-1}}$, and initial iron core radius of $R_0 \sim 1500\,\mathrm{km}$. For the second case of mixed $^{56}$Ni, we adopt typical values of $M_{\mathrm{Ni}} = 0.1\,M_{\odot}$, $M_{\mathrm{ej}} = 3\,M_{\odot}$, $m_i = 3.2 \times 10^{-23}\,\mathrm{g}$ (corresponding to $A \sim 19$), $v_{\mathrm{ej}} = 10{,}000\,\mathrm{km\,s^{-1}}$, and initial progenitor radius of $15\,\mathrm{R_{\odot}}$ in Eq.~\ref{eq:n_i} to obtain $n_i (t)$ for SE~SNe.

Figure~\ref{fig:B_sat_prog}c illustrates the evolution of the magnetic field strength in SE~SNe. The dark cyan band shows the range of magnetic field produced by plasma instabilities under various combinations of $M_{\mathrm{ej}}$, $M_{\mathrm{Ni}}$, $v_{\mathrm{ej}}$, and $R_0$. The dashed line represents the magnetic field evolution for a typical SE~SN with centrally concentrated $^{56}$Ni and the dash-dotted line for mixed $^{56}$Ni. While the saturated magnetic field qualitatively behaves similarly to that in SNe~II, it has a tighter range and stronger difference compared to the progenitor fields. Additionally, the spread in the progenitor magnetic field is wider, reflecting the broad range of potential progenitor scenarios for SE~SNe.

Given the uncertainty surrounding SE~SNe progenitors, their surface magnetic fields are even more uncertain. For this reason, we assume fields similar to those of RSGs (Gauss-level, see Section~\ref{subsubsec:II_Bfields}), and scale them radially by conserving magnetic flux to estimate surface fields for other progenitor cases. Assuming magnetic energy density the same as RSGs at radii of RSGs, YSGs will have $<500\,\mathrm{G}$ surface fields at radii of $\sim60\,\mathrm{R_{\odot}}$. Rigel is a lower-mass BSG compared to typical SE~SNe progenitor masses \citep[][]{Smartt15}, but one of the only BSGs with measurable surface magnetic fields. It shows no significant magnetic field detected, with an upper limit of $50\,\mathrm{G}$ \citep{Shultz11}. WR stars have measured magnetic fields reaching up to $\sim500\,\mathrm{G}$ \citep[][]{delaChevrotiere14}. Consequently, we estimate a wide range of initial surface magnetic fields, from $\sim500\,\mathrm{G}$ for YSGs ($\sim60\,\mathrm{R_{\odot}}$) to $\sim500\,\mathrm{G}$ for WR stars ($\sim1\,\mathrm{R_{\odot}}$). 

By comparing plasma-generated fields with progenitor magnetic fields (violet band in Figure~\ref{fig:B_sat_prog}c), which decay as $B_{\mathrm{prog}} \propto t^{-2}$, we find that plasma-generated fields surpass typical progenitor fields (violet dotted line) within $10^{-1}$ days in our typical cases. Even in the most extreme scenario, plasma-generated fields dominate after $\sim10^{2}$ days post-explosion. Subsequently, the instantaneous plasma-generated field follows a power-law index of $-1.5$, similar to SNe~Ia and SNe~II. Due to the higher $^{56}$Ni mass and faster expansion velocities of SE~SNe, the resulting magnetic field strengths are comparable to those in SNe~II and significantly exceed progenitor fields throughout the early phases. Plasma-generated fields remain dominant over the ISM magnetic field until $\sim10^{4}-10^{5}$ days post-explosion, at which point the ISM field becomes comparable in strength.

\subsection{Kilonovae} \label{subsec:KNe}
\subsubsection{Overview and Physical Properties} \label{subsubsec:KN_overview}
KNe are electromagnetic transients powered by the merger of binary NS or an NS and a black hole \citep[BH;][]{LiPaczynski98,Metzger10,Roberts11,BarnesKasen13}. Observations of the gravitational wave event GW170817 \citep{Abbott17a} have provided compelling evidence of an optical/infrared KN, consistent with theoretical predictions for the electromagnetic counterpart of a binary NS merger \citep[e.g.,][]{Abbott17b,Arcavi17,Chornock17,Coulter17,Cowperthwaite17,Drout17,Kasen17,Kasliwal17,Kilpatrick17,McCully17,Nicholl17,Shappee17a,Smartt17,Soares-Santos17,Tanaka17,Tanvir17,PiroKollmeier18}. NS radii, including magnetic and rotating NSs, typically range from $6-20\,\mathrm{km}$, with a common value around $12\,\mathrm{km}$ \citep[e.g.,][]{Lattimer04,Steiner13}.

During these mergers, neutron-rich material is ejected and undergoes rapid neutron capture ($r$-process), synthesizing heavy elements \citep{Arnould07}. Radioactive decay of these r-process elements powers the KNe light curve, typically peaking within $\lesssim1$ day post-merger \citep{Metzger10,BarnesKasen13,TanakaHotokezaka13,Grossman14}. Compared to the iron-rich ejecta of SNe, the heavier and more neutron-rich ejecta in KNe exhibit distinct opacity and energy transport properties due to lanthanides and actinides, which must be factored into radiative transfer models for accurate light curves \citep{Kasen13}. The ejected mass is estimated at $M_{\mathrm{ej}} = 10^{-3} - 10^{-1}\,\mathrm{M_{\odot}}$ \citep{BarnesKasen13}, with ejecta velocities on the order of NS escape speeds, $v_{\mathrm{ej}} = 0.1 - 0.3\,c$ \citep{Metzger10,Kasen13}. For a common $r$-process nucleus mass ($m_i \sim 130\, \text{amu}$), the evolving ion number density in the homogeneously expanding ejecta can be expressed as:
\begin{align} \label{eq:n_KN}
n_{i,\mathrm{KN}} (t) =\,&1.58 \times 10^{8} \, \text{cm}^{-3} \, \left(\frac{M_{\text{ej}}}{10^{-2} \, M_{\odot}}\right) \left(\frac{m_i}{130\, \text{amu}}\right)^{-1} \notag \\
&\times \left(\frac{R_0 + 86400\,v_{\text{ej}}\,t}{12 \, \text{km} + (6.0 \times 10^{4}) \, \text{km s}^{-1} \cdot 1 \, \text{d}}\right)^{-3} .
\end{align}
This equation is relevant for modeling magnetic field generation in KNe. We acknowledge the fact that uncertainties in nuclear physics and diversity in the models produce a broad range of possible physical parameters for KNe \citep{Barnes21}, and that Eq.~\ref{eq:n_KN} is only a crude approximation.

The $r$-process yields a broad distribution of nuclei with mass numbers $A\sim110 - 210$. NS mergers create nuclei with solar-like abundances, peaking near $A\sim130$ and $A\sim195$, with the first peak dominating $\beta$-decay heating rates \citep[e.g.,][]{Metzger10,Roberts11,Korobkin12}. As \citet{LiPaczynski98} predicted, the precise distribution of heavy nuclei does not critically affect the total radioactive heating rate, provided the heating is not dominated by a few decay chains. This allows for statistical modeling, leading to an approximate power-law radioactive heating profile.

\citet{Korobkin12} show that the NS merger energy generation rate can be fit by
\begin{equation}
    \dot{\epsilon}(t) = \epsilon_0 \left( \frac{1}{2} - \frac{1}{\pi} \arctan{\frac{t - t_0}{\sigma}} \right)^\alpha \times \left( \frac{\epsilon_{\text{th}}}{0.5} \right) ,
    \label{eq:nuc_energy_gen}
\end{equation}
where $\epsilon_0 = 2 \times 10^{18}\, \mathrm{ergs\,s^{-1}\,g^{-1}}$, $t_0 = 1.3\,\mathrm{s}$, $\sigma = 0.11\,\mathrm{s}$, and $\alpha=1.3$. Eq.~\ref{eq:nuc_energy_gen} transitions smoothly from a constant rate at early times ($t < t_0$) due to the $r$-process to a power-law decay, $\dot\epsilon \propto \epsilon_0 t^{-\alpha}$ at later times ($t \gtrsim 5\,\mathrm{s}$) as synthesized isotopes decay back to stability. The power-law index $\alpha$ has been estimated between $1.1$ and $1.4$ \citep{Metzger10,Goriely11,Roberts11,Korobkin12}. 

\citet{Barnes16} report that most of the total radioactive energy in KNe comes from $\beta$-decays from 1 second after the merger up to 100 days, with the $\beta$-decay contribution ranging from $f_{\beta} = 50 - 90\,\%$ over this time span. Beyond 100 days, we extrapolate to a constant fraction of $f_{\beta} = 80\,\%$, and $f_{\beta} = 50\,\%$ before 1 second. Additionally, \citet{Barnes16} estimate that about $f_{\beta, \mathrm{part}} = 20\,\%$ of this $\beta$-decay energy is released as $\beta$-particles (high-energy electrons), with an average particle energy of $E_{\beta} = 0.5\,\mathrm{MeV}$. Consequently, the total $\beta$-particle energy generation rate is:
\begin{equation}
    \dot{\epsilon}_{\beta}(t) = f_{\beta} \, f_{\beta,\mathrm{part}} \, \dot{\epsilon}(t) .
    \label{eq:eps_beta_dot}
\end{equation} 
The $\beta$-particle (high-energy electron) number density over time is then
\begin{equation}
    n_{\beta}(t) = \rho_{\mathrm{KN}}(t) \int_0^t \frac{\dot{\epsilon}_{\beta}(t')}{E_{\beta}} \, dt',
\end{equation}
where $\rho_{\mathrm{KN}}(t)$ is the mass density of KNe as a function of time, approximated by the ejecta mass $M_{\mathrm{ej}}$ divided by the homologously expanding ejecta volume. Finally, assuming the same volume for electrons and ions, the fraction of high-energy electrons is 
\begin{equation}
    f_e(t) = n_{\beta}(t)/n_{i,\mathrm{KN}}(t) ,    
    \label{eq:f_e_KN}
\end{equation}
which will be used in Eq.~\ref{eq:Bsat_vs_f_p} for the calculation of the magnetic field.

In addition to the primary KN emission from the $r$-process ejecta, there is a distinct signal expected within the first hours. This arises from the outermost ejecta layers with velocities $> 0.6\,c$ and masses in the range $10^{-5} - 10^{-4} M_\odot$, powered by the beta decay of free neutrons (half-life of about 10 minutes) in these fast-moving regions \citep[e.g.,][]{Metzger15}. The resulting electrons are also expected to drive streaming instabilities, though the dynamics may differ due to the comparable velocities of the free neutrons and decay electrons. As illustrated in Figure~\ref{fig:free_n} (Appendix~\ref{app:free_n}), magnetic field strengths generated by free neutron decay, while lower than those from $r$-process decay, could still surpass progenitor fields.

\subsubsection{Magnetic Fields} \label{subsubsec:KN_Bfields}
Unlike SNe, which produce high-energy positrons via $\beta^+$-decay of isotopes such as $^{56}$Ni, KNe primarily generate high-energy electrons through the $\beta^-$-decay of $r$-process elements. This fundamental distinction requires adapting the magnetic field generation models originally developed for SNe. Specifically, the high-energy electron fraction $f_e$ replaces the high-energy positron fraction $f_p$ used in SNe models (Eq.~\ref{eq:Bsat_vs_f_p}), to reflect the shift in dominant decay mechanisms.
 
In KNe, filamentation instabilities arise as high-energy electrons stream through the ejecta, generating localized magnetic fields. As discussed in Section~\ref{sec:PIC}, plasma instabilities driven by charged particle velocity anisotropies function similarly in KNe as in SNe, despite the replacement of high-energy positrons with electrons. The MeV-range electrons in KNe trigger these instability mechanisms, leading to magnetic field amplification via similar processes. These magnetic fields are generated following Eq.~\ref{eq:Bsat_vs_f_p}, with the high-energy electron fraction provided by Eq.~\ref{eq:f_e_KN}.

Another source of magnetic fields in KNe is the progenitor NSs. Typical surface magnetic field strengths of NSs span a broad range, from $10^7$ to $10^{15}\,\mathrm{G}$. Dipolar magnetic fields range from $10^8$ to $10^{14}\,\mathrm{G}$, as inferred from radio pulsars, accreting X-ray pulsars, and low-mass X-ray binaries \citep[e.g,][]{Reisenegger01,Bhattacharya02,Igoshev21}. In some cases, these fields are formed through the flux conservation of fossil fields in O and B stars \citep{Igoshev21}. The surface magnetic fields of millisecond pulsars typically lie between $10^7$ and $10^{11}\,\mathrm{G}$ \citep[e.g.,][]{Miller98,Psaltis99,LambBoutloukos08}, while magnetars--highly magnetic NSs--exhibit stronger fields, ranging from $10^{11}$ to $10^{15}\,\mathrm{G}$ \citep[see review by][]{KaspiBeloborodov17}. Although magnetars may harbor even stronger internal magnetic fields (up to $>10^{16}\,\mathrm{G}$), these ultra-strong fields are transient and decay within $\sim 1$ kyr \citep{BeloborodovLi16}. Given their rarity and short lifetimes, such extreme fields are unlikely to be present on merging NSs and are therefore excluded from our range of considered values.

A comparison between plasma-generated magnetic fields (dark cyan band) and progenitor fields (violet band) is shown in Figure~\ref{fig:B_sat_prog}d. While progenitor fields decay as $B_{\mathrm{prog}} \propto t^{-2}$ due to the assumption of homologous expansion, starting from values between $10^7$ and $10^{15}\,\mathrm{G}$ with a typical value of $10^{12}\,\mathrm{G}$, the instantaneous saturated magnetic fields follow an almost pure power law, approximately $B_{\mathrm{plas}} \propto t^{-1.5}$. The shallower slope of plasma-generated fields ensures they remain stronger than progenitor fields throughout the KN’s evolution. This dominance highlights the importance of plasma instabilities in shaping the magnetic field environment of KNe.

It has long been suspected that magnetic fields influence the dynamics of charged particles in KNe \citep[e.g.,][]{Barnes16,Ciolfi20,Palenzuela22}. Some studies suggest that magnetic field amplification can occur during accretion, where turbulence driven by the merger or in the resultant accretion disk can amplify initial fields of $10^{13}\,\mathrm{G}$ by a factor of $\sim 10^3$, saturating at a magnetic field energy of $\gtrsim 4 \times 10^{50}\,\mathrm{erg}$ \citep{Kiuchi14,Kiuchi15}. This is approximately an order of magnitude stronger than the strongest assumed progenitor fields but still subdominant to the typical plasma-generated fields. These fields, whether originating from progenitors, from plasma instabilities, or from merger-driven amplification, play a critical role in confining high-energy electrons within the ejecta and regulating energy transport. 

In Figure~\ref{fig:B_sat_prog}d, we show that magnetic fields generated by filamentation-like plasma instabilities consistently exceed progenitor fields in strength. This suggests that even in the absence of pre-existing fields, plasma instabilities alone can generate and amplify magnetic fields sufficiently to dominate the dynamics of charged particles, thereby influencing the energy transport and physical properties of KNe.

\section{Conclusion and Summary} \label{sec:conclusion}
In this work, we explore the fundamental role of plasma instabilities in shaping the behavior of high-energy leptons within the ejecta of radioactive transients, including SNe~Ia, SNe~II, SE~SNe, and KNe, where the radioactive decay of synthesized isotopes releases high-energy leptons through $\beta^+$ and $\beta^-$ decay. These leptons significantly impact the transient's evolution by depositing energy into the expanding ejecta. When the ejecta becomes collisionless, electromagnetic forces dominate particle dynamics. This work represents the first step in exploring the potential significance of plasma streaming instabilities in astrophysical transients, aiming to provide an order-of-magnitude estimate of their effects.

Using fully kinetic PIC simulations, we demonstrate that plasma streaming instabilities amplify magnetic fields within the ejecta almost instantaneously (within microseconds), even in initially unmagnetized environments. These self-generated fields slow lepton diffusion by confining high-energy leptons and redirecting their energy into heating the thermal electrons and ions. Our results show that plasma-generated magnetic fields dominate any residual progenitor magnetic fields after $\lesssim 1$ day.

The confinement of high-energy leptons by plasma instabilities provides a natural explanation for the sustained positron trapping observed in late-time light curves of SNe~Ia. Unlike models requiring pre-existing strong magnetic fields, plasma instabilities alone can drive significant magnetic field amplification, thereby altering particle transport and heating dynamics. This instability-driven mechanism likely plays a ubiquitous role across diverse radioactive transients, affecting their thermal and magnetic field evolution. Furthermore, observed positron trapping has implications for the Galactic positron problem, potentially refining our understanding of the contribution of SNe to Galactic positron populations \citep[e.g.,][]{MeraEvans22}. Our work provides theoretical support for observational studies, contributing to a deeper understanding of the physical processes that govern SNe and KNe.

On the observational front, our results open up new pathways for studying the influence of magnetic fields in radioactive transients. Prompt synchrotron emission from trapped positrons, driven by interactions with self-generated magnetic fields, could serve as a direct observational signature of plasma instabilities. Under our highly optimistic scenario, nearby extragalactic SNe~Ia, like SN~2011fe and SN~2014J, are unlikely to be detected by LOFAR, SKAO, and ngVLA (next-generation Very Large Array). For most SNe, CSM interaction will likely dominate the synchrotron emission. Therefore, only a Galactic or local-group SN~Ia in a very clean environment may be detectable with LOFAR and SKAO. Further theoretical work, luck (Galactic SN with no CSM), and future generations of radio telescopes are needed to realize the potential of this diagnostic to directly probe the magnetic fields from plasma instabilities in SN ejecta.

While this study establishes the relevance of plasma streaming instabilities, several simplifying assumptions were made, which may impact the first-order details of the results. These assumptions come from numerous uncertainties in both the fields and particles of the systems. First, we assumed that leptons escape as a collimated beam into an unmagnetized environment and are instantaneously injected, which oversimplifies the complexity of ejecta conditions. Nonetheless, the saturated magnetic field should still depend on the free energy of the drifting leptons. Furthermore, mechanisms such as continuous particle injection are expected to slightly enhance the magnetic field saturation level and represent an important avenue for future investigation. Pre-existing magnetic fields, if present, would alter the instability's specifics without negating its self-confining nature. Second, while the instability growth rate consistently exceeds collisional relaxation times, collisions--neglected in this work--would influence the energy partition between fields and particles at extremely early times. Third, our PIC simulations employed a reduced ion-to-electron mass ratio, which likely affects the final energy distribution, but not the instability's growth or initial saturation. 

Additional uncertainties arise from the assumed initial ejecta temperature and ionic composition, as well as the neglect of wave damping, which may influence the long-term evolution of electromagnetic fields. Moreover, the structure of the magnetic power spectrum in $k$-space, critical for lepton escape rates and energy transfer, depends on the details of the initial leptonic distribution function, which warrants further investigation. Future work should incorporate transport models for lepton escape based on diffusion coefficients derived from self-generated fields.

Despite these limitations, our findings underscore that plasmas inherently abhor anisotropies in the distribution function. If high-energy leptons escape a localized region, plasma streaming instabilities are likely to be generated to confine them. This study represents a foundational step toward understanding the significance of plasma instabilities in astrophysical transients. While we leave the refinement of specific details to future work, our results strongly suggest that these instabilities play a crucial role in transient evolution across a wide range of events.

\section*{Acknowledgments}
We thank Federica Chiti, Brian Metzger, Christopher Kochanek, J. J. Hermes, and Anthony Piro for helpful comments.

The Shappee group at the University of Hawai'i is supported with funds from NSF (grants AST-1908952, AST-1911074, \& AST-1920392) and NASA (grants HST-GO-17087, 80NSSC24K0521, 80NSSC24K0490, 80NSSC24K0508, 80NSSC23K0058, \& 80NSSC23K1431).

The work of CCH is supported by the NSF/DOE Grant PHY-2205991, NSF-FDSS Grant AGS-1936393, NSF-CAREER Grant AGS-2338131 and NASA grant HTMS-80NSSC24K0173. Simulations were performed on TACC’s Stampede 2 and Purdue’s ANVIL. With allocations through NSF-ACCESS (formally XSEDE) PHY220089 and AST180008.

CA is supported by STScI grants (JWST-GO-02114, JWST-GO-02122, JWST-GO-04522, JWST-GO-03726, JWST-GO-6582, HST-AR-17555, JWST-GO-04217, JWST-GO-6023, JWST-GO-5290, JWST-GO-5057, JWST-GO-6677) and JPL-1717705.

LC is grateful for support from NSF grant AST-2107070.

AB acknowledges support by the Ministerio de Econom\'ia y Competitividad of Spain (Grant No. PID2021-125550OB-I00).

\section*{Data Availability}
The PIC code \texttt{tristan-mp v2} is publicly available and the simulated data underlying this paper can be shared upon reasonable request of the corresponding authors.




\bibliographystyle{mnras}
\bibliography{plasma} 

\begin{thebibliography}{}
\makeatletter
\relax
\def\mn@urlcharsother{\let\do\@makeother \do\$\do\&\do\#\do\^\do\_\do\%\do\~}
\def\mn@doi{\begingroup\mn@urlcharsother \@ifnextchar [ {\mn@doi@} {\mn@doi@[]}}
\def\mn@doi@[#1]#2{\def\@tempa{#1}\ifx\@tempa\@empty \href {http://dx.doi.org/#2} {doi:#2}\else \href {http://dx.doi.org/#2} {#1}\fi \endgroup}
\def\mn@eprint#1#2{\mn@eprint@#1:#2::\@nil}
\def\mn@eprint@arXiv#1{\href {http://arxiv.org/abs/#1} {{\tt arXiv:#1}}}
\def\mn@eprint@dblp#1{\href {http://dblp.uni-trier.de/rec/bibtex/#1.xml} {dblp:#1}}
\def\mn@eprint@#1:#2:#3:#4\@nil{\def\@tempa {#1}\def\@tempb {#2}\def\@tempc {#3}\ifx \@tempc \@empty \let \@tempc \@tempb \let \@tempb \@tempa \fi \ifx \@tempb \@empty \def\@tempb {arXiv}\fi \@ifundefined {mn@eprint@\@tempb}{\@tempb:\@tempc}{\expandafter \expandafter \csname mn@eprint@\@tempb\endcsname \expandafter{\@tempc}}}

\bibitem[\protect\citeauthoryear{{Abbott} et~al.,}{{Abbott} et~al.}{2017a}]{Abbott17a}
{Abbott} B.~P.,  et~al., 2017a, \mn@doi [\prl] {10.1103/PhysRevLett.119.161101}, \href {https://ui.adsabs.harvard.edu/abs/2017PhRvL.119p1101A} {119, 161101}

\bibitem[\protect\citeauthoryear{{Abbott} et~al.,}{{Abbott} et~al.}{2017b}]{Abbott17b}
{Abbott} B.~P.,  et~al., 2017b, \mn@doi [\apjl] {10.3847/2041-8213/aa91c9}, \href {https://ui.adsabs.harvard.edu/abs/2017ApJ...848L..12A} {848, L12}

\bibitem[\protect\citeauthoryear{{Aldering}, {Humphreys}  \& {Richmond}}{{Aldering} et~al.}{1994}]{Aldering94}
{Aldering} G.,  {Humphreys} R.~M.,   {Richmond} M.,  1994, \mn@doi [\aj] {10.1086/116886}, \href {https://ui.adsabs.harvard.edu/abs/1994AJ....107..662A} {107, 662}

\bibitem[\protect\citeauthoryear{{Anderson}}{{Anderson}}{2019}]{Anderson19}
{Anderson} J.~P.,  2019, \mn@doi [\aap] {10.1051/0004-6361/201935027}, \href {https://ui.adsabs.harvard.edu/abs/2019A&A...628A...7A} {628, A7}

\bibitem[\protect\citeauthoryear{{Arcavi} et~al.,}{{Arcavi} et~al.}{2017}]{Arcavi17}
{Arcavi} I.,  et~al., 2017, \mn@doi [\nat] {10.1038/nature24291}, \href {https://ui.adsabs.harvard.edu/abs/2017Natur.551...64A} {551, 64}

\bibitem[\protect\citeauthoryear{{Arnett}}{{Arnett}}{1982}]{Arnett82}
{Arnett} W.~D.,  1982, \mn@doi [\apj] {10.1086/159681}, \href {https://ui.adsabs.harvard.edu/abs/1982ApJ...253..785A} {253, 785}

\bibitem[\protect\citeauthoryear{{Arnett}}{{Arnett}}{1996}]{Arnett96}
{Arnett} D.,  1996, {Supernovae and Nucleosynthesis: An Investigation of the History of Matter from the Big Bang to the Present}.
Princeton University Press

\bibitem[\protect\citeauthoryear{{Arnould}, {Goriely}  \& {Takahashi}}{{Arnould} et~al.}{2007}]{Arnould07}
{Arnould} M.,  {Goriely} S.,   {Takahashi} K.,  2007, \mn@doi [\physrep] {10.1016/j.physrep.2007.06.002}, \href {https://ui.adsabs.harvard.edu/abs/2007PhR...450...97A} {450, 97}

\bibitem[\protect\citeauthoryear{{Ashall}, {Mazzali}, {Bersier}, {Hachinger}, {Phillips}, {Percival}, {James}  \& {Maguire}}{{Ashall} et~al.}{2014}]{Ashall14}
{Ashall} C.,  {Mazzali} P.,  {Bersier} D.,  {Hachinger} S.,  {Phillips} M.,  {Percival} S.,  {James} P.,   {Maguire} K.,  2014, \mn@doi [\mnras] {10.1093/mnras/stu1995}, \href {https://ui.adsabs.harvard.edu/abs/2014MNRAS.445.4427A} {445, 4427}

\bibitem[\protect\citeauthoryear{{Ashall}, {Mazzali}, {Pian}  \& {James}}{{Ashall} et~al.}{2016}]{Ashall16}
{Ashall} C.,  {Mazzali} P.~A.,  {Pian} E.,   {James} P.~A.,  2016, \mn@doi [\mnras] {10.1093/mnras/stw2114}, \href {https://ui.adsabs.harvard.edu/abs/2016MNRAS.463.1891A} {463, 1891}

\bibitem[\protect\citeauthoryear{{Ashall} et~al.,}{{Ashall} et~al.}{2019}]{Ashall19}
{Ashall} C.,  et~al., 2019, \mn@doi [\apjl] {10.3847/2041-8213/ab1654}, \href {https://ui.adsabs.harvard.edu/abs/2019ApJ...875L..14A} {875, L14}

\bibitem[\protect\citeauthoryear{{Ashall} et~al.,}{{Ashall} et~al.}{2024}]{Ashall24}
{Ashall} C.,  et~al., 2024, \mn@doi [\apj] {10.3847/1538-4357/ad6608}, \href {https://ui.adsabs.harvard.edu/abs/2024ApJ...975..203A} {975, 203}

\bibitem[\protect\citeauthoryear{{Baer-Way} et~al.,}{{Baer-Way} et~al.}{2024}]{Baer-Way24}
{Baer-Way} R.,  et~al., 2024, \mn@doi [\apj] {10.3847/1538-4357/ad2175}, \href {https://ui.adsabs.harvard.edu/abs/2024ApJ...964..172B} {964, 172}

\bibitem[\protect\citeauthoryear{{Barnes} \& {Kasen}}{{Barnes} \& {Kasen}}{2013}]{BarnesKasen13}
{Barnes} J.,  {Kasen} D.,  2013, \mn@doi [\apj] {10.1088/0004-637X/775/1/18}, \href {https://ui.adsabs.harvard.edu/abs/2013ApJ...775...18B} {775, 18}

\bibitem[\protect\citeauthoryear{{Barnes}, {Kasen}, {Wu}  \& {Mart{\'\i}nez-Pinedo}}{{Barnes} et~al.}{2016}]{Barnes16}
{Barnes} J.,  {Kasen} D.,  {Wu} M.-R.,   {Mart{\'\i}nez-Pinedo} G.,  2016, \mn@doi [\apj] {10.3847/0004-637X/829/2/110}, \href {https://ui.adsabs.harvard.edu/abs/2016ApJ...829..110B} {829, 110}

\bibitem[\protect\citeauthoryear{{Barnes}, {Zhu}, {Lund}, {Sprouse}, {Vassh}, {McLaughlin}, {Mumpower}  \& {Surman}}{{Barnes} et~al.}{2021}]{Barnes21}
{Barnes} J.,  {Zhu} Y.~L.,  {Lund} K.~A.,  {Sprouse} T.~M.,  {Vassh} N.,  {McLaughlin} G.~C.,  {Mumpower} M.~R.,   {Surman} R.,  2021, \mn@doi [\apj] {10.3847/1538-4357/ac0aec}, \href {https://ui.adsabs.harvard.edu/abs/2021ApJ...918...44B} {918, 44}

\bibitem[\protect\citeauthoryear{{Bell}}{{Bell}}{1978}]{Bell78}
{Bell} A.~R.,  1978, \mn@doi [\mnras] {10.1093/mnras/182.2.147}, \href {https://ui.adsabs.harvard.edu/abs/1978MNRAS.182..147B} {182, 147}

\bibitem[\protect\citeauthoryear{{Bell}}{{Bell}}{2004}]{Bell04}
{Bell} A.~R.,  2004, \mn@doi [\mnras] {10.1111/j.1365-2966.2004.08097.x}, \href {https://ui.adsabs.harvard.edu/abs/2004MNRAS.353..550B} {353, 550}

\bibitem[\protect\citeauthoryear{{Beloborodov} \& {Li}}{{Beloborodov} \& {Li}}{2016}]{BeloborodovLi16}
{Beloborodov} A.~M.,  {Li} X.,  2016, \mn@doi [\apj] {10.3847/1538-4357/833/2/261}, \href {https://ui.adsabs.harvard.edu/abs/2016ApJ...833..261B} {833, 261}

\bibitem[\protect\citeauthoryear{{Benvenuto}, {Bersten}  \& {Nomoto}}{{Benvenuto} et~al.}{2013}]{Benvenuto13}
{Benvenuto} O.~G.,  {Bersten} M.~C.,   {Nomoto} K.,  2013, \mn@doi [\apj] {10.1088/0004-637X/762/2/74}, \href {https://ui.adsabs.harvard.edu/abs/2013ApJ...762...74B} {762, 74}

\bibitem[\protect\citeauthoryear{{Bhattacharya}}{{Bhattacharya}}{2002}]{Bhattacharya02}
{Bhattacharya} D.,  2002, \mn@doi [Journal of Astrophysics and Astronomy] {10.1007/BF02702467}, \href {https://ui.adsabs.harvard.edu/abs/2002JApA...23...67B} {23, 67}

\bibitem[\protect\citeauthoryear{{Blondin}, {Dessart}  \& {Hillier}}{{Blondin} et~al.}{2018}]{Blondin18}
{Blondin} S.,  {Dessart} L.,   {Hillier} D.~J.,  2018, \mn@doi [\mnras] {10.1093/mnras/stx3058}, \href {https://ui.adsabs.harvard.edu/abs/2018MNRAS.474.3931B} {474, 3931}

\bibitem[\protect\citeauthoryear{{Braun}, {Bonaldi}, {Bourke}, {Keane}  \& {Wagg}}{{Braun} et~al.}{2019}]{Braun19}
{Braun} R.,  {Bonaldi} A.,  {Bourke} T.,  {Keane} E.,   {Wagg} J.,  2019, \mn@doi [arXiv e-prints] {10.48550/arXiv.1912.12699}, \href {https://ui.adsabs.harvard.edu/abs/2019arXiv191212699B} {p. arXiv:1912.12699}

\bibitem[\protect\citeauthoryear{{Bret}}{{Bret}}{2007}]{BretCPC}
{Bret} A.,  2007, \mn@doi [Computer Physics Communications] {10.1016/j.cpc.2006.11.006}, \href {https://ui.adsabs.harvard.edu/abs/2007CoPhC.176..362B} {176, 362}

\bibitem[\protect\citeauthoryear{{Bret}}{{Bret}}{2009}]{bret09}
{Bret} A.,  2009, \mn@doi [\apj] {10.1088/0004-637X/699/2/990}, \href {https://ui.adsabs.harvard.edu/abs/2009ApJ...699..990B} {699, 990}

\bibitem[\protect\citeauthoryear{Bret, Gremillet  \& Dieckmann}{Bret et~al.}{2010}]{BretPoP2010}
Bret A.,  Gremillet L.,   Dieckmann M.~E.,  2010, \mn@doi [Physics of Plasmas] {http://dx.doi.org/10.1063/1.3514586}, 17, 120501

\bibitem[\protect\citeauthoryear{{Camacho-Neves} et~al.,}{{Camacho-Neves} et~al.}{2023}]{Camacho-Neves23}
{Camacho-Neves} Y.,  et~al., 2023, \mn@doi [\apj] {10.3847/1538-4357/acd558}, \href {https://ui.adsabs.harvard.edu/abs/2023ApJ...951...67C} {951, 67}

\bibitem[\protect\citeauthoryear{{Caprioli} \& {Spitkovsky}}{{Caprioli} \& {Spitkovsky}}{2013}]{Caprioli13}
{Caprioli} D.,  {Spitkovsky} A.,  2013, \mn@doi [\apjl] {10.1088/2041-8205/765/1/L20}, \href {https://ui.adsabs.harvard.edu/abs/2013ApJ...765L..20C} {765, L20}

\bibitem[\protect\citeauthoryear{{Cattell} et~al.,}{{Cattell} et~al.}{2021}]{Cattell11}
{Cattell} C.,  et~al., 2021, \mn@doi [\apjl] {10.3847/2041-8213/abefdd}, \href {https://ui.adsabs.harvard.edu/abs/2021ApJ...911L..29C} {911, L29}

\bibitem[\protect\citeauthoryear{{Chan} \& {Lingenfelter}}{{Chan} \& {Lingenfelter}}{1993}]{ChanLingenfelter93}
{Chan} K.-W.,  {Lingenfelter} R.~E.,  1993, \mn@doi [\apj] {10.1086/172393}, \href {https://ui.adsabs.harvard.edu/abs/1993ApJ...405..614C} {405, 614}

\bibitem[\protect\citeauthoryear{{Chen} et~al.,}{{Chen} et~al.}{2023}]{Chen23}
{Chen} N.~M.,  et~al., 2023, \mn@doi [\apjl] {10.3847/2041-8213/acb6d8}, \href {https://ui.adsabs.harvard.edu/abs/2023ApJ...944L..28C} {944, L28}

\bibitem[\protect\citeauthoryear{{Childress} et~al.,}{{Childress} et~al.}{2015}]{Childress15}
{Childress} M.~J.,  et~al., 2015, \mn@doi [\mnras] {10.1093/mnras/stv2173}, \href {https://ui.adsabs.harvard.edu/abs/2015MNRAS.454.3816C} {454, 3816}

\bibitem[\protect\citeauthoryear{{Chomiuk} et~al.,}{{Chomiuk} et~al.}{2016}]{Chomiuk_2016}
{Chomiuk} L.,  et~al., 2016, \mn@doi [\apj] {10.3847/0004-637X/821/2/119}, \href {https://ui.adsabs.harvard.edu/abs/2016ApJ...821..119C} {821, 119}

\bibitem[\protect\citeauthoryear{{Chornock} et~al.,}{{Chornock} et~al.}{2017}]{Chornock17}
{Chornock} R.,  et~al., 2017, \mn@doi [\apjl] {10.3847/2041-8213/aa905c}, \href {https://ui.adsabs.harvard.edu/abs/2017ApJ...848L..19C} {848, L19}

\bibitem[\protect\citeauthoryear{{Ciolfi}}{{Ciolfi}}{2020}]{Ciolfi20}
{Ciolfi} R.,  2020, \mn@doi [General Relativity and Gravitation] {10.1007/s10714-020-02714-x}, \href {https://ui.adsabs.harvard.edu/abs/2020GReGr..52...59C} {52, 59}

\bibitem[\protect\citeauthoryear{{Clocchiatti}, {Wheeler}, {Benetti}  \& {Frueh}}{{Clocchiatti} et~al.}{1996}]{clocchiatti96}
{Clocchiatti} A.,  {Wheeler} J.~C.,  {Benetti} S.,   {Frueh} M.,  1996, \mn@doi [\apj] {10.1086/176919}, \href {https://ui.adsabs.harvard.edu/abs/1996ApJ...459..547C} {459, 547}

\bibitem[\protect\citeauthoryear{{Colgate}, {Petschek}  \& {Kriese}}{{Colgate} et~al.}{1980}]{Colgate80}
{Colgate} S.~A.,  {Petschek} A.~G.,   {Kriese} J.~T.,  1980, \mn@doi [\apjl] {10.1086/183239}, \href {https://ui.adsabs.harvard.edu/abs/1980ApJ...237L..81C} {237, L81}

\bibitem[\protect\citeauthoryear{{Coulter} et~al.,}{{Coulter} et~al.}{2017}]{Coulter17}
{Coulter} D.~A.,  et~al., 2017, \mn@doi [Science] {10.1126/science.aap9811}, \href {https://ui.adsabs.harvard.edu/abs/2017Sci...358.1556C} {358, 1556}

\bibitem[\protect\citeauthoryear{{Cowperthwaite} et~al.,}{{Cowperthwaite} et~al.}{2017}]{Cowperthwaite17}
{Cowperthwaite} P.~S.,  et~al., 2017, \mn@doi [\apjl] {10.3847/2041-8213/aa8fc7}, \href {https://ui.adsabs.harvard.edu/abs/2017ApJ...848L..17C} {848, L17}

\bibitem[\protect\citeauthoryear{{Crockett} et~al.,}{{Crockett} et~al.}{2008}]{Crockett08}
{Crockett} R.~M.,  et~al., 2008, \mn@doi [\mnras] {10.1111/j.1745-3933.2008.00540.x}, \href {https://ui.adsabs.harvard.edu/abs/2008MNRAS.391L...5C} {391, L5}

\bibitem[\protect\citeauthoryear{{Crusius} \& {Schlickeiser}}{{Crusius} \& {Schlickeiser}}{1986}]{CRUSIUS1986}
{Crusius} A.,  {Schlickeiser} R.,  1986, Astronomy and Astrophysics, 164, L16

\bibitem[\protect\citeauthoryear{{Davis}, {Rueda-Becerril}  \& {Giannios}}{{Davis} et~al.}{2024}]{Davis2024}
{Davis} Z.,  {Rueda-Becerril} J.~M.,   {Giannios} D.,  2024, \mn@doi [\apj] {10.3847/1538-4357/ad8bc2}, \href {https://ui.adsabs.harvard.edu/abs/2024ApJ...976..182D} {976, 182}

\bibitem[\protect\citeauthoryear{{Desai} et~al.,}{{Desai} et~al.}{2023}]{Desai23}
{Desai} D.~D.,  et~al., 2023, \mn@doi [\mnras] {10.1093/mnras/stad1932}, \href {https://ui.adsabs.harvard.edu/abs/2023MNRAS.524..767D} {524, 767}

\bibitem[\protect\citeauthoryear{{Desai} et~al.,}{{Desai} et~al.}{2024}]{Desai24}
{Desai} D.~D.,  et~al., 2024, \mn@doi [\mnras] {10.1093/mnras/stae606}, \href {https://ui.adsabs.harvard.edu/abs/2024MNRAS.530.5016D} {530, 5016}

\bibitem[\protect\citeauthoryear{{Dessart} \& {Hillier}}{{Dessart} \& {Hillier}}{2020}]{Dessart20}
{Dessart} L.,  {Hillier} D.~J.,  2020, \mn@doi [\aap] {10.1051/0004-6361/202038148}, \href {https://ui.adsabs.harvard.edu/abs/2020A&A...642A..33D} {642, A33}

\bibitem[\protect\citeauthoryear{{Dessart}, {Hillier}, {Livne}, {Yoon}, {Woosley}, {Waldman}  \& {Langer}}{{Dessart} et~al.}{2011}]{Dessart11}
{Dessart} L.,  {Hillier} D.~J.,  {Livne} E.,  {Yoon} S.-C.,  {Woosley} S.,  {Waldman} R.,   {Langer} N.,  2011, \mn@doi [\mnras] {10.1111/j.1365-2966.2011.18598.x}, \href {https://ui.adsabs.harvard.edu/abs/2011MNRAS.414.2985D} {414, 2985}

\bibitem[\protect\citeauthoryear{{Dessart}, {Hillier}, {Sukhbold}, {Woosley}  \& {Janka}}{{Dessart} et~al.}{2021}]{Dessart21}
{Dessart} L.,  {Hillier} D.~J.,  {Sukhbold} T.,  {Woosley} S.~E.,   {Janka} H.~T.,  2021, \mn@doi [\aap] {10.1051/0004-6361/202141927}, \href {https://ui.adsabs.harvard.edu/abs/2021A&A...656A..61D} {656, A61}

\bibitem[\protect\citeauthoryear{{Dessart}, {Guti{\'e}rrez}, {Kuncarayakti}, {Fox}  \& {Filippenko}}{{Dessart} et~al.}{2023a}]{Dessart23}
{Dessart} L.,  {Guti{\'e}rrez} C.~P.,  {Kuncarayakti} H.,  {Fox} O.~D.,   {Filippenko} A.~V.,  2023a, \mn@doi [\aap] {10.1051/0004-6361/202345969}, \href {https://ui.adsabs.harvard.edu/abs/2023A&A...675A..33D} {675, A33}

\bibitem[\protect\citeauthoryear{{Dessart}, {Hillier}, {Woosley}  \& {Kuncarayakti}}{{Dessart} et~al.}{2023b}]{Dessart23b}
{Dessart} L.,  {Hillier} D.~J.,  {Woosley} S.~E.,   {Kuncarayakti} H.,  2023b, \mn@doi [\aap] {10.1051/0004-6361/202346626}, \href {https://ui.adsabs.harvard.edu/abs/2023A&A...677A...7D} {677, A7}

\bibitem[\protect\citeauthoryear{{Drewes}, {McDonald}, {Sablon}  \& {Vitagliano}}{{Drewes} et~al.}{2022}]{Drewes22}
{Drewes} M.,  {McDonald} J.,  {Sablon} L.,   {Vitagliano} E.,  2022, \mn@doi [\apj] {10.3847/1538-4357/ac7874}, \href {https://ui.adsabs.harvard.edu/abs/2022ApJ...934...99D} {934, 99}

\bibitem[\protect\citeauthoryear{{Drout} et~al.,}{{Drout} et~al.}{2017}]{Drout17}
{Drout} M.~R.,  et~al., 2017, \mn@doi [Science] {10.1126/science.aaq0049}, \href {https://ui.adsabs.harvard.edu/abs/2017Sci...358.1570D} {358, 1570}

\bibitem[\protect\citeauthoryear{{Ertini} et~al.,}{{Ertini} et~al.}{2023}]{Ertini23}
{Ertini} K.,  et~al., 2023, \mn@doi [\mnras] {10.1093/mnras/stad2705}, \href {https://ui.adsabs.harvard.edu/abs/2023MNRAS.526..279E} {526, 279}

\bibitem[\protect\citeauthoryear{{Ferrario}, {de Martino}  \& {G{\"a}nsicke}}{{Ferrario} et~al.}{2015}]{Ferrario15}
{Ferrario} L.,  {de Martino} D.,   {G{\"a}nsicke} B.~T.,  2015, \mn@doi [\ssr] {10.1007/s11214-015-0152-0}, \href {https://ui.adsabs.harvard.edu/abs/2015SSRv..191..111F} {191, 111}

\bibitem[\protect\citeauthoryear{{Filippenko}}{{Filippenko}}{1997}]{filippenko97}
{Filippenko} A.~V.,  1997, \mn@doi [\araa] {10.1146/annurev.astro.35.1.309}, \href {https://ui.adsabs.harvard.edu/abs/1997ARA\&A..35..309F} {35, 309}

\bibitem[\protect\citeauthoryear{{Fl{\"o}rs} et~al.,}{{Fl{\"o}rs} et~al.}{2020}]{Flors20}
{Fl{\"o}rs} A.,  et~al., 2020, \mn@doi [\mnras] {10.1093/mnras/stz3013}, \href {https://ui.adsabs.harvard.edu/abs/2020MNRAS.491.2902F} {491, 2902}

\bibitem[\protect\citeauthoryear{{Foglizzo} et~al.,}{{Foglizzo} et~al.}{2015}]{Foglizzo15}
{Foglizzo} T.,  et~al., 2015, \mn@doi [\pasa] {10.1017/pasa.2015.9}, \href {https://ui.adsabs.harvard.edu/abs/2015PASA...32....9F} {32, e009}

\bibitem[\protect\citeauthoryear{{Fontaine}, {Brassard}  \& {Bergeron}}{{Fontaine} et~al.}{2001}]{Fontaine01}
{Fontaine} G.,  {Brassard} P.,   {Bergeron} P.,  2001, \mn@doi [\pasp] {10.1086/319535}, \href {https://ui.adsabs.harvard.edu/abs/2001PASP..113..409F} {113, 409}

\bibitem[\protect\citeauthoryear{{Garasev} \& {Derishev}}{{Garasev} \& {Derishev}}{2016}]{GarasevDerishev16}
{Garasev} M.,  {Derishev} E.,  2016, \mn@doi [\mnras] {10.1093/mnras/stw1345}, \href {https://ui.adsabs.harvard.edu/abs/2016MNRAS.461..641G} {461, 641}

\bibitem[\protect\citeauthoryear{{Georgy}, {Ekstr{\"o}m}, {Meynet}, {Massey}, {Levesque}, {Hirschi}, {Eggenberger}  \& {Maeder}}{{Georgy} et~al.}{2012}]{Georgy12}
{Georgy} C.,  {Ekstr{\"o}m} S.,  {Meynet} G.,  {Massey} P.,  {Levesque} E.~M.,  {Hirschi} R.,  {Eggenberger} P.,   {Maeder} A.,  2012, \mn@doi [\aap] {10.1051/0004-6361/201118340}, \href {https://ui.adsabs.harvard.edu/abs/2012A&A...542A..29G} {542, A29}

\bibitem[\protect\citeauthoryear{{Golant}, {Vanthieghem}, {Gro{\v{s}}elj}  \& {Sironi}}{{Golant} et~al.}{2025}]{Golant25}
{Golant} R.,  {Vanthieghem} A.,  {Gro{\v{s}}elj} D.,   {Sironi} L.,  2025, \mn@doi [\apj] {10.3847/1538-4357/add404}, \href {https://ui.adsabs.harvard.edu/abs/2025ApJ...986..211G} {986, 211}

\bibitem[\protect\citeauthoryear{{Goriely}, {Bauswein}  \& {Janka}}{{Goriely} et~al.}{2011}]{Goriely11}
{Goriely} S.,  {Bauswein} A.,   {Janka} H.-T.,  2011, \mn@doi [\apjl] {10.1088/2041-8205/738/2/L32}, \href {https://ui.adsabs.harvard.edu/abs/2011ApJ...738L..32G} {738, L32}

\bibitem[\protect\citeauthoryear{{Gould}}{{Gould}}{1979}]{gould1979}
{Gould} R.~J.,  1979, \aap, \href {https://ui.adsabs.harvard.edu/abs/1979A&A....76..306G} {76, 306}

\bibitem[\protect\citeauthoryear{{Gr{\"a}fener}, {Owocki}  \& {Vink}}{{Gr{\"a}fener} et~al.}{2012}]{Grafener12}
{Gr{\"a}fener} G.,  {Owocki} S.~P.,   {Vink} J.~S.,  2012, \mn@doi [\aap] {10.1051/0004-6361/201117497}, \href {https://ui.adsabs.harvard.edu/abs/2012A&A...538A..40G} {538, A40}

\bibitem[\protect\citeauthoryear{{Graur}}{{Graur}}{2019}]{Graur19}
{Graur} O.,  2019, \mn@doi [\apj] {10.3847/1538-4357/aaf1c3}, \href {https://ui.adsabs.harvard.edu/abs/2019ApJ...870...14G} {870, 14}

\bibitem[\protect\citeauthoryear{{Graur}, {Zurek}, {Shara}, {Riess}, {Seitenzahl}  \& {Rest}}{{Graur} et~al.}{2016}]{Graur16}
{Graur} O.,  {Zurek} D.,  {Shara} M.~M.,  {Riess} A.~G.,  {Seitenzahl} I.~R.,   {Rest} A.,  2016, \mn@doi [\apj] {10.3847/0004-637X/819/1/31}, \href {https://ui.adsabs.harvard.edu/abs/2016ApJ...819...31G} {819, 31}

\bibitem[\protect\citeauthoryear{{Graur} et~al.,}{{Graur} et~al.}{2018}]{Graur18}
{Graur} O.,  et~al., 2018, \mn@doi [\apj] {10.3847/1538-4357/aabe25}, \href {https://ui.adsabs.harvard.edu/abs/2018ApJ...859...79G} {859, 79}

\bibitem[\protect\citeauthoryear{{Grossman}, {Korobkin}, {Rosswog}  \& {Piran}}{{Grossman} et~al.}{2014}]{Grossman14}
{Grossman} D.,  {Korobkin} O.,  {Rosswog} S.,   {Piran} T.,  2014, \mn@doi [\mnras] {10.1093/mnras/stt2503}, \href {https://ui.adsabs.harvard.edu/abs/2014MNRAS.439..757G} {439, 757}

\bibitem[\protect\citeauthoryear{{Gupta}, {Caprioli}  \& {Haggerty}}{{Gupta} et~al.}{2021}]{Gupta21}
{Gupta} S.,  {Caprioli} D.,   {Haggerty} C.~C.,  2021, \mn@doi [\apj] {10.3847/1538-4357/ac23cf}, \href {https://ui.adsabs.harvard.edu/abs/2021ApJ...923..208G} {923, 208}

\bibitem[\protect\citeauthoryear{{Guti{\'e}rrez} et~al.,}{{Guti{\'e}rrez} et~al.}{2017}]{Gutierrez17}
{Guti{\'e}rrez} C.~P.,  et~al., 2017, \mn@doi [\apj] {10.3847/1538-4357/aa8f52}, \href {https://ui.adsabs.harvard.edu/abs/2017ApJ...850...89G} {850, 89}

\bibitem[\protect\citeauthoryear{Hakobyan, Spitkovsky, Chernoglazov, Philippov, Groselj  \& Mahlmann}{Hakobyan et~al.}{2023}]{tristanv2}
Hakobyan H.,  Spitkovsky A.,  Chernoglazov A.,  Philippov A.,  Groselj D.,   Mahlmann J.,  2023, PrincetonUniversity/tristan-mp-v2: v2.6, \mn@doi{10.5281/zenodo.7566725}, \url {https://doi.org/10.5281/zenodo.7566725}

\bibitem[\protect\citeauthoryear{{Hamuy}}{{Hamuy}}{2003}]{Hamuy03}
{Hamuy} M.,  2003, \mn@doi [\apj] {10.1086/344689}, \href {https://ui.adsabs.harvard.edu/abs/2003ApJ...582..905H} {582, 905}

\bibitem[\protect\citeauthoryear{{Han}}{{Han}}{2017}]{Han17}
{Han} J.~L.,  2017, \mn@doi [\araa] {10.1146/annurev-astro-091916-055221}, \href {https://ui.adsabs.harvard.edu/abs/2017ARA&A..55..111H} {55, 111}

\bibitem[\protect\citeauthoryear{{Harris}, {Sarbadhicary}, {Chomiuk}, {Piro}, {Sand}  \& {Valenti}}{{Harris} et~al.}{2023}]{Harris23}
{Harris} C.~E.,  {Sarbadhicary} S.~K.,  {Chomiuk} L.,  {Piro} A.~L.,  {Sand} D.~J.,   {Valenti} S.,  2023, \mn@doi [\apj] {10.3847/1538-4357/acd84f}, \href {https://ui.adsabs.harvard.edu/abs/2023ApJ...952...24H} {952, 24}

\bibitem[\protect\citeauthoryear{{Hatt} et~al.,}{{Hatt} et~al.}{2024}]{Hatt24}
{Hatt} E.~J.,  et~al., 2024, \mn@doi [\mnras] {10.1093/mnras/stae2053}, \href {https://ui.adsabs.harvard.edu/abs/2024MNRAS.534.1060H} {534, 1060}

\bibitem[\protect\citeauthoryear{{Healy}, {Horiuchi}  \& {Ashall}}{{Healy} et~al.}{2024}]{Healy24}
{Healy} S.,  {Horiuchi} S.,   {Ashall} C.,  2024, \mn@doi [arXiv e-prints] {10.48550/arXiv.2412.04386}, \href {https://ui.adsabs.harvard.edu/abs/2024arXiv241204386H} {p. arXiv:2412.04386}

\bibitem[\protect\citeauthoryear{{Heger}, {Fryer}, {Woosley}, {Langer}  \& {Hartmann}}{{Heger} et~al.}{2003}]{heger03}
{Heger} A.,  {Fryer} C.~L.,  {Woosley} S.~E.,  {Langer} N.,   {Hartmann} D.~H.,  2003, \mn@doi [\apj] {10.1086/375341}, \href {https://ui.adsabs.harvard.edu/abs/2003ApJ...591..288H} {591, 288}

\bibitem[\protect\citeauthoryear{{Holmbo} et~al.,}{{Holmbo} et~al.}{2023}]{Holmbo23}
{Holmbo} S.,  et~al., 2023, \mn@doi [\aap] {10.1051/0004-6361/202245334}, \href {https://ui.adsabs.harvard.edu/abs/2023A&A...675A..83H} {675, A83}

\bibitem[\protect\citeauthoryear{{Hoyle} \& {Fowler}}{{Hoyle} \& {Fowler}}{1960}]{hoyle60}
{Hoyle} F.,  {Fowler} W.~A.,  1960, \mn@doi [\apj] {10.1086/146963}, \href {https://ui.adsabs.harvard.edu/abs/1960ApJ...132..565H} {132, 565}

\bibitem[\protect\citeauthoryear{{Hristov}, {Hoeflich}  \& {Collins}}{{Hristov} et~al.}{2021}]{Hristov21}
{Hristov} B.,  {Hoeflich} P.,   {Collins} D.~C.,  2021, \mn@doi [\apj] {10.3847/1538-4357/ac0ef8}, \href {https://ui.adsabs.harvard.edu/abs/2021ApJ...923..210H} {923, 210}

\bibitem[\protect\citeauthoryear{{Igoshev}, {Popov}  \& {Hollerbach}}{{Igoshev} et~al.}{2021}]{Igoshev21}
{Igoshev} A.~P.,  {Popov} S.~B.,   {Hollerbach} R.,  2021, \mn@doi [Universe] {10.3390/universe7090351}, \href {https://ui.adsabs.harvard.edu/abs/2021Univ....7..351I} {7, 351}

\bibitem[\protect\citeauthoryear{{Iwamoto}, {Brachwitz}, {Nomoto}, {Kishimoto}, {Umeda}, {Hix}  \& {Thielemann}}{{Iwamoto} et~al.}{1999}]{Iwamoto99}
{Iwamoto} K.,  {Brachwitz} F.,  {Nomoto} K.,  {Kishimoto} N.,  {Umeda} H.,  {Hix} W.~R.,   {Thielemann} F.-K.,  1999, \mn@doi [\apjs] {10.1086/313278}, \href {https://ui.adsabs.harvard.edu/abs/1999ApJS..125..439I} {125, 439}

\bibitem[\protect\citeauthoryear{{Jerkstrand}, {Fransson}  \& {Kozma}}{{Jerkstrand} et~al.}{2011}]{Jerkstrand11}
{Jerkstrand} A.,  {Fransson} C.,   {Kozma} C.,  2011, \mn@doi [\aap] {10.1051/0004-6361/201015937}, \href {https://ui.adsabs.harvard.edu/abs/2011A&A...530A..45J} {530, A45}

\bibitem[\protect\citeauthoryear{{Jerkstrand}, {Fransson}, {Maguire}, {Smartt}, {Ergon}  \& {Spyromilio}}{{Jerkstrand} et~al.}{2012}]{Jerkstrand12}
{Jerkstrand} A.,  {Fransson} C.,  {Maguire} K.,  {Smartt} S.,  {Ergon} M.,   {Spyromilio} J.,  2012, \mn@doi [\aap] {10.1051/0004-6361/201219528}, \href {https://ui.adsabs.harvard.edu/abs/2012A&A...546A..28J} {546, A28}

\bibitem[\protect\citeauthoryear{{Jerkstrand}, {Ergon}, {Smartt}, {Fransson}, {Sollerman}, {Taubenberger}, {Bersten}  \& {Spyromilio}}{{Jerkstrand} et~al.}{2015}]{Jerkstrand15}
{Jerkstrand} A.,  {Ergon} M.,  {Smartt} S.~J.,  {Fransson} C.,  {Sollerman} J.,  {Taubenberger} S.,  {Bersten} M.,   {Spyromilio} J.,  2015, \mn@doi [\aap] {10.1051/0004-6361/201423983}, \href {https://ui.adsabs.harvard.edu/abs/2015A&A...573A..12J} {573, A12}

\bibitem[\protect\citeauthoryear{{Jha}}{{Jha}}{2017}]{Jha17}
{Jha} S.~W.,  2017, in {Alsabti} A.~W.,  {Murdin} P.,  eds, , Handbook of Supernovae.
Springer International Publishing AG, p.~375, \mn@doi{10.1007/978-3-319-21846-5_42}

\bibitem[\protect\citeauthoryear{{Karachentsev} \& {Kashibadze}}{{Karachentsev} \& {Kashibadze}}{2006}]{KarachentsevKashibadze06}
{Karachentsev} I.~D.,  {Kashibadze} O.~G.,  2006, \mn@doi [Astrophysics] {10.1007/s10511-006-0002-6}, \href {https://ui.adsabs.harvard.edu/abs/2006Ap.....49....3K} {49, 3}

\bibitem[\protect\citeauthoryear{{Karinkuzhi}, {Mukhopadhyay}, {Wickramasinghe}  \& {Tout}}{{Karinkuzhi} et~al.}{2024}]{Karinkuzhi24}
{Karinkuzhi} D.,  {Mukhopadhyay} B.,  {Wickramasinghe} D.,   {Tout} C.~A.,  2024, \mn@doi [\mnras] {10.1093/mnras/stae829}, \href {https://ui.adsabs.harvard.edu/abs/2024MNRAS.529.4577K} {529, 4577}

\bibitem[\protect\citeauthoryear{{Kasen} \& {Barnes}}{{Kasen} \& {Barnes}}{2019}]{KasenBarnes19}
{Kasen} D.,  {Barnes} J.,  2019, \mn@doi [\apj] {10.3847/1538-4357/ab06c2}, \href {https://ui.adsabs.harvard.edu/abs/2019ApJ...876..128K} {876, 128}

\bibitem[\protect\citeauthoryear{{Kasen}, {Badnell}  \& {Barnes}}{{Kasen} et~al.}{2013}]{Kasen13}
{Kasen} D.,  {Badnell} N.~R.,   {Barnes} J.,  2013, \mn@doi [\apj] {10.1088/0004-637X/774/1/25}, \href {https://ui.adsabs.harvard.edu/abs/2013ApJ...774...25K} {774, 25}

\bibitem[\protect\citeauthoryear{{Kasen}, {Metzger}, {Barnes}, {Quataert}  \& {Ramirez-Ruiz}}{{Kasen} et~al.}{2017}]{Kasen17}
{Kasen} D.,  {Metzger} B.,  {Barnes} J.,  {Quataert} E.,   {Ramirez-Ruiz} E.,  2017, \mn@doi [\nat] {10.1038/nature24453}, \href {https://ui.adsabs.harvard.edu/abs/2017Natur.551...80K} {551, 80}

\bibitem[\protect\citeauthoryear{{Kasliwal} et~al.,}{{Kasliwal} et~al.}{2017}]{Kasliwal17}
{Kasliwal} M.~M.,  et~al., 2017, \mn@doi [Science] {10.1126/science.aap9455}, \href {https://ui.adsabs.harvard.edu/abs/2017Sci...358.1559K} {358, 1559}

\bibitem[\protect\citeauthoryear{{Kaspi} \& {Beloborodov}}{{Kaspi} \& {Beloborodov}}{2017}]{KaspiBeloborodov17}
{Kaspi} V.~M.,  {Beloborodov} A.~M.,  2017, \mn@doi [\araa] {10.1146/annurev-astro-081915-023329}, \href {https://ui.adsabs.harvard.edu/abs/2017ARA&A..55..261K} {55, 261}

\bibitem[\protect\citeauthoryear{{Kilpatrick} et~al.,}{{Kilpatrick} et~al.}{2017}]{Kilpatrick17}
{Kilpatrick} C.~D.,  et~al., 2017, \mn@doi [Science] {10.1126/science.aaq0073}, \href {https://ui.adsabs.harvard.edu/abs/2017Sci...358.1583K} {358, 1583}

\bibitem[\protect\citeauthoryear{{Kiuchi}, {Kyutoku}, {Sekiguchi}, {Shibata}  \& {Wada}}{{Kiuchi} et~al.}{2014}]{Kiuchi14}
{Kiuchi} K.,  {Kyutoku} K.,  {Sekiguchi} Y.,  {Shibata} M.,   {Wada} T.,  2014, \mn@doi [\prd] {10.1103/PhysRevD.90.041502}, \href {https://ui.adsabs.harvard.edu/abs/2014PhRvD..90d1502K} {90, 041502}

\bibitem[\protect\citeauthoryear{{Kiuchi}, {Cerd{\'a}-Dur{\'a}n}, {Kyutoku}, {Sekiguchi}  \& {Shibata}}{{Kiuchi} et~al.}{2015}]{Kiuchi15}
{Kiuchi} K.,  {Cerd{\'a}-Dur{\'a}n} P.,  {Kyutoku} K.,  {Sekiguchi} Y.,   {Shibata} M.,  2015, \mn@doi [\prd] {10.1103/PhysRevD.92.124034}, \href {https://ui.adsabs.harvard.edu/abs/2015PhRvD..92l4034K} {92, 124034}

\bibitem[\protect\citeauthoryear{{Korobkin}, {Rosswog}, {Arcones}  \& {Winteler}}{{Korobkin} et~al.}{2012}]{Korobkin12}
{Korobkin} O.,  {Rosswog} S.,  {Arcones} A.,   {Winteler} C.,  2012, \mn@doi [\mnras] {10.1111/j.1365-2966.2012.21859.x}, \href {https://ui.adsabs.harvard.edu/abs/2012MNRAS.426.1940K} {426, 1940}

\bibitem[\protect\citeauthoryear{{K{\"u}lebi}, {Jordan}, {Euchner}, {G{\"a}nsicke}  \& {Hirsch}}{{K{\"u}lebi} et~al.}{2009}]{Kulebi09}
{K{\"u}lebi} B.,  {Jordan} S.,  {Euchner} F.,  {G{\"a}nsicke} B.~T.,   {Hirsch} H.,  2009, \mn@doi [\aap] {10.1051/0004-6361/200912570}, \href {https://ui.adsabs.harvard.edu/abs/2009A&A...506.1341K} {506, 1341}

\bibitem[\protect\citeauthoryear{{Kushnir} \& {Waxman}}{{Kushnir} \& {Waxman}}{2020}]{KushnirWaxman20}
{Kushnir} D.,  {Waxman} E.,  2020, \mn@doi [\mnras] {10.1093/mnras/staa690}, \href {https://ui.adsabs.harvard.edu/abs/2020MNRAS.493.5617K} {493, 5617}

\bibitem[\protect\citeauthoryear{{Lamb} \& {Boutloukos}}{{Lamb} \& {Boutloukos}}{2008}]{LambBoutloukos08}
{Lamb} F.~K.,  {Boutloukos} S.,  2008, in {Milone} E.~F.,  {Leahy} D.~A.,   {Hobill} D.~W.,  eds,  Astrophysics and Space Science Library Vol. 352, Astrophysics and Space Science Library. Springer, p.~87 (\mn@eprint {arXiv} {0705.0155}), \mn@doi{10.1007/978-1-4020-6544-6_5}

\bibitem[\protect\citeauthoryear{{Lattimer} \& {Prakash}}{{Lattimer} \& {Prakash}}{2004}]{Lattimer04}
{Lattimer} J.~M.,  {Prakash} M.,  2004, \mn@doi [Science] {10.1126/science.1090720}, \href {https://ui.adsabs.harvard.edu/abs/2004Sci...304..536L} {304, 536}

\bibitem[\protect\citeauthoryear{{Leloudas} et~al.,}{{Leloudas} et~al.}{2009}]{Leloudas09}
{Leloudas} G.,  et~al., 2009, \mn@doi [\aap] {10.1051/0004-6361/200912364}, \href {https://ui.adsabs.harvard.edu/abs/2009A&A...505..265L} {505, 265}

\bibitem[\protect\citeauthoryear{{Levesque}}{{Levesque}}{2010}]{Levesque10}
{Levesque} E.~M.,  2010, \mn@doi [\nar] {10.1016/j.newar.2009.10.002}, \href {https://ui.adsabs.harvard.edu/abs/2010NewAR..54....1L} {54, 1}

\bibitem[\protect\citeauthoryear{{Levesque}}{{Levesque}}{2017}]{Levesque17}
{Levesque} E.~M.,  2017, {Astrophysics of Red Supergiants}.
IOP Publishing, \mn@doi{10.1088/978-0-7503-1329-2}

\bibitem[\protect\citeauthoryear{{Levesque}, {Massey}, {Olsen}, {Plez}, {Josselin}, {Maeder}  \& {Meynet}}{{Levesque} et~al.}{2005}]{Levesque05}
{Levesque} E.~M.,  {Massey} P.,  {Olsen} K.~A.~G.,  {Plez} B.,  {Josselin} E.,  {Maeder} A.,   {Meynet} G.,  2005, \mn@doi [\apj] {10.1086/430901}, \href {https://ui.adsabs.harvard.edu/abs/2005ApJ...628..973L} {628, 973}

\bibitem[\protect\citeauthoryear{{Li} \& {Paczy{\'n}ski}}{{Li} \& {Paczy{\'n}ski}}{1998}]{LiPaczynski98}
{Li} L.-X.,  {Paczy{\'n}ski} B.,  1998, \mn@doi [\apjl] {10.1086/311680}, \href {https://ui.adsabs.harvard.edu/abs/1998ApJ...507L..59L} {507, L59}

\bibitem[\protect\citeauthoryear{{Li}, {Hillier}  \& {Dessart}}{{Li} et~al.}{2012}]{Li12}
{Li} C.,  {Hillier} D.~J.,   {Dessart} L.,  2012, \mn@doi [\mnras] {10.1111/j.1365-2966.2012.21198.x}, \href {https://ui.adsabs.harvard.edu/abs/2012MNRAS.426.1671L} {426, 1671}

\bibitem[\protect\citeauthoryear{{Li} et~al.,}{{Li} et~al.}{2019}]{Li19}
{Li} W.,  et~al., 2019, \mn@doi [\apj] {10.3847/1538-4357/ab2b49}, \href {https://ui.adsabs.harvard.edu/abs/2019ApJ...882...30L} {882, 30}

\bibitem[\protect\citeauthoryear{{Liu}, {Modjaz}, {Bianco}  \& {Graur}}{{Liu} et~al.}{2016}]{Liu16}
{Liu} Y.-Q.,  {Modjaz} M.,  {Bianco} F.~B.,   {Graur} O.,  2016, \mn@doi [\apj] {10.3847/0004-637X/827/2/90}, \href {https://ui.adsabs.harvard.edu/abs/2016ApJ...827...90L} {827, 90}

\bibitem[\protect\citeauthoryear{{Liu} et~al.,}{{Liu} et~al.}{2023}]{Liu23}
{Liu} J.,  et~al., 2023, \mn@doi [\mnras] {10.1093/mnras/stad2851}, \href {https://ui.adsabs.harvard.edu/abs/2023MNRAS.526.1268L} {526, 1268}

\bibitem[\protect\citeauthoryear{{Magkotsios}, {Timmes}, {Hungerford}, {Fryer}, {Young}  \& {Wiescher}}{{Magkotsios} et~al.}{2010}]{Magkotsios10}
{Magkotsios} G.,  {Timmes} F.~X.,  {Hungerford} A.~L.,  {Fryer} C.~L.,  {Young} P.~A.,   {Wiescher} M.,  2010, \mn@doi [\apjs] {10.1088/0067-0049/191/1/66}, \href {https://ui.adsabs.harvard.edu/abs/2010ApJS..191...66M} {191, 66}

\bibitem[\protect\citeauthoryear{{Maoz}, {Mannucci}  \& {Nelemans}}{{Maoz} et~al.}{2014}]{Maoz14}
{Maoz} D.,  {Mannucci} F.,   {Nelemans} G.,  2014, \mn@doi [\araa] {10.1146/annurev-astro-082812-141031}, \href {https://ui.adsabs.harvard.edu/abs/2014ARA&A..52..107M} {52, 107}

\bibitem[\protect\citeauthoryear{{Martinez} et~al.,}{{Martinez} et~al.}{2022}]{Martinez22}
{Martinez} L.,  et~al., 2022, \mn@doi [\aap] {10.1051/0004-6361/202142076}, \href {https://ui.adsabs.harvard.edu/abs/2022A&A...660A..41M} {660, A41}

\bibitem[\protect\citeauthoryear{{Matheson}, {Filippenko}, {Li}, {Leonard}  \& {Shields}}{{Matheson} et~al.}{2001}]{matheson01}
{Matheson} T.,  {Filippenko} A.~V.,  {Li} W.,  {Leonard} D.~C.,   {Shields} J.~C.,  2001, \mn@doi [\aj] {10.1086/319390}, \href {https://ui.adsabs.harvard.edu/abs/2001AJ....121.1648M} {121, 1648}

\bibitem[\protect\citeauthoryear{{Mathias} et~al.,}{{Mathias} et~al.}{2018}]{Mathias18}
{Mathias} P.,  et~al., 2018, \mn@doi [\aap] {10.1051/0004-6361/201732542}, \href {https://ui.adsabs.harvard.edu/abs/2018A&A...615A.116M} {615, A116}

\bibitem[\protect\citeauthoryear{{Maund}}{{Maund}}{2017}]{Maund17}
{Maund} J.~R.,  2017, \mn@doi [\mnras] {10.1093/mnras/stx879}, \href {https://ui.adsabs.harvard.edu/abs/2017MNRAS.469.2202M} {469, 2202}

\bibitem[\protect\citeauthoryear{{Maund}, {Smartt}, {Kudritzki}, {Podsiadlowski}  \& {Gilmore}}{{Maund} et~al.}{2004}]{Maund04}
{Maund} J.~R.,  {Smartt} S.~J.,  {Kudritzki} R.~P.,  {Podsiadlowski} P.,   {Gilmore} G.~F.,  2004, \mn@doi [\nat] {10.1038/nature02161}, \href {https://ui.adsabs.harvard.edu/abs/2004Natur.427..129M} {427, 129}

\bibitem[\protect\citeauthoryear{{Maund} et~al.,}{{Maund} et~al.}{2011}]{Maund11}
{Maund} J.~R.,  et~al., 2011, \mn@doi [\apjl] {10.1088/2041-8205/739/2/L37}, \href {https://ui.adsabs.harvard.edu/abs/2011ApJ...739L..37M} {739, L37}

\bibitem[\protect\citeauthoryear{{Mazzali}, {R{\"o}pke}, {Benetti}  \& {Hillebrandt}}{{Mazzali} et~al.}{2007}]{Mazzali07}
{Mazzali} P.~A.,  {R{\"o}pke} F.~K.,  {Benetti} S.,   {Hillebrandt} W.,  2007, \mn@doi [Science] {10.1126/science.1136259}, \href {https://ui.adsabs.harvard.edu/abs/2007Sci...315..825M} {315, 825}

\bibitem[\protect\citeauthoryear{{Mazzali} et~al.,}{{Mazzali} et~al.}{2015}]{Mazzali15}
{Mazzali} P.~A.,  et~al., 2015, \mn@doi [\mnras] {10.1093/mnras/stv761}, \href {https://ui.adsabs.harvard.edu/abs/2015MNRAS.450.2631M} {450, 2631}

\bibitem[\protect\citeauthoryear{{McCully} et~al.,}{{McCully} et~al.}{2017}]{McCully17}
{McCully} C.,  et~al., 2017, \mn@doi [\apjl] {10.3847/2041-8213/aa9111}, \href {https://ui.adsabs.harvard.edu/abs/2017ApJ...848L..32M} {848, L32}

\bibitem[\protect\citeauthoryear{{Mera Evans}, {Hoeflich}  \& {Diehl}}{{Mera Evans} et~al.}{2022}]{MeraEvans22}
{Mera Evans} T.~B.,  {Hoeflich} P.,   {Diehl} R.,  2022, \mn@doi [\apj] {10.3847/1538-4357/ac5253}, \href {https://ui.adsabs.harvard.edu/abs/2022ApJ...930..107M} {930, 107}

\bibitem[\protect\citeauthoryear{{Metzger} et~al.,}{{Metzger} et~al.}{2010}]{Metzger10}
{Metzger} B.~D.,  et~al., 2010, \mn@doi [\mnras] {10.1111/j.1365-2966.2010.16864.x}, \href {https://ui.adsabs.harvard.edu/abs/2010MNRAS.406.2650M} {406, 2650}

\bibitem[\protect\citeauthoryear{{Metzger}, {Bauswein}, {Goriely}  \& {Kasen}}{{Metzger} et~al.}{2015}]{Metzger15}
{Metzger} B.~D.,  {Bauswein} A.,  {Goriely} S.,   {Kasen} D.,  2015, \mn@doi [\mnras] {10.1093/mnras/stu2225}, \href {https://ui.adsabs.harvard.edu/abs/2015MNRAS.446.1115M} {446, 1115}

\bibitem[\protect\citeauthoryear{{Miller}, {Lamb}  \& {Psaltis}}{{Miller} et~al.}{1998}]{Miller98}
{Miller} M.~C.,  {Lamb} F.~K.,   {Psaltis} D.,  1998, \mn@doi [\apj] {10.1086/306408}, \href {https://ui.adsabs.harvard.edu/abs/1998ApJ...508..791M} {508, 791}

\bibitem[\protect\citeauthoryear{{Milne}, {The}  \& {Leising}}{{Milne} et~al.}{1999}]{Milne99}
{Milne} P.~A.,  {The} L.~S.,   {Leising} M.~D.,  1999, \mn@doi [\apjs] {10.1086/313262}, \href {https://ui.adsabs.harvard.edu/abs/1999ApJS..124..503M} {124, 503}

\bibitem[\protect\citeauthoryear{{Modjaz} et~al.,}{{Modjaz} et~al.}{2009}]{Modjaz09}
{Modjaz} M.,  et~al., 2009, \mn@doi [\apj] {10.1088/0004-637X/702/1/226}, \href {https://ui.adsabs.harvard.edu/abs/2009ApJ...702..226M} {702, 226}

\bibitem[\protect\citeauthoryear{{M{\"u}ller}}{{M{\"u}ller}}{2020}]{Muller20}
{M{\"u}ller} B.,  2020, \mn@doi [Living Reviews in Computational Astrophysics] {10.1007/s41115-020-0008-5}, \href {https://ui.adsabs.harvard.edu/abs/2020LRCA....6....3M} {6, 3}

\bibitem[\protect\citeauthoryear{{M{\"u}ller}, {Prieto}, {Pejcha}  \& {Clocchiatti}}{{M{\"u}ller} et~al.}{2017}]{Muller17}
{M{\"u}ller} T.,  {Prieto} J.~L.,  {Pejcha} O.,   {Clocchiatti} A.,  2017, \mn@doi [\apj] {10.3847/1538-4357/aa72f1}, \href {https://ui.adsabs.harvard.edu/abs/2017ApJ...841..127M} {841, 127}

\bibitem[\protect\citeauthoryear{{Murphy} et~al.,}{{Murphy} et~al.}{2018}]{Murphy18}
{Murphy} E.~J.,  et~al., 2018, in {Murphy} E.,  ed.,  Astronomical Society of the Pacific Conference Series Vol. 517, Science with a Next Generation Very Large Array. p.~3 (\mn@eprint {arXiv} {1810.07524}), \mn@doi{10.48550/arXiv.1810.07524}

\bibitem[\protect\citeauthoryear{{Nagao} et~al.,}{{Nagao} et~al.}{2024}]{Nagao24}
{Nagao} T.,  et~al., 2024, \mn@doi [\aap] {10.1051/0004-6361/202346715}, \href {https://ui.adsabs.harvard.edu/abs/2024A&A...681A..11N} {681, A11}

\bibitem[\protect\citeauthoryear{{Nicholl} et~al.,}{{Nicholl} et~al.}{2017}]{Nicholl17}
{Nicholl} M.,  et~al., 2017, \mn@doi [\apjl] {10.3847/2041-8213/aa9029}, \href {https://ui.adsabs.harvard.edu/abs/2017ApJ...848L..18N} {848, L18}

\bibitem[\protect\citeauthoryear{{Palenzuela}, {Aguilera-Miret}, {Carrasco}, {Ciolfi}, {Kalinani}, {Kastaun}, {Mi{\~n}ano}  \& {Vigan{\`o}}}{{Palenzuela} et~al.}{2022}]{Palenzuela22}
{Palenzuela} C.,  {Aguilera-Miret} R.,  {Carrasco} F.,  {Ciolfi} R.,  {Kalinani} J.~V.,  {Kastaun} W.,  {Mi{\~n}ano} B.,   {Vigan{\`o}} D.,  2022, \mn@doi [\prd] {10.1103/PhysRevD.106.023013}, \href {https://ui.adsabs.harvard.edu/abs/2022PhRvD.106b3013P} {106, 023013}

\bibitem[\protect\citeauthoryear{{Pan} et~al.,}{{Pan} et~al.}{2024}]{Pan24}
{Pan} Y.~C.,  et~al., 2024, \mn@doi [\mnras] {10.1093/mnras/stae1618}, \href {https://ui.adsabs.harvard.edu/abs/2024MNRAS.532.1887P} {532, 1887}

\bibitem[\protect\citeauthoryear{{Pauldrach}, {Vanbeveren}  \& {Hoffmann}}{{Pauldrach} et~al.}{2012}]{pauldrach12}
{Pauldrach} A.~W.~A.,  {Vanbeveren} D.,   {Hoffmann} T.~L.,  2012, \mn@doi [\aap] {10.1051/0004-6361/201117621}, \href {https://ui.adsabs.harvard.edu/abs/2012A\&A...538A..75P} {538, A75}

\bibitem[\protect\citeauthoryear{{Penney} \& {Hoeflich}}{{Penney} \& {Hoeflich}}{2014}]{Penney14}
{Penney} R.,  {Hoeflich} P.,  2014, \mn@doi [\apj] {10.1088/0004-637X/795/1/84}, \href {https://ui.adsabs.harvard.edu/abs/2014ApJ...795...84P} {795, 84}

\bibitem[\protect\citeauthoryear{{Piro} \& {Kollmeier}}{{Piro} \& {Kollmeier}}{2018}]{PiroKollmeier18}
{Piro} A.~L.,  {Kollmeier} J.~A.,  2018, \mn@doi [\apj] {10.3847/1538-4357/aaaab3}, \href {https://ui.adsabs.harvard.edu/abs/2018ApJ...855..103P} {855, 103}

\bibitem[\protect\citeauthoryear{{Piro} \& {Morozova}}{{Piro} \& {Morozova}}{2016}]{Piro16}
{Piro} A.~L.,  {Morozova} V.~S.,  2016, \mn@doi [\apj] {10.3847/0004-637X/826/1/96}, \href {https://ui.adsabs.harvard.edu/abs/2016ApJ...826...96P} {826, 96}

\bibitem[\protect\citeauthoryear{{Podsiadlowski}, {Hsu}, {Joss}  \& {Ross}}{{Podsiadlowski} et~al.}{1993}]{podsiadlowski93}
{Podsiadlowski} P.,  {Hsu} J.~J.~L.,  {Joss} P.~C.,   {Ross} R.~R.,  1993, \mn@doi [\nat] {10.1038/364509a0}, \href {https://ui.adsabs.harvard.edu/abs/1993Natur.364..509P} {364, 509}

\bibitem[\protect\citeauthoryear{{Podsiadlowski}, {Langer}, {Poelarends}, {Rappaport}, {Heger}  \& {Pfahl}}{{Podsiadlowski} et~al.}{2004}]{podsiadlowski04}
{Podsiadlowski} P.,  {Langer} N.,  {Poelarends} A.~J.~T.,  {Rappaport} S.,  {Heger} A.,   {Pfahl} E.,  2004, \mn@doi [\apj] {10.1086/421713}, \href {https://ui.adsabs.harvard.edu/abs/2004ApJ...612.1044P} {612, 1044}

\bibitem[\protect\citeauthoryear{{Popov}}{{Popov}}{1993}]{Popov93}
{Popov} D.~V.,  1993, \mn@doi [\apj] {10.1086/173117}, \href {https://ui.adsabs.harvard.edu/abs/1993ApJ...414..712P} {414, 712}

\bibitem[\protect\citeauthoryear{{Prentice} et~al.,}{{Prentice} et~al.}{2016}]{Prentice16}
{Prentice} S.~J.,  et~al., 2016, \mn@doi [\mnras] {10.1093/mnras/stw299}, \href {https://ui.adsabs.harvard.edu/abs/2016MNRAS.458.2973P} {458, 2973}

\bibitem[\protect\citeauthoryear{{Psaltis} \& {Chakrabarty}}{{Psaltis} \& {Chakrabarty}}{1999}]{Psaltis99}
{Psaltis} D.,  {Chakrabarty} D.,  1999, \mn@doi [\apj] {10.1086/307525}, \href {https://ui.adsabs.harvard.edu/abs/1999ApJ...521..332P} {521, 332}

\bibitem[\protect\citeauthoryear{{Rauscher}, {Heger}, {Hoffman}  \& {Woosley}}{{Rauscher} et~al.}{2002}]{Rauscher02}
{Rauscher} T.,  {Heger} A.,  {Hoffman} R.~D.,   {Woosley} S.~E.,  2002, \mn@doi [\apj] {10.1086/341728}, \href {https://ui.adsabs.harvard.edu/abs/2002ApJ...576..323R} {576, 323}

\bibitem[\protect\citeauthoryear{Razin}{Razin}{1960}]{razin1960}
Razin V.,  1960, Izvestiya Vysshikh Uchebnykh Zavedenii. Radiofizika, 3, 921

\bibitem[\protect\citeauthoryear{{Reisenegger}}{{Reisenegger}}{2001}]{Reisenegger01}
{Reisenegger} A.,  2001, in {Mathys} G.,  {Solanki} S.~K.,   {Wickramasinghe} D.~T.,  eds,  Astronomical Society of the Pacific Conference Series Vol. 248, Magnetic Fields Across the Hertzsprung-Russell Diagram. p.~469 (\mn@eprint {arXiv} {astro-ph/0103010}), \mn@doi{10.48550/arXiv.astro-ph/0103010}

\bibitem[\protect\citeauthoryear{{Reville} \& {Bell}}{{Reville} \& {Bell}}{2012}]{Reville12}
{Reville} B.,  {Bell} A.~R.,  2012, \mn@doi [\mnras] {10.1111/j.1365-2966.2011.19892.x}, \href {https://ui.adsabs.harvard.edu/abs/2012MNRAS.419.2433R} {419, 2433}

\bibitem[\protect\citeauthoryear{{Riquelme} \& {Spitkovsky}}{{Riquelme} \& {Spitkovsky}}{2011}]{Riquelme11}
{Riquelme} M.~A.,  {Spitkovsky} A.,  2011, \mn@doi [\apj] {10.1088/0004-637X/733/1/63}, \href {https://ui.adsabs.harvard.edu/abs/2011ApJ...733...63R} {733, 63}

\bibitem[\protect\citeauthoryear{{Rizzo Smith}, {Kochanek}  \& {Neustadt}}{{Rizzo Smith} et~al.}{2023}]{RizzoSmith23}
{Rizzo Smith} M.,  {Kochanek} C.~S.,   {Neustadt} J.~M.~M.,  2023, \mn@doi [\mnras] {10.1093/mnras/stad1483}, \href {https://ui.adsabs.harvard.edu/abs/2023MNRAS.523.1474R} {523, 1474}

\bibitem[\protect\citeauthoryear{{Roberg-Clark}, {Drake}, {Swisdak}  \& {Reynolds}}{{Roberg-Clark} et~al.}{2018}]{RobergClark18}
{Roberg-Clark} G.~T.,  {Drake} J.~F.,  {Swisdak} M.,   {Reynolds} C.~S.,  2018, \mn@doi [\apj] {10.3847/1538-4357/aae393}, \href {https://ui.adsabs.harvard.edu/abs/2018ApJ...867..154R} {867, 154}

\bibitem[\protect\citeauthoryear{{Roberts}, {Kasen}, {Lee}  \& {Ramirez-Ruiz}}{{Roberts} et~al.}{2011}]{Roberts11}
{Roberts} L.~F.,  {Kasen} D.,  {Lee} W.~H.,   {Ramirez-Ruiz} E.,  2011, \mn@doi [\apjl] {10.1088/2041-8205/736/1/L21}, \href {https://ui.adsabs.harvard.edu/abs/2011ApJ...736L..21R} {736, L21}

\bibitem[\protect\citeauthoryear{{Rodr{\'\i}guez}, {Meza}, {Pineda-Garc{\'\i}a}  \& {Ramirez}}{{Rodr{\'\i}guez} et~al.}{2021}]{Rodriguez21}
{Rodr{\'\i}guez} {\'O}.,  {Meza} N.,  {Pineda-Garc{\'\i}a} J.,   {Ramirez} M.,  2021, \mn@doi [\mnras] {10.1093/mnras/stab1335}, \href {https://ui.adsabs.harvard.edu/abs/2021MNRAS.505.1742R} {505, 1742}

\bibitem[\protect\citeauthoryear{{Rodr{\'\i}guez}, {Maoz}  \& {Nakar}}{{Rodr{\'\i}guez} et~al.}{2023}]{Rodriguez23}
{Rodr{\'\i}guez} {\'O}.,  {Maoz} D.,   {Nakar} E.,  2023, \mn@doi [\apj] {10.3847/1538-4357/ace2bd}, \href {https://ui.adsabs.harvard.edu/abs/2023ApJ...955...71R} {955, 71}

\bibitem[\protect\citeauthoryear{{Rodr{\'\i}guez}, {Nakar}  \& {Maoz}}{{Rodr{\'\i}guez} et~al.}{2024}]{Rodriguez24}
{Rodr{\'\i}guez} {\'O}.,  {Nakar} E.,   {Maoz} D.,  2024, \mn@doi [\nat] {10.1038/s41586-024-07262-x}, \href {https://ui.adsabs.harvard.edu/abs/2024Natur.628..733R} {628, 733}

\bibitem[\protect\citeauthoryear{{Ruiter}}{{Ruiter}}{2020}]{Ruiter20}
{Ruiter} A.~J.,  2020, \mn@doi [IAU Symposium] {10.1017/S1743921320000587}, \href {https://ui.adsabs.harvard.edu/abs/2020IAUS..357....1R} {357, 1}

\bibitem[\protect\citeauthoryear{{Sasdelli}, {Mazzali}, {Pian}, {Nomoto}, {Hachinger}, {Cappellaro}  \& {Benetti}}{{Sasdelli} et~al.}{2014}]{Sasdelli14}
{Sasdelli} M.,  {Mazzali} P.~A.,  {Pian} E.,  {Nomoto} K.,  {Hachinger} S.,  {Cappellaro} E.,   {Benetti} S.,  2014, \mn@doi [\mnras] {10.1093/mnras/stu1777}, \href {https://ui.adsabs.harvard.edu/abs/2014MNRAS.445..711S} {445, 711}

\bibitem[\protect\citeauthoryear{{Saumon}, {Blouin}  \& {Tremblay}}{{Saumon} et~al.}{2022}]{Saumon22}
{Saumon} D.,  {Blouin} S.,   {Tremblay} P.-E.,  2022, \mn@doi [\physrep] {10.1016/j.physrep.2022.09.001}, \href {https://ui.adsabs.harvard.edu/abs/2022PhR...988....1S} {988, 1}

\bibitem[\protect\citeauthoryear{{Scalzo}, {Ruiter}  \& {Sim}}{{Scalzo} et~al.}{2014}]{Scalzo14}
{Scalzo} R.~A.,  {Ruiter} A.~J.,   {Sim} S.~A.,  2014, \mn@doi [\mnras] {10.1093/mnras/stu1808}, \href {https://ui.adsabs.harvard.edu/abs/2014MNRAS.445.2535S} {445, 2535}

\bibitem[\protect\citeauthoryear{{Scalzo} et~al.,}{{Scalzo} et~al.}{2019}]{Scalzo19}
{Scalzo} R.~A.,  et~al., 2019, \mn@doi [\mnras] {10.1093/mnras/sty3178}, \href {https://ui.adsabs.harvard.edu/abs/2019MNRAS.483..628S} {483, 628}

\bibitem[\protect\citeauthoryear{{Schmidt} et~al.,}{{Schmidt} et~al.}{2003}]{Schmidt03}
{Schmidt} G.~D.,  et~al., 2003, \mn@doi [\apj] {10.1086/377476}, \href {https://ui.adsabs.harvard.edu/abs/2003ApJ...595.1101S} {595, 1101}

\bibitem[\protect\citeauthoryear{{Schroer}, {Pezzi}, {Caprioli}, {Haggerty}  \& {Blasi}}{{Schroer} et~al.}{2021}]{Schroer21}
{Schroer} B.,  {Pezzi} O.,  {Caprioli} D.,  {Haggerty} C.,   {Blasi} P.,  2021, \mn@doi [\apjl] {10.3847/2041-8213/ac02cd}, \href {https://ui.adsabs.harvard.edu/abs/2021ApJ...914L..13S} {914, L13}

\bibitem[\protect\citeauthoryear{{Seitenzahl}, {Taubenberger}  \& {Sim}}{{Seitenzahl} et~al.}{2009}]{Seitenzahl09}
{Seitenzahl} I.~R.,  {Taubenberger} S.,   {Sim} S.~A.,  2009, \mn@doi [\mnras] {10.1111/j.1365-2966.2009.15478.x}, \href {https://ui.adsabs.harvard.edu/abs/2009MNRAS.400..531S} {400, 531}

\bibitem[\protect\citeauthoryear{{Seitenzahl} et~al.,}{{Seitenzahl} et~al.}{2013}]{Seitenzahl13}
{Seitenzahl} I.~R.,  et~al., 2013, \mn@doi [\mnras] {10.1093/mnras/sts402}, \href {https://ui.adsabs.harvard.edu/abs/2013MNRAS.429.1156S} {429, 1156}

\bibitem[\protect\citeauthoryear{{Seitenzahl}, {Timmes}  \& {Magkotsios}}{{Seitenzahl} et~al.}{2014}]{Seitenzahl14}
{Seitenzahl} I.~R.,  {Timmes} F.~X.,   {Magkotsios} G.,  2014, \mn@doi [\apj] {10.1088/0004-637X/792/1/10}, \href {https://ui.adsabs.harvard.edu/abs/2014ApJ...792...10S} {792, 10}

\bibitem[\protect\citeauthoryear{{Shappee} \& {Stanek}}{{Shappee} \& {Stanek}}{2011}]{Shappee2011}
{Shappee} B.~J.,  {Stanek} K.~Z.,  2011, \mn@doi [\apj] {10.1088/0004-637X/733/2/124}, \href {https://ui.adsabs.harvard.edu/abs/2011ApJ...733..124S} {733, 124}

\bibitem[\protect\citeauthoryear{{Shappee} et~al.,}{{Shappee} et~al.}{2017a}]{Shappee17a}
{Shappee} B.~J.,  et~al., 2017a, \mn@doi [Science] {10.1126/science.aaq0186}, \href {https://ui.adsabs.harvard.edu/abs/2017Sci...358.1574S} {358, 1574}

\bibitem[\protect\citeauthoryear{{Shappee}, {Stanek}, {Kochanek}  \& {Garnavich}}{{Shappee} et~al.}{2017b}]{Shappee17b}
{Shappee} B.~J.,  {Stanek} K.~Z.,  {Kochanek} C.~S.,   {Garnavich} P.~M.,  2017b, \mn@doi [\apj] {10.3847/1538-4357/aa6eab}, \href {https://ui.adsabs.harvard.edu/abs/2017ApJ...841...48S} {841, 48}

\bibitem[\protect\citeauthoryear{{Shultz} et~al.,}{{Shultz} et~al.}{2011}]{Shultz11}
{Shultz} M.,  et~al., 2011, in {Neiner} C.,  {Wade} G.,  {Meynet} G.,   {Peters} G.,  eds,  IAU Symposium Vol. 272, Active OB Stars: Structure, Evolution, Mass Loss, and Critical Limits. pp 212--213 (\mn@eprint {arXiv} {1009.2516}), \mn@doi{10.1017/S1743921311010398}

\bibitem[\protect\citeauthoryear{{Silverman} et~al.,}{{Silverman} et~al.}{2017}]{Silverman17}
{Silverman} J.~M.,  et~al., 2017, \mn@doi [\mnras] {10.1093/mnras/stx058}, \href {https://ui.adsabs.harvard.edu/abs/2017MNRAS.467..369S} {467, 369}

\bibitem[\protect\citeauthoryear{{Sironi} \& {Spitkovsky}}{{Sironi} \& {Spitkovsky}}{2011}]{Sironi11}
{Sironi} L.,  {Spitkovsky} A.,  2011, \mn@doi [\apj] {10.1088/0004-637X/726/2/75}, \href {https://ui.adsabs.harvard.edu/abs/2011ApJ...726...75S} {726, 75}

\bibitem[\protect\citeauthoryear{{Sironi}, {Spitkovsky}  \& {Arons}}{{Sironi} et~al.}{2013}]{Sironi13}
{Sironi} L.,  {Spitkovsky} A.,   {Arons} J.,  2013, \mn@doi [\apj] {10.1088/0004-637X/771/1/54}, \href {https://ui.adsabs.harvard.edu/abs/2013ApJ...771...54S} {771, 54}

\bibitem[\protect\citeauthoryear{{Smartt}}{{Smartt}}{2009}]{Smartt09}
{Smartt} S.~J.,  2009, \mn@doi [\araa] {10.1146/annurev-astro-082708-101737}, \href {https://ui.adsabs.harvard.edu/abs/2009ARA&A..47...63S} {47, 63}

\bibitem[\protect\citeauthoryear{{Smartt}}{{Smartt}}{2015}]{Smartt15}
{Smartt} S.~J.,  2015, \mn@doi [\pasa] {10.1017/pasa.2015.17}, \href {https://ui.adsabs.harvard.edu/abs/2015PASA...32...16S} {32, e016}

\bibitem[\protect\citeauthoryear{{Smartt} et~al.,}{{Smartt} et~al.}{2017}]{Smartt17}
{Smartt} S.~J.,  et~al., 2017, \mn@doi [\nat] {10.1038/nature24303}, \href {https://ui.adsabs.harvard.edu/abs/2017Natur.551...75S} {551, 75}

\bibitem[\protect\citeauthoryear{{Smith}, {Hinkle}  \& {Ryde}}{{Smith} et~al.}{2009}]{Smith09}
{Smith} N.,  {Hinkle} K.~H.,   {Ryde} N.,  2009, \mn@doi [\aj] {10.1088/0004-6256/137/3/3558}, \href {https://ui.adsabs.harvard.edu/abs/2009AJ....137.3558S} {137, 3558}

\bibitem[\protect\citeauthoryear{{Soares-Santos} et~al.,}{{Soares-Santos} et~al.}{2017}]{Soares-Santos17}
{Soares-Santos} M.,  et~al., 2017, \mn@doi [\apjl] {10.3847/2041-8213/aa9059}, \href {https://ui.adsabs.harvard.edu/abs/2017ApJ...848L..16S} {848, L16}

\bibitem[\protect\citeauthoryear{{Spitkovsky}}{{Spitkovsky}}{2005}]{Spitkovsky05}
{Spitkovsky} A.,  2005, in {Bulik} T.,  {Rudak} B.,   {Madejski} G.,  eds,  American Institute of Physics Conference Series Vol. 801, Astrophysical Sources of High Energy Particles and Radiation. pp 345--350 (\mn@eprint {arXiv} {astro-ph/0603211}), \mn@doi{10.1063/1.2141897}

\bibitem[\protect\citeauthoryear{{Steiner}, {Lattimer}  \& {Brown}}{{Steiner} et~al.}{2013}]{Steiner13}
{Steiner} A.~W.,  {Lattimer} J.~M.,   {Brown} E.~F.,  2013, \mn@doi [\apjl] {10.1088/2041-8205/765/1/L5}, \href {https://ui.adsabs.harvard.edu/abs/2013ApJ...765L...5S} {765, L5}

\bibitem[\protect\citeauthoryear{{Stritzinger} \& {Sollerman}}{{Stritzinger} \& {Sollerman}}{2007}]{Stritzinger07}
{Stritzinger} M.,  {Sollerman} J.,  2007, \mn@doi [\aap] {10.1051/0004-6361:20066999}, \href {https://ui.adsabs.harvard.edu/abs/2007A&A...470L...1S} {470, L1}

\bibitem[\protect\citeauthoryear{{Svensson}}{{Svensson}}{1982}]{Svensson82}
{Svensson} R.,  1982, \mn@doi [\apj] {10.1086/160081}, \href {https://ui.adsabs.harvard.edu/abs/1982ApJ...258..321S} {258, 321}

\bibitem[\protect\citeauthoryear{{Taddia} et~al.,}{{Taddia} et~al.}{2018}]{Taddia18}
{Taddia} F.,  et~al., 2018, \mn@doi [\aap] {10.1051/0004-6361/201730844}, \href {https://ui.adsabs.harvard.edu/abs/2018A&A...609A.136T} {609, A136}

\bibitem[\protect\citeauthoryear{{Tanaka} \& {Hotokezaka}}{{Tanaka} \& {Hotokezaka}}{2013}]{TanakaHotokezaka13}
{Tanaka} M.,  {Hotokezaka} K.,  2013, \mn@doi [\apj] {10.1088/0004-637X/775/2/113}, \href {https://ui.adsabs.harvard.edu/abs/2013ApJ...775..113T} {775, 113}

\bibitem[\protect\citeauthoryear{{Tanaka} et~al.,}{{Tanaka} et~al.}{2017}]{Tanaka17}
{Tanaka} M.,  et~al., 2017, \mn@doi [\pasj] {10.1093/pasj/psx121}, \href {https://ui.adsabs.harvard.edu/abs/2017PASJ...69..102T} {69, 102}

\bibitem[\protect\citeauthoryear{{Tanvir} et~al.,}{{Tanvir} et~al.}{2017}]{Tanvir17}
{Tanvir} N.~R.,  et~al., 2017, \mn@doi [\apjl] {10.3847/2041-8213/aa90b6}, \href {https://ui.adsabs.harvard.edu/abs/2017ApJ...848L..27T} {848, L27}

\bibitem[\protect\citeauthoryear{{Tessore}, {L{\`e}bre}, {Morin}, {Mathias}, {Josselin}  \& {Auri{\`e}re}}{{Tessore} et~al.}{2017}]{Tessore17}
{Tessore} B.,  {L{\`e}bre} A.,  {Morin} J.,  {Mathias} P.,  {Josselin} E.,   {Auri{\`e}re} M.,  2017, \mn@doi [\aap] {10.1051/0004-6361/201730473}, \href {https://ui.adsabs.harvard.edu/abs/2017A&A...603A.129T} {603, A129}

\bibitem[\protect\citeauthoryear{{Thielemann}, {Nomoto}  \& {Hashimoto}}{{Thielemann} et~al.}{1996}]{Thielemann96}
{Thielemann} F.-K.,  {Nomoto} K.,   {Hashimoto} M.-A.,  1996, \mn@doi [\apj] {10.1086/176980}, \href {https://ui.adsabs.harvard.edu/abs/1996ApJ...460..408T} {460, 408}

\bibitem[\protect\citeauthoryear{{Tiwari}, {Graur}, {Fisher}, {Seitenzahl}, {Leung}, {Nomoto}, {Perets}  \& {Shen}}{{Tiwari} et~al.}{2022}]{Tiwari22}
{Tiwari} V.,  {Graur} O.,  {Fisher} R.,  {Seitenzahl} I.,  {Leung} S.-C.,  {Nomoto} K.,  {Perets} H.~B.,   {Shen} K.,  2022, \mn@doi [\mnras] {10.1093/mnras/stac1618}, \href {https://ui.adsabs.harvard.edu/abs/2022MNRAS.515.3703T} {515, 3703}

\bibitem[\protect\citeauthoryear{{Tremblay} et~al.,}{{Tremblay} et~al.}{2017}]{Tremblay17}
{Tremblay} P.~E.,  et~al., 2017, \mn@doi [\mnras] {10.1093/mnras/stw2854}, \href {https://ui.adsabs.harvard.edu/abs/2017MNRAS.465.2849T} {465, 2849}

\bibitem[\protect\citeauthoryear{{Tucker}, {Shappee}, {Kochanek}, {Stanek}, {Ashall}, {Anand}  \& {Garnavich}}{{Tucker} et~al.}{2022}]{Tucker22}
{Tucker} M.~A.,  {Shappee} B.~J.,  {Kochanek} C.~S.,  {Stanek} K.~Z.,  {Ashall} C.,  {Anand} G.~S.,   {Garnavich} P.,  2022, \mn@doi [\mnras] {10.1093/mnras/stac2873}, \href {https://ui.adsabs.harvard.edu/abs/2022MNRAS.517.4119T} {517, 4119}

\bibitem[\protect\citeauthoryear{{Utrobin}}{{Utrobin}}{2004}]{Utrobin04}
{Utrobin} V.~P.,  2004, \mn@doi [Astronomy Letters] {10.1134/1.1738152}, \href {https://ui.adsabs.harvard.edu/abs/2004AstL...30..293U} {30, 293}

\bibitem[\protect\citeauthoryear{{Van Dyk} et~al.,}{{Van Dyk} et~al.}{2011}]{VanDyk11}
{Van Dyk} S.~D.,  et~al., 2011, \mn@doi [\apjl] {10.1088/2041-8205/741/2/L28}, \href {https://ui.adsabs.harvard.edu/abs/2011ApJ...741L..28V} {741, L28}

\bibitem[\protect\citeauthoryear{{Van Dyk} et~al.,}{{Van Dyk} et~al.}{2012a}]{VanDyk12a}
{Van Dyk} S.~D.,  et~al., 2012a, \mn@doi [\aj] {10.1088/0004-6256/143/1/19}, \href {https://ui.adsabs.harvard.edu/abs/2012AJ....143...19V} {143, 19}

\bibitem[\protect\citeauthoryear{{Van Dyk} et~al.,}{{Van Dyk} et~al.}{2012b}]{VanDyk12b}
{Van Dyk} S.~D.,  et~al., 2012b, \mn@doi [\apj] {10.1088/0004-637X/756/2/131}, \href {https://ui.adsabs.harvard.edu/abs/2012ApJ...756..131V} {756, 131}

\bibitem[\protect\citeauthoryear{{Van Dyk} et~al.,}{{Van Dyk} et~al.}{2014}]{VanDyk14}
{Van Dyk} S.~D.,  et~al., 2014, \mn@doi [\aj] {10.1088/0004-6256/147/2/37}, \href {https://ui.adsabs.harvard.edu/abs/2014AJ....147...37V} {147, 37}

\bibitem[\protect\citeauthoryear{{Vasylyev} et~al.,}{{Vasylyev} et~al.}{2023}]{Vasylyev23}
{Vasylyev} S.~S.,  et~al., 2023, \mn@doi [\apjl] {10.3847/2041-8213/acf1a3}, \href {https://ui.adsabs.harvard.edu/abs/2023ApJ...955L..37V} {955, L37}

\bibitem[\protect\citeauthoryear{{Vlasov}}{{Vlasov}}{1961}]{Vlasov61}
{Vlasov} A.~A.,  1961, {Many-particle theory and its application to plasma}.
Gordon and Breach

\bibitem[\protect\citeauthoryear{{Weibel}}{{Weibel}}{1959}]{Weibel59}
{Weibel} E.~S.,  1959, \mn@doi [\prl] {10.1103/PhysRevLett.2.83}, \href {https://ui.adsabs.harvard.edu/abs/1959PhRvL...2...83W} {2, 83}

\bibitem[\protect\citeauthoryear{{Wellstein} \& {Langer}}{{Wellstein} \& {Langer}}{1999}]{wellstein99}
{Wellstein} S.,  {Langer} N.,  1999, \aap, \href {https://ui.adsabs.harvard.edu/abs/1999A\&A...350..148W} {350, 148}

\bibitem[\protect\citeauthoryear{{Wellstein}, {Langer}  \& {Braun}}{{Wellstein} et~al.}{2001}]{wellstein01}
{Wellstein} S.,  {Langer} N.,   {Braun} H.,  2001, \mn@doi [\aap] {10.1051/0004-6361:20010151}, \href {https://ui.adsabs.harvard.edu/abs/2001A\&A...369..939W} {369, 939}

\bibitem[\protect\citeauthoryear{{Wilk}, {Hillier}  \& {Dessart}}{{Wilk} et~al.}{2020}]{Wilk20}
{Wilk} K.~D.,  {Hillier} D.~J.,   {Dessart} L.,  2020, \mn@doi [\mnras] {10.1093/mnras/staa640}, \href {https://ui.adsabs.harvard.edu/abs/2020MNRAS.494.2221W} {494, 2221}

\bibitem[\protect\citeauthoryear{{Woosley} \& {Weaver}}{{Woosley} \& {Weaver}}{1986}]{WoosleyWeaver86}
{Woosley} S.~E.,  {Weaver} T.~A.,  1986, \mn@doi [\araa] {10.1146/annurev.aa.24.090186.001225}, \href {https://ui.adsabs.harvard.edu/abs/1986ARA&A..24..205W} {24, 205}

\bibitem[\protect\citeauthoryear{{Woosley} \& {Weaver}}{{Woosley} \& {Weaver}}{1995}]{WoosleyWeaver95}
{Woosley} S.~E.,  {Weaver} T.~A.,  1995, \mn@doi [\apjs] {10.1086/192237}, \href {https://ui.adsabs.harvard.edu/abs/1995ApJS..101..181W} {101, 181}

\bibitem[\protect\citeauthoryear{{Woosley}, {Langer}  \& {Weaver}}{{Woosley} et~al.}{1995}]{woosley95}
{Woosley} S.~E.,  {Langer} N.,   {Weaver} T.~A.,  1995, \mn@doi [\apj] {10.1086/175963}, \href {https://ui.adsabs.harvard.edu/abs/1995ApJ...448..315W} {448, 315}

\bibitem[\protect\citeauthoryear{{Woosley}, {Heger}  \& {Weaver}}{{Woosley} et~al.}{2002}]{Woosley02}
{Woosley} S.~E.,  {Heger} A.,   {Weaver} T.~A.,  2002, \mn@doi [Reviews of Modern Physics] {10.1103/RevModPhys.74.1015}, \href {https://ui.adsabs.harvard.edu/abs/2002RvMP...74.1015W} {74, 1015}

\bibitem[\protect\citeauthoryear{{Yoon}}{{Yoon}}{2017}]{Yoon17}
{Yoon} S.-C.,  2017, \mn@doi [\mnras] {10.1093/mnras/stx1496}, \href {https://ui.adsabs.harvard.edu/abs/2017MNRAS.470.3970Y} {470, 3970}

\bibitem[\protect\citeauthoryear{{Yoon}, {Gr{\"a}fener}, {Vink}, {Kozyreva}  \& {Izzard}}{{Yoon} et~al.}{2012}]{Yoon12}
{Yoon} S.~C.,  {Gr{\"a}fener} G.,  {Vink} J.~S.,  {Kozyreva} A.,   {Izzard} R.~G.,  2012, \mn@doi [\aap] {10.1051/0004-6361/201219790}, \href {https://ui.adsabs.harvard.edu/abs/2012A&A...544L..11Y} {544, L11}

\bibitem[\protect\citeauthoryear{{Zhang} et~al.,}{{Zhang} et~al.}{2020}]{Zhang20}
{Zhang} K.~D.,  et~al., 2020, \mn@doi [\mnras] {10.1093/mnras/staa3191}, \href {https://ui.adsabs.harvard.edu/abs/2020MNRAS.499.5325Z} {499, 5325}

\bibitem[\protect\citeauthoryear{{de la Chevroti{\`e}re}, {St-Louis}, {Moffat}  \& {MiMeS Collaboration}}{{de la Chevroti{\`e}re} et~al.}{2014}]{delaChevrotiere14}
{de la Chevroti{\`e}re} A.,  {St-Louis} N.,  {Moffat} A.~F.~J.,   {MiMeS Collaboration} 2014, \mn@doi [\apj] {10.1088/0004-637X/781/2/73}, \href {https://ui.adsabs.harvard.edu/abs/2014ApJ...781...73D} {781, 73}

\bibitem[\protect\citeauthoryear{{van Haarlem} et~al.,}{{van Haarlem} et~al.}{2013}]{vanHaarlem13}
{van Haarlem} M.~P.,  et~al., 2013, \mn@doi [\aap] {10.1051/0004-6361/201220873}, \href {https://ui.adsabs.harvard.edu/abs/2013A&A...556A...2V} {556, A2}

\makeatother
\end{thebibliography}




\appendix
\section{Assumption of beam-like positrons} \label{app:beam_ass}
In Section~\ref{sec:PIC}, we have assumed a simplified geometry where a stream of high-velocity positrons propagates into a medium of ions and electrons. This is not a fundamental representation of the physical system, but a deliberate simplifying assumption to explore the most extreme case of positron escape. This configuration explores the consequences of having high-velocity positrons trying to escape through the region of nuclear decay. In reality, positrons from radioactive decay are emitted isotropically. However, a radially expanding shell of positrons will naturally lead to localized regions at the shell's leading edge where their distribution appears beam-like. This anisotropy would drive streaming instabilities predominantly at these edges.

While a detailed treatment of the full distribution function is beyond the scope of this work, we conducted additional simulations to explore the impact of isotropic positron distributions on the magnetic field. In the first simulation, $50\%$ of the positrons were beam-like, while the remaining $50\%$ followed a relativistic Maxwell-J\"uttner distribution, with a temperature approximated from the decay energy ($k_B T_0 = m_e c^2$). In the second simulation, all positrons were initialized isotropically, according to the Maxwell-J\"uttner distribution, approximating isotropic emission. 

\begin{figure}
    \centering
    \includegraphics[width=1\linewidth]{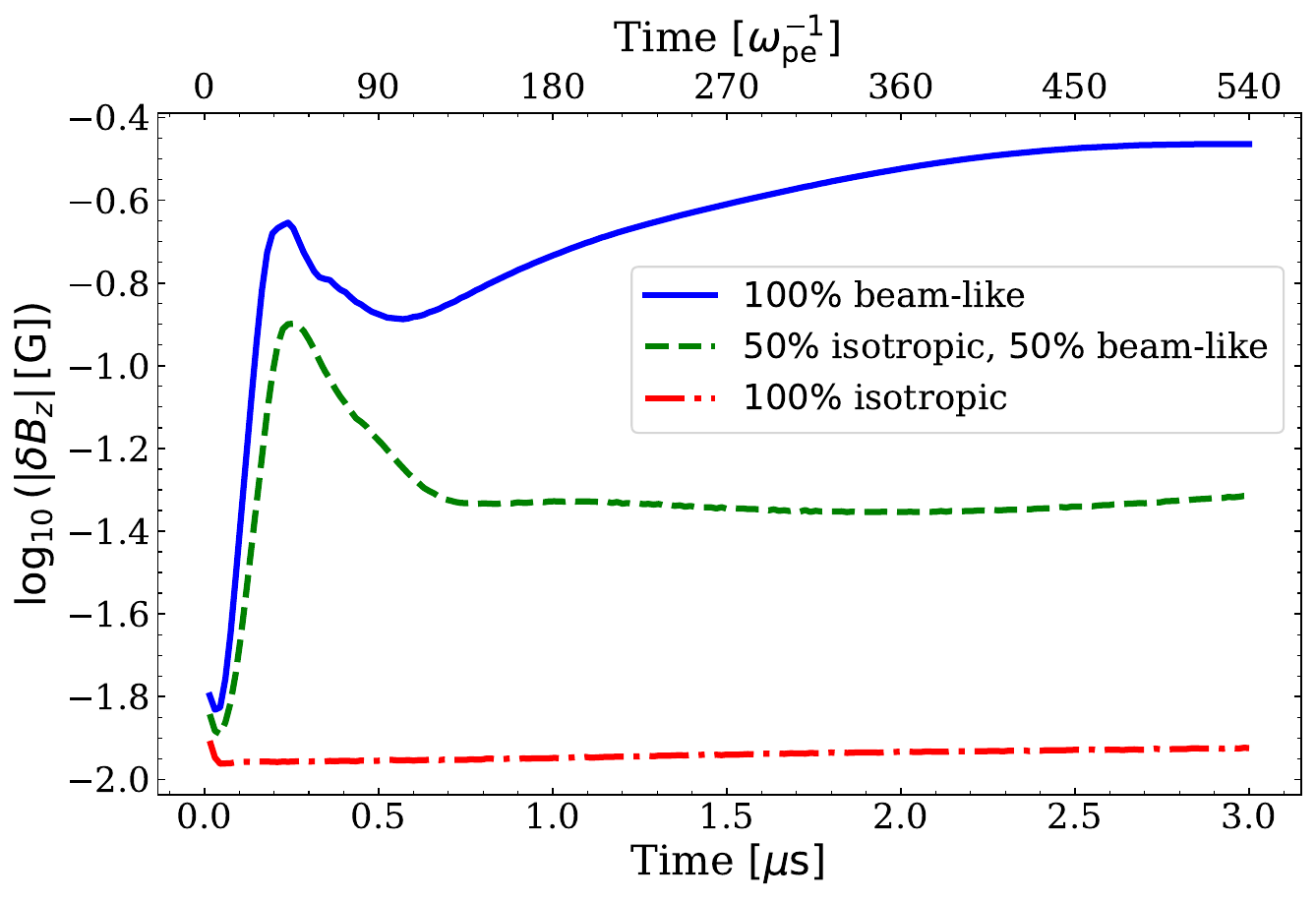}
    \caption{Comparison of the beam-like positron assumption with (i) all positrons beam-like (solid blue line, identical to Figure~\ref{fig:Bz_vs_t}), (ii) half beam-like and half isotropically distributed positrons (dashed green line), and (iii) fully isotropic positrons (dash-dotted red line). The isotropic positrons follow a relativistic Maxwell-J\"uttner distribution, with a temperature approximated from the decay energy ($k_B T_0 = m_e c^2$).}
    \label{fig:beam_like}
\end{figure}

The results, shown in Figure~\ref{fig:beam_like}, indicate that as long as a non-zero fraction of positrons remains beam-like, the streaming instability develops and peaks at a magnetic field strength proportional to the anisotropy. The growth rate is comparable, albeit slightly slower, when only a fraction of the positrons are beam-like. Conversely, when all positrons are isotropic, the instability does not develop, confirming the critical role of anisotropy in driving the instability.

\section{Prompt Synchrotron Emission from SNe~Ia Positrons} \label{app:synch}
\begin{figure}
    \centering
    \includegraphics[width=1\linewidth]{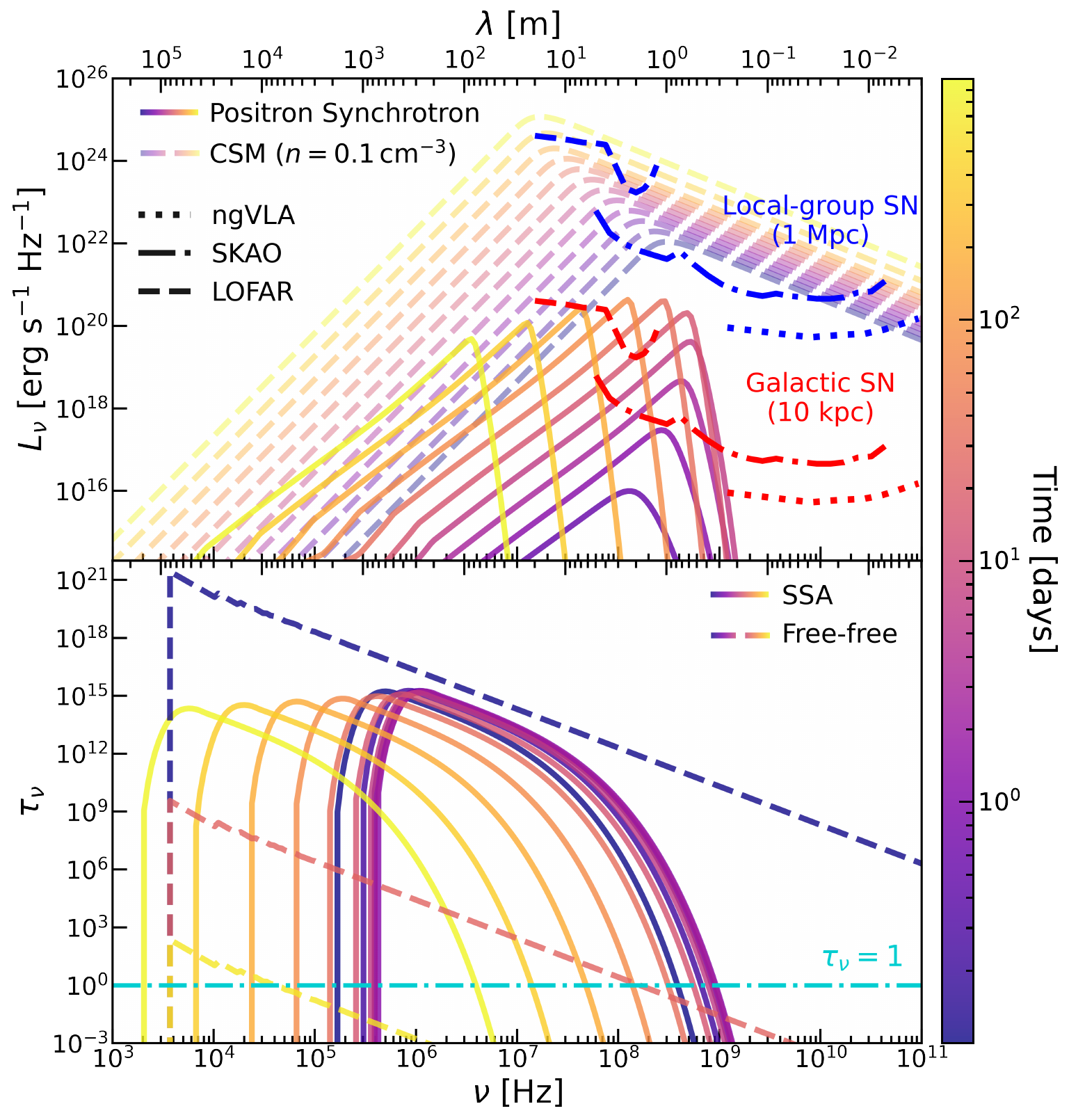}
    \caption{\textit{Top}: Synchrotron emission spectral luminosity ($L_\nu$) from SNe~Ia positrons as a function of frequency ($\nu$) for different times (solid lines) and synchrotron emission from shocked CSM using lower-frequency extension of models from \citet[dashed lines;][]{Chomiuk_2016}. Overlaid are the sensitivity curves for ngVLA (dotted line), SKAO (dash-dotted line), and LOFAR (dashed line), assuming 10 hours of integration time. Sensitivity curves are shown for two example distances in different colors: a local-group SN at $1\,\mathrm{Mpc}$ and a Galactic SN at $10\,\mathrm{kpc}$. \textit{Bottom}: Optical depth ($\tau_{\nu}$) as a function of frequency ($\nu$) at different times from SSA (solid lines) and free-free absorption (dashed lines). The color bar indicates the time in days. The cyan dash-dotted line shows where $\tau_\nu = 1$.}
    \label{fig:syn_sed_tau}
\end{figure}

Relativistic leptons produce synchrotron emission as they interact with magnetic fields. Previous studies have searched for synchrotron emission in nearby SNe~Ia from ejecta interaction with CSM \citep[e.g.,][]{Chomiuk_2016,Harris23}. Taking the assumptions directly from \citet{Chomiuk_2016} and extending them to lower frequencies, in Figure~\ref{fig:syn_sed_tau} top panel, we plot the modeled synchrotron emission from the shocked CSM in SNe~Ia for a modest CSM density of $n = 0.1\,\mathrm{cm^{-3}}$ in a constant density medium with a fiducial $10\%$ post-shock energy density in magnetic fields. 

Now, opening an avenue for directly detecting confined positrons from their radio luminosity, we calculate the synchrotron emission from relativistic positrons, whose distribution is shown in Figure~\ref{fig:init_final_vel_energy}. Assuming no CSM, we focus only on the positron distribution because it exhibits a robust profile that maintains a consistent qualitative appearance across all simulations and can be easily rescaled with the positron number density. The positron number density $n_p (t)$ is derived from Eq.~\ref{eq:n_p} using standard assumptions for a typical SN Ia, as detailed in Section~\ref{subsubsec:Ia_overview}. The magnetic field strength $B(t)$ follows from Eq.~\ref{eq:Bsat_vs_f_p}. To calculate the synchrotron emissivity, we assume the distribution of positrons is isotropic and integrate over the synchrotron power spectrum were the pitch angle is assumed random for each particle and averaged for the distribution \citep[$P_\nu(\nu_c,\gamma)$;][]{CRUSIUS1986},
\begin{equation}
      j_{\nu} = \frac{1}{4 \pi} \int_{1}^{\infty} {\rm d}\gamma P_{\nu}(\nu_c,\gamma) n(\gamma),
\end{equation}
where $\gamma$ is the particle's Lorentz factor and $\nu_c \equiv 3e B \gamma^2 / 4\pi m_e c$ is the synchrotron critical frequency. $P_\nu(\nu_c,\gamma)$ does not include the Razin effect \citep{razin1960}, which may significantly reduce the low frequency emission because of reduced beaming from the background plasma. To calculate the luminosity ($L_\nu$), we approximate the source as a homogeneous spherical object \citep{gould1979} with radius $R_{\mathrm{ej}} = v_{\mathrm{ej}} t$, where t is the time since explosion, and $v_{\mathrm{ej}}$ is the ejecta velocity. For this estimate, we use $v_{\mathrm{ej}} = 11{,}000\,\mathrm{km\,s^{-1}}$ as a high-end conservative estimate for the bulk of the Fe and Ni distribution. 

As the photons travel through the source, a significant portion of the spectrum is expected to be absorbed via free-free absorption or synchrotron self-absorption \citep[SSA;][]{Chomiuk_2016}. 
The luminosity per unit frequency is given by
\begin{equation}
    L_\nu = 3 \frac{f(\tau_\nu)}{\tau_\nu}\left(\frac{4 \pi }{3} \right) R_{\mathrm{ej}}^3 j_\nu,
\end{equation}
where  $\tau_\nu = 2 R_{\mathrm{ej}} \left(k_s + k_f\right)$ is the optical depth, $k_s$ is the SSA absorption coefficient, $k_f$ is the free-free absorption coefficient, and 
\begin{equation}
    f(\tau_\nu) = \frac{1}{2} + \frac{e^{-\tau_\nu}}{\tau_\nu}  - \frac{1 - e^{-\tau_\nu}}{\tau_\nu^2}
\end{equation}
is the optical depth function.
We numerically compute $L_\nu$ using \texttt{Tleco} \citep{Davis2024}, with the results shown in the top panel of Figure~\ref{fig:syn_sed_tau}. The optical depth as a function of frequency is shown in the bottom panel of Figure~\ref{fig:syn_sed_tau}.

\begin{figure}
    \centering
    \includegraphics[width=1\linewidth]{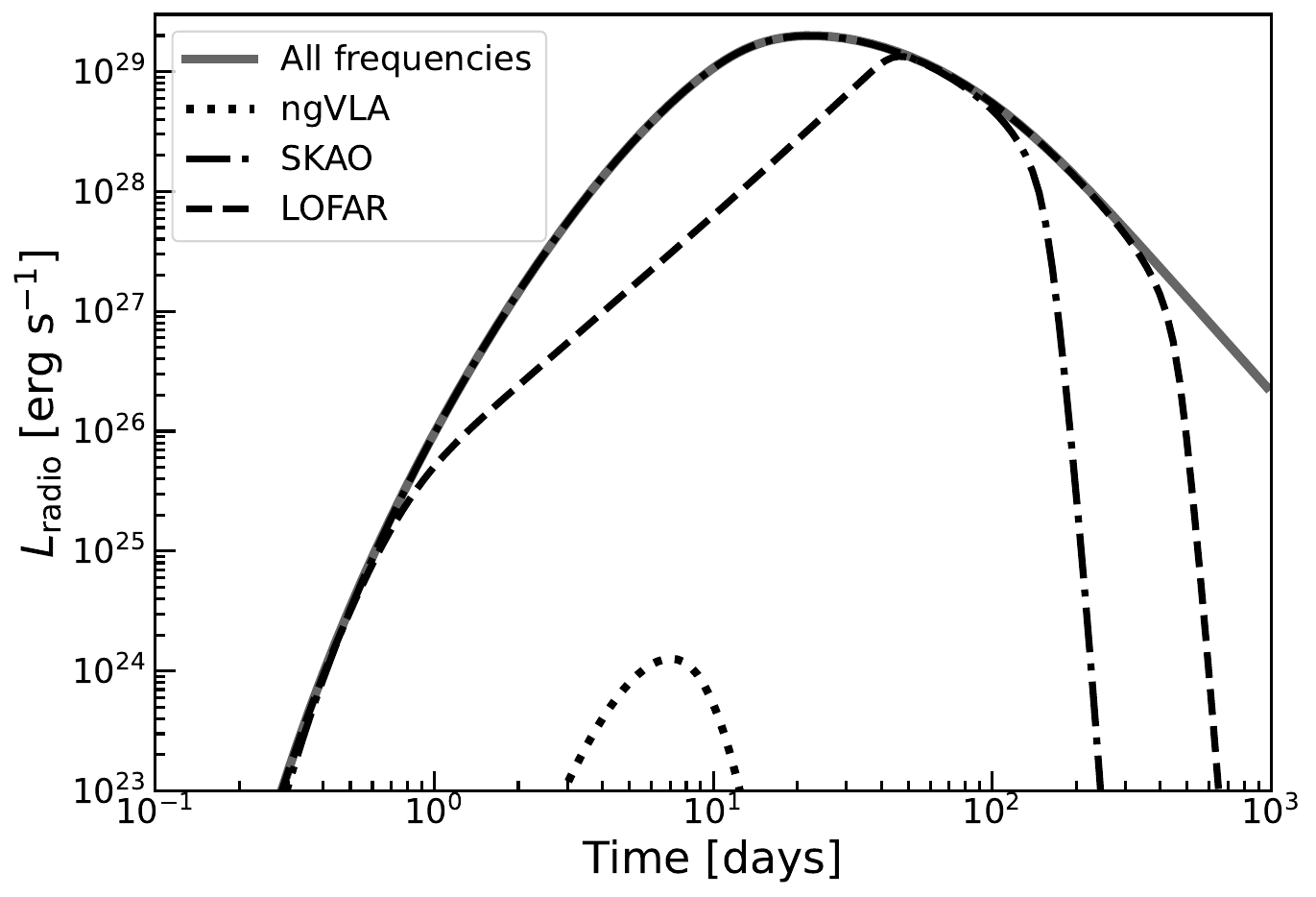}
    \caption{Frequency-integrated light curves showing the evolution of the total emitted synchrotron luminosity over time. The solid line is integration over all frequencies, while the dotted, dash-dotted, and dashed lines correspond to luminosity in the ngVLA, SKAO, and LOFAR frequency ranges, respectively.}
    \label{fig:syn_light_curves}
\end{figure}

Free-free absorption is dominant and shapes the overall spectrum at early times ($<10$ days), while SSA becomes significant at later times. Due to the heavy absorption, the spectrum peaks well above $\nu_c$ but with highly suppressed values of $L_\nu$. These radio luminosities are not excluded by the radio non-detections of SNe~Ia by \citet{Chomiuk_2016}. 
Furthermore, the simplified assumption of instantaneous positron injection and $3\,\mu s$-later distribution measurements neglects significant evolutionary processes such as diffusion and thermalization, necessitating cautious interpretation of the resulting radio $L_\nu$ estimates, particularly regarding potential positron energy overestimation. Figure~\ref{fig:syn_light_curves} shows the frequency-integrated light curve over all radio frequencies (solid line) as well as the contributions from specific frequency ranges corresponding to ngVLA and LOFAR (dotted, dashed, and dash-dotted lines). The total emission peaks around 20 days post-explosion near $\nu_{pk} \approx 1\, \mathrm{GHz}$.

To assess detectability of SNe~Ia, we show the sensitivity curves for ngVLA \citep[$1.2-116\,\mathrm{GHz}$;][]{Murphy18}\footnote{ngVLA sensitivity data were gathered from: \\\url{https://ngvla.nrao.edu/page/performance}.}, SKAO \citep[$50\,\mathrm{MHz}-50\,\mathrm{GHz}$;][]{Braun19}\footnote{SKAO sensitivity data were gathered from: \\ \url{https://www.skao.int/en/science-users/ska-tools/493/ska-sensitivity-calculators}.}, and the LOFAR \citep[][]{vanHaarlem13}, with low-band ($10-80\,\mathrm{MHz}$) and high-band ($110-250\,\mathrm{MHz}$)\footnote{LOFAR sensitivity data were gathered from: \\\url{https://science.astron.nl/telescopes/lofar/lofar-system-overview/observing-modes/lofar-imaging-capabilities-and-sensitivity/}} coverage in Figure~\ref{fig:syn_sed_tau} as well, assuming a total integration time of 10 hours. We compare the detectability for distances of $1\,\mathrm{Mpc}$ (a hypothetical local-group SN) and $10\,\mathrm{kpc}$ (a hypothetical Galactic SN). We emphasize that our calculations represent a highly optimistic scenario for synchrotron emission from high-energy positrons, and even a small amount of CSM will dominate the emission by orders of magnitude. 

Under such conditions and in extremely clean environments with no CSM, a Galactic SN could be detectable with LOFAR and SKAO with less than 10 hr integration time, and unlikely to be detected with ngVLA. With SKAO, it would be detectable from roughly $10$ days to $100$ days after explosion with 10 hr integration time. However, nearby extragalactic SNe~Ia, such as SN~2011fe \citep[$6.4$ Mpc;][]{Shappee2011} and SN~2014J \citep[$3.5$ Mpc;][]{KarachentsevKashibadze06}, are unlikely to be detected by these radio facilities, even in clean environments. While their emission is close to the ngVLA sensitivity limits at the lowest frequencies, the spectrum is expected to rise rapidly at lower frequencies, making detailed modeling crucial to these estimates. A local-group SN may only be marginally detectable with a 10 hr integration in this highly optimistic scenario with no CSM. 

It is important to reiterate that the emission results are based on specific simplifying assumptions and optimistic scenario, as detailed above. Although these results do not fully evaluate the detectability of nearby SNe~Ia, they nonetheless underscore the importance of future work to model the expected radio emission from SNe~Ia, particularly at frequencies relevant to future radio observatories. This modeling offers a promising avenue for constraining their synchrotron emission mechanisms.

\section{Free neutron decay in kilonovae} \label{app:free_n}
\citet{Metzger15} show that $\beta$-decay of free neutrons that avoid capture in the outermost layers of KN ejecta power a precursor to the main KN emission on a time scale of hours. Here, we explore the generation of magnetic fields through plasma instabilities in this neutron-rich ejecta, focusing on a simplified model with free neutron decay as the sole energy source.

Given a free neutron e-folding time of $\tau_n = 14.9\,\mathrm{min}$, the $\beta$-decay electron number density is 
\begin{equation}
    n_{\beta,n}(t) = N_{\beta,n}(t)(1 - e^{-t/\tau_n}) / V_n(t),
\end{equation}
where $N_{\beta,n}(t)$ is the total number of free neutrons given by their total mass ($M_n$), and $V_n(t)$ is the expanding volume of free neutrons with velocity $v_n$. The mass $M_n$ ranges between $10^{-5} - 10^{-4}\,\mathrm{M_\odot}$ and the speed $v_n$ ranges between $0.2 - 0.9 \, c$ \citep{Metzger15}. The ion number density $n_{i,n}(t)$ is determined directly from the free neutron mass $M_n$ and velocity $v_n$. These densities are then replaced into Eq.~\ref{eq:f_e_KN} to obtain the new $f_e(t)$ for free neutron decay. Subsequent magnetic field estimation follows the established procedure outlined in Eq.~\ref{eq:Bsat_vs_f_p}.

\begin{figure}
    \centering
    \includegraphics[width=1\linewidth]{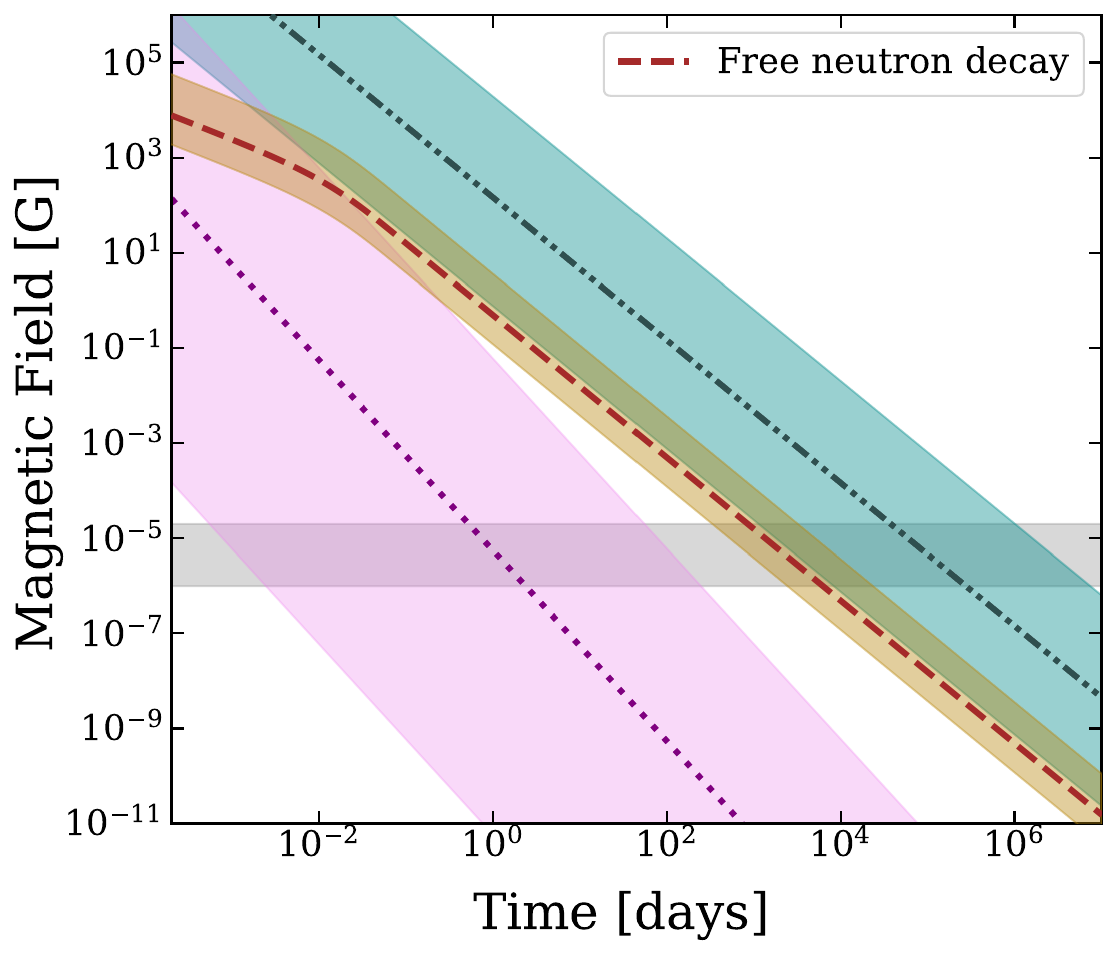}
    \caption{Same as Figure~\ref{fig:B_sat_prog}d with the addition of median (dashed brown line) and range (brown shaded region) of magnetic fields from free neutron decay.}
    \label{fig:free_n}
\end{figure}

Figure ~\ref{fig:free_n} shows that the magnetic field generated by free neutron decay is typically $2-3$ orders of magnitude weaker than that arising from $r$-process decay. Nevertheless, this field strength remains significantly above progenitor field levels. It is important to acknowledge the simplicity of this free neutron model, highlighting the need for more sophisticated future investigations.

\bsp	
\label{lastpage}
\end{document}